\documentclass[fleqn,usenatbib]{mnras}

\usepackage{newtxtext,newtxmath}

\usepackage[T1]{fontenc}
\usepackage{ae,aecompl}
\usepackage{subcaption} 
\hypersetup{draft}

\usepackage{url}
\usepackage{xspace}
\usepackage{graphicx}	
\usepackage{amsmath}	
\usepackage{amssymb}	

\newcommand{\ox}{$\left[\ion{O}{\,\sc II}\right]$\xspace}
\newcommand{\oem}{$\left[\ion{O}{\,\sc II}\right]$~emitters\xspace}
\newcommand{\gl}{{\sc galform}\xspace}

\newcommand{\Ha}{H${\alpha}$\xspace}
\newcommand{\Hb}{H${\beta}$\xspace}

\newcommand{\oiii}{$\left[\ion{O}{\,\sc III}\right]$\xspace}
\newcommand{\HII}{$\mathrm{H\textrm{\textsc{II}}}$\xspace}
\newcommand{\Mpc}{~h^{-1}{\rm Mpc}}

\newcommand{\LSE}{large-scale environment\xspace}
\newcommand{\vweb}{{\sc Vweb}\xspace}
\newcommand{\pweb}{{\sc Pweb}\xspace}

\title[Do ELGs live in filaments?]{Do model emission line galaxies live in filaments at z$\sim$1?}

\author[V. Gonzalez-Perez et al.]{
\parbox[c]{0.92\textwidth}{
\vspace{-0.8cm}
V. Gonzalez-Perez$^{1,2,3}$\thanks{E-mail: violetagp@protonmail.com},
W. Cui$^{4,5}$\thanks{E-mail: cuiweiguang@gmail.com},
S. Contreras$^{6}$,
C. M. Baugh$^{7}$, 
J. Comparat$^{8}$,
A. J. Griffin$^{7}$,
J. Helly$^{7}$,
A. Knebe$^{4,9,10}$,
C. Lacey$^{7}$,
P. Norberg$^{7,11}$.
}
\vspace*{6pt} 
\\
$^{1}$Astrophysics Research Institute, Liverpool John Moores University, 146 Brownlow Hill, Liverpool L3 5RF, UK.\\
$^{2}$Energy Lancaster, Lancaster University, Lancaster LA14YB, UK.\\
$^{3}$Institute of Cosmology \& Gravitation, University of Portsmouth, Dennis Sciama Building, Portsmouth, PO1 3FX, UK.\\
$^{4}$Departamento de F\'isica Te\'orica, M\'odulo 15, Facultad de Ciencias, Universidad Aut\'onoma de Madrid, E-28049 Madrid, Spain.\\
$^{5}$Institute for Astronomy, University of Edinburgh, Royal Observatory, EH9 3HJ Edinburgh, United Kingdom\\
$^{6}$Donostia International Physics Center (DIPC), Manuel Lardizabal Pasealekua 4, E-20018 Donostia, Basque Country, Spain.\\
$^{7}$Institute for Computational Cosmology, Department of Physics, Durham University, South Road, Durham, DH1 3LE, UK. \\
$^{8}$Max-Planck Institut fur extraterrestrische Physik, Postfach 1312, D-85741 Garching bei Munchen, Germany.\\
$^{9}$Centro de Investigaci\'{o}n Avanzada en F\'isica Fundamental, Facultad de Ciencias, Universidad Aut\'{o}noma de Madrid, 28049 Madrid, Spain.\\
$^{10}$International Centre for Radio Astronomy Research, University of Western Australia, 35 Stirling Highway, Crawley, Western Australia 6009.\\
$^{11}$ Centre for Extragalactic Astronomy, Department of Physics, Durham University, South Road, Durham, DH1 3LE, UK.\\
}

\date{Accepted XXX. Received YYY; in original form ZZZ}

\pubyear{2020}

\begin{document}
\label{firstpage}
\pagerange{\pageref{firstpage}--\pageref{lastpage}}
\maketitle

\begin{abstract}
Current and future cosmological surveys are targeting star-forming galaxies at $z\sim 1$ with nebular emission lines. 
We use a state-of-the-art semi-analytical model of galaxy formation and evolution to explore the large scale environment of star-forming emission line galaxies (ELGs). Model ELGs are selected 
such that they can be  compared directly with the DEEP2, VVDS, eBOSS-SGC and DESI surveys. The large scale environment of the ELGs is classified using  velocity-shear-tensor and  tidal-tensor algorithms. Half of the model ELGs live in filaments and about a third in sheets. Model ELGs which reside in knots have the largest satellite fractions. We find that the shape of the mean halo occupation distribution of model ELGs varies widely for different large scale environments. To interpret our results, we also study  fixed number density samples of ELGs and galaxies 
selected using simpler criteria, with single cuts in stellar mass, star formation rate and \ox luminosity. 
The fixed number density ELG selection produces samples that are close to L\ox and SFR selected samples for densities above $10^{-4.2}h^{3}{\rm Mpc}^{-3}$. ELGs with an extra cut in stellar mass applied to fix their number density, present differences in sheets and knots with respect to the other samples. ELGs, SFR and L\ox selected samples with equal number density have similar large scale bias but their clustering below separations of $1h^{-1}$Mpc is different.
\end{abstract}

\begin{keywords}
galaxies: evolution -- galaxies: formation -- cosmology: large-scale of the Universe
\end{keywords}



\section{Introduction}\label{sec:intro}


The distribution of matter in the Universe is highly inhomogeneous on megaparsec scales, on which the filamentary structure of the cosmic web arises~\citep[e.g.][]{colless2001,gott2005,Cui2018}. In the prevalent theory of hierarchical formation of structure, gas cools following the cosmic web~\citep[e.g.][]{whiterees1978}. Galaxies at different distances from the filamentary structures have been found to have different properties that cannot be explained by density alone~(e.g. \citealt{laigle2018,kraljic2018}), although this might not be a universal result~\citep[e.g.][]{goh2019}. 

Here we aim to study the large scale environment of star-forming emission line galaxies (hereafter ELGs). These galaxies have spectra characterised by strong nebular emission lines, which allow the robust determination of their redshift. Cosmological surveys have started to target ELGs to study the epoch when the expansion of the Universe first became dominated by dark energy, $z\sim 1$~\citep[e.g.][]{comparat2013elgsdss}. Understanding the connection between ELGs and their host dark matter haloes is a crucial step to maximally exploit these surveys. Cosmological surveys that have targetted or plan to target ELGs include: 
ATLAS Probe\footnote{Astrophysics Telescope for Large Area Spectroscopy, \url{http://atlas-probe.ipac.caltech.edu/} \citep{wang18}.},
DESI\footnote{Dark Energy Spectroscopic Instrument, \url{http://desi.lbl.gov/} \citep{levi13}.}, 
Euclid\footnote{\url{https://www.euclid-ec.org/} \citep{laureijs11}.},
Hetdex\footnote{Hobby-Eberly Telescope Dark Energy Experiment \url{http://hetdex.org/} \citep{hetdex}.},
MSE\footnote{MaunaKea Spectroscopic Explorer \url{https://mse.cfht.hawaii.edu/} \citep{percival2019}.},
PFS\footnote{Prime Focus Spectrograph,\url{http://sumire.ipmu.jp/en/2652} \citep{takada14}.}, 
SDSS-IV/eBOSS\footnote{extended Baryon Oscillation Spectroscopic Survey,  \url{http://www.sdss.org/surveys/eboss/} \citep{dr16}.},
WFIRST\footnote{Wide Field Infrared Survey Telescope \url{https://www.nasa.gov/wfirst} \citep{wfirst}.} WiggleZ\footnote{WiggleZ Dark Energy Survey \url{http://wigglez.swin.edu.au/site/} \citep{drinkwater2010,drinkwater18}.}, 
4MOST\footnote{4-metre Multi-Object Spectroscopic Telescope, \url{https://www.4most.eu/} \citep{deJong14}.}, etc. Most of these surveys have either optical or infrared detectors and thus, they will focus on detecting \Ha$\lambda6563$\AA, \Hb$\lambda4861$\AA, \ox$\lambda\lambda3726,3729$\AA~and \oiii$\lambda\lambda4959,5007$\AA~nebular emission lines at redshifts between 0.5 and 2~\citep[e.g.][]{sobral12}. Here we focus on \oem, which are the prevalent ELGs detected at $z\sim 1$ by optical instruments, such as those from SDSS-IV/eBOSS~\citep{comparat15eboss,raichoor17,delubac17}.

The nebular emission lines in ELGs are produced by ionised gas in the interstellar medium. The gas can be heated by either newly formed stars or by the nuclear activity following mass accretion into a supermassive black hole. In this study we focus only on star-forming ELGs, as this is the population targeted by the cosmological surveys mentioned above. Only a small fraction, less than 1 per cent, of eBOSS ELGs are expected to be AGNs~\citep{comparat2013}. Other observational studies have found that between 8 per cent to 17 per cent of their ELGs were AGNs~\citep{sobral16,valentino2017}. We will refer to star-forming ELGs simply as ELGs and thus, here we study a sub-sample of star-forming galaxies. 

Star-forming and less massive galaxies have been found closer to axis of filaments than more massive or quiescent galaxies are by a range of observational studies including: GAMA at $0.01\leq z \leq0.25$~\citep{alpaslan2016,kraljic2018}, CHILES at $z \sim0.45$~\citep{luber2019} SDSS at $z\leq0.7$~(\citealt{chen2017},~\citealt{poudel2017}), VIPERS at $z\sim0.7$~\citep{malavasi2017},  VIS$^3$COS at $z\sim 0.89$~\citep{paulino-afonso2019}, COSMOS at $z<0.9$~\citep{laigle2018}, etc. The HiZELS survey also found that galaxies in filaments at z$\sim0.53$ and z$\sim0.84$ have a low average electron density and high metallicity, compared to galaxies in the field, with a larger presence of \Ha emitters within filaments~\citep{darvish2014,darvish2015}. Thus, \oem, which are found in a large range of different environments~\citep{hayashi2020}, are also expected to be more common in filaments than in the field.

Detailed observations from MUSE suggest that filaments assist gas cooling, enhancing the star formation in galaxies~\citep{vulcani2019}. Hydrodynamical simulations have shown similar results~\citep[e.g.][]{liao2019}. These results could be a consequence of the outskirts of filaments being vorticity rich regions in simulations~\citep[e.g.][]{laigle2015}, dominated by smooth accretion~\citep[e.g.][]{kraljic2019}.  
Using cosmological hydrodynamic simulations \citep{Cui2012,Cui2014}, \citet{Cui2019} quantified that, at $z = 1$, $\sim 68$ per cent of the gas is cold. About 48 per cent of this cold gas, $T<10^5$K, was found in sheets and 28 per cent in filaments.
Furthermore, at z = 1 most halos ($\sim$70 - 100 per cent) with masses in the range ${\rm 10^{11} \lesssim M_{halo}(}h{\rm^{-1} M_\odot) \lesssim 3\times 10^{13} }$ live in filaments. Haloes in sheets are on average less massive than those found in filaments. The percentage of haloes in filaments with masses of $\sim 10^{10} h^{-1} M_\odot$ drops to $\rm \sim45\ per\ cent$ with sheet halos increase to $\sim 40$ per cent \citep{Cautun2014}. As cold, dense gas that locates in halos is needed for star formation to happen, we expect to find more model star-forming galaxies in filaments and sheets at $z=1$.

star-forming ELGs at $z \sim 1$, have been found to populate haloes of masses $\sim 10^{12}h^{-1}{\rm M}_{\odot}$~\citep{favole16,khostovan2018,guo2018} and have a linear bias around 1.5~\citep{comparat2013elglensing,guo2018}. Using our previous semi-analytical model of galaxy formation~\citep[][hereafter GP18]{gp18} we found model ELGs living in haloes with masses consistent with the observations but less clustered on large scales. As we found the percentage of model satellite ELGs to be below that in~\citet{favole16}, we argued that this could drive the differences found in the clustering of these objects. The fraction of model star-forming satellite galaxies is related to the modelling of the gas cycle (cooling, accretion, star formation, death of stars, etc). Nevertheless, the lack of assembly bias in the models used to interpret the observations of star-forming ELGs might be partly responsible for this difference~\citep{contreras2019}.

Both observations and models show that star-forming galaxies in general, and ELGs in particular, populate dark matter haloes in a different way than mass selected samples~\citep[e.g.][]{zheng05,cochrane2018eagle,favole16,guo2018,gp18,alam2019,contreras2019}.

Here our objectives are to (i) characterise how ELGs trace the cosmic web, and (ii) put the model \oem populations into context, by comparing their properties with those of star-formation rate and stellar mass selected samples. The star formation histories of galaxies are affected by their environment. In turn, this can have an impact on their relation with their host haloes and thus, their assembly bias. Our first objective, as stated above, is to start to understand these connections and their relevance both for modelling galaxy formation and in terms of their influence on the estimation of cosmological parameters. Our second objective is to contrast ELGs with other cosmological tracers. We expect the properties of star-forming ELGs to follow many of the trends found for the population of star-forming galaxies and thus, it will be useful to know when special care is needed to model these tracers, beyond assuming that a simple proxy, such as a cut in star-formation rate, is adequate.

Here we use the results from a semi-analytical model of galaxy formation and evolution. We use an updated version of the model presented in GP18. In Fig. 12 from that paper we showed that, qualitatively, model \oem trace filaments better than mass selected galaxies.The distribution of matter on large scales can be segmented into four dynamically distinct environments: knots, filaments, sheets, and voids \citep[e.g.][]{Klypin1983,Geller1989, Bond1996}. These four different cosmological structures are a natural outcome of gravitational collapse. Many methods have been developed to classify/identify these cosmological structures and we refer the reader to \citet{Libeskind2018} for a detailed description. Here, we measure the \LSE of the dark matter simulation using two algorithms, one that uses a shear tensor, \vweb, and the other a tidal one, \pweb \cite{Cui2018,Cui2019}. Both methods use the eigenvalues of the Hessian matrix for the indicator field (velocity and potential respectively) to spatially separate out these structures.

The plan of this paper is as follows. In \S~\ref{sec:model} we introduce an updated version of the \gl semi-analytical model for galaxy formation and evolution. The \vweb and \pweb methods, used to characterize the \LSE are described in \S~\ref{sec:LSE} and in Appendix~\ref{App:check}. The selection of model \oem is presented in \S~\ref{sec:modelelgs}. The results on how \oem populate the cosmic web can be found in \S~\ref{sec:elgcw}. Fixed number density samples are defined in \S~\ref{sec:nd} and their properties in different \LSE are presented in \S~\ref{sec:props}. In \S~\ref{sec:conclusions} we summarise our results.

\section{Methods}\label{sec:methods}

In this work we analyse the $z=0.83$ and $z=1$ outputs from the MS-W7 N-body simulation \citep{guo13,jiang14,gp14}. This N-body simulation is run within a box of 500 $h^{-1}$Mpc comoving side and assumes a cosmology consistent with the 7$^{\rm th}$ year release from WMAP \citep{wmap7}: matter density $\Omega_{m, 0}=0.272$, cosmological constant $\Omega_{\Lambda , 0}=0.728$, baryon density $\Omega_{b,0}=0.0455$, a normalization of density fluctuations given by $\sigma_{8, 0}=0.810$, $n_{s}=0.967$ and a Hubble constant today of $H(z=0) = 100 h\,{\rm km\, s}^{-1}{\rm Mpc}^{-1}$ with $h=0.704$.

The MS-W7 simulation has been populated with galaxies using a semi-analytical model for galaxy formation and evolution, based on that described in \citet{gp18}. This model is introduced in \S~\ref{sec:model}.

The \LSE of the whole simulation box has been classified into knots, filaments, sheets and voids using two algorithms: \vweb, which uses a velocity-shear-tensor, and \pweb, which uses a tidal-tensor \cite{Cui2018,Cui2019}. These algorithms are described in \S~\ref{sec:LSE} and their resolution and threshold setting are further discussed in Appendix \S~\ref{App:check}.

\subsection{The \gl galaxy model}\label{sec:model}
\begin{figure*}
%
\includegraphics[width=0.33\textwidth]{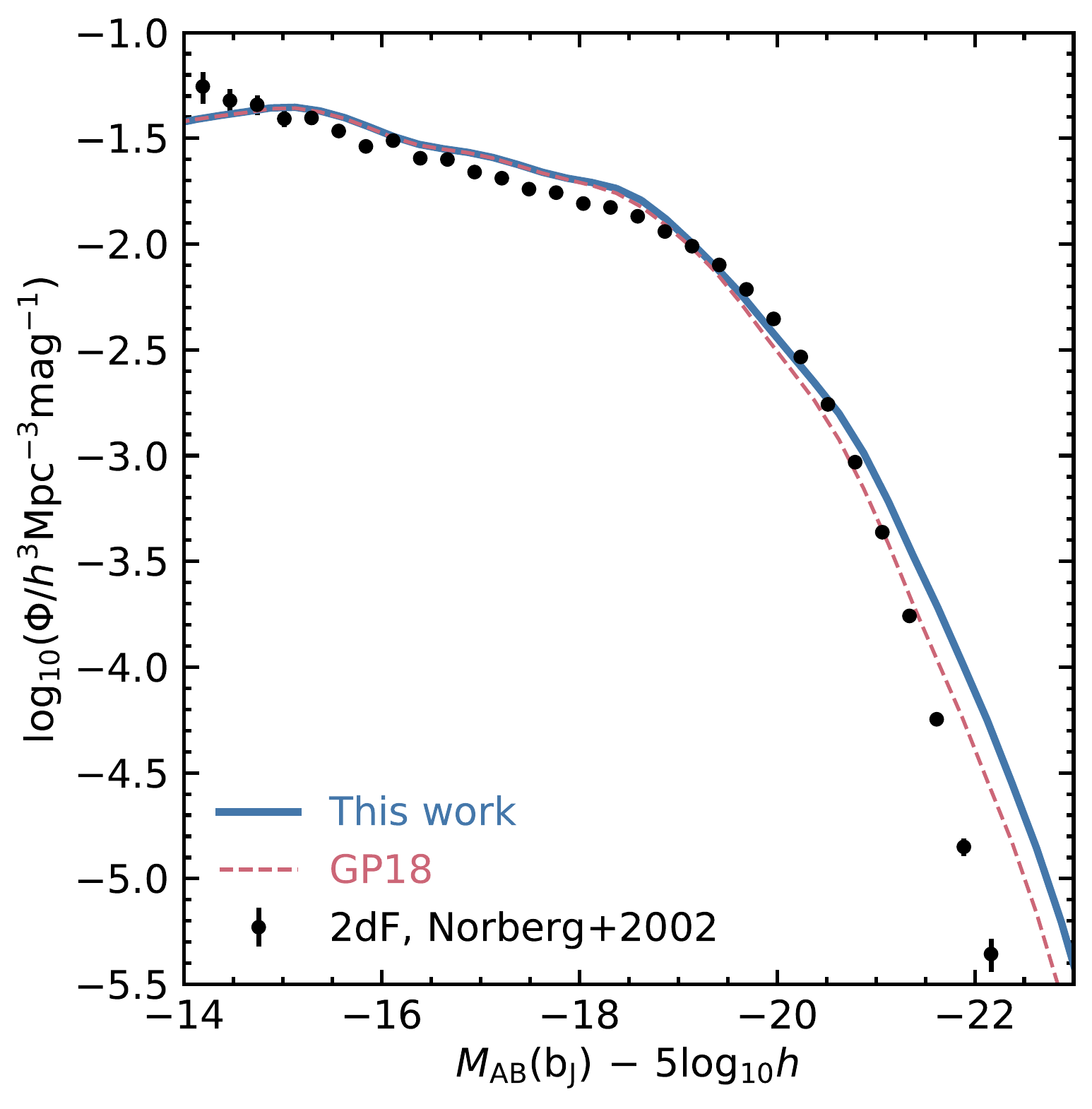} 
\includegraphics[width=0.33\textwidth]{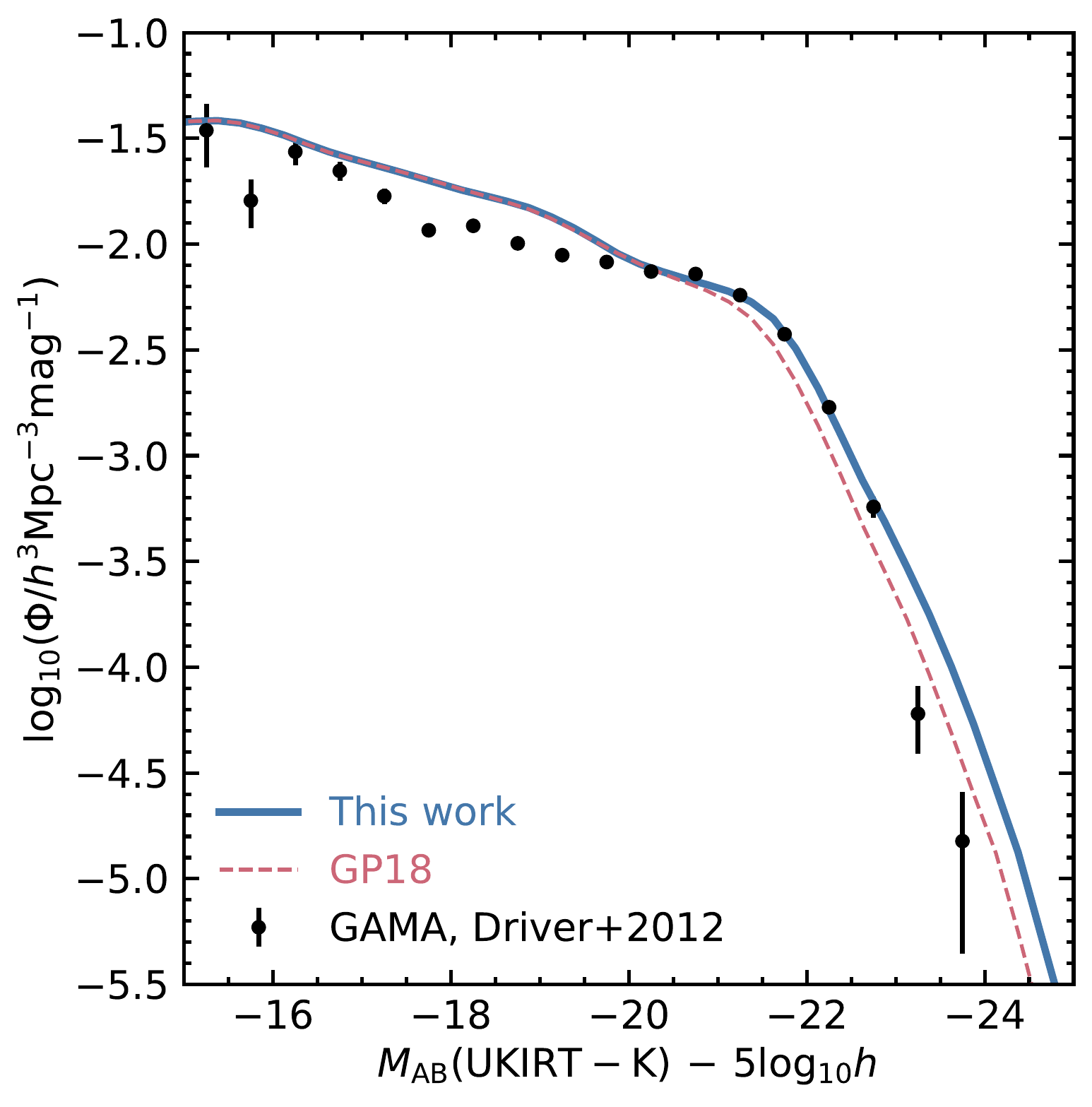}  
\includegraphics[width=0.32\textwidth]{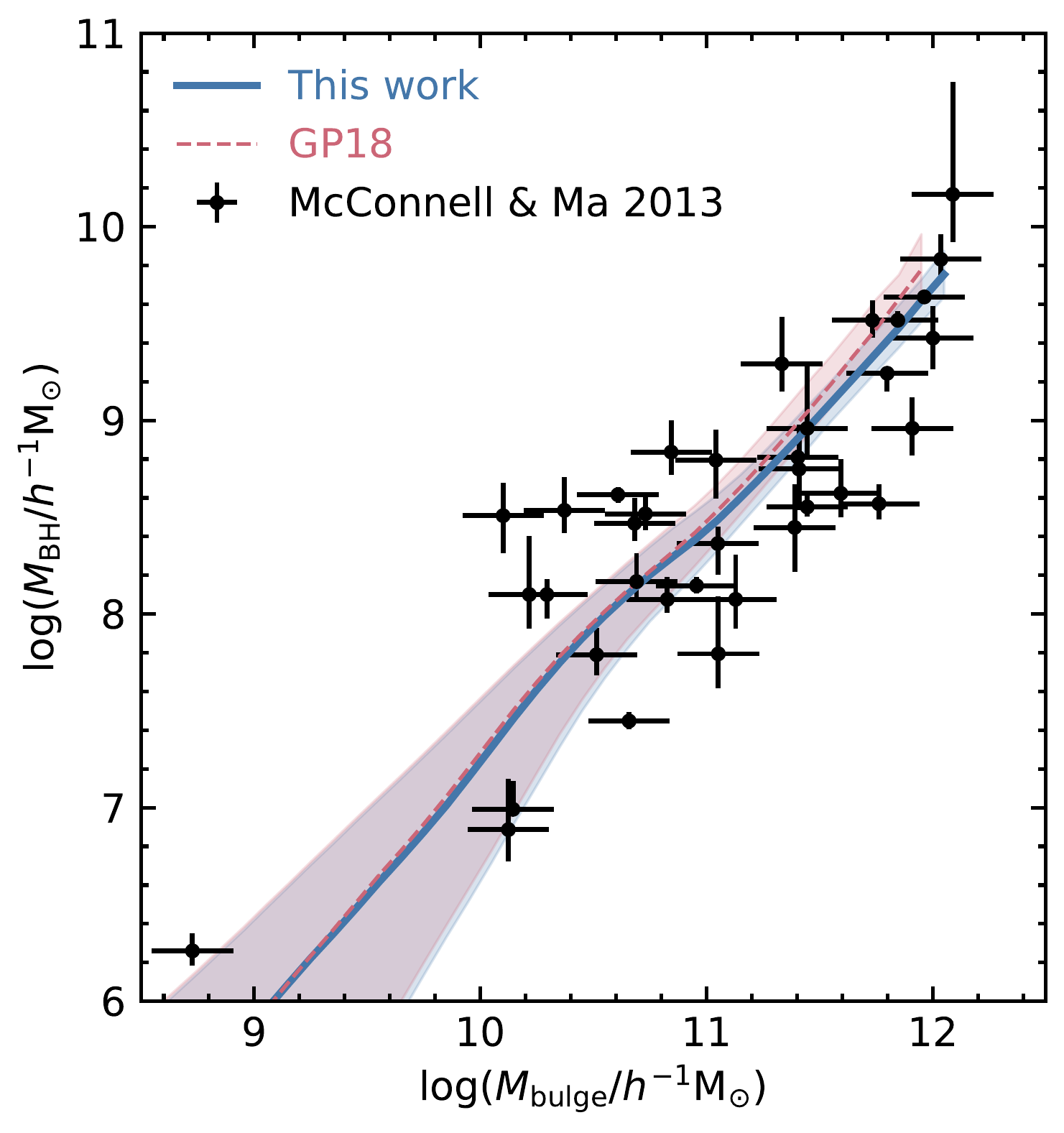}
\caption{\label{fig:cal} The predicted luminosity functions at $z=0$, in the b$_{\rm J}$-band ($\lambda _{\rm eff}= 4500$\AA, left) and in the K-band ($\lambda _{\rm eff}=2.2\mu$m, middle), compared with observations from \citet{norberg02} and \citet{driver12}, respectively. The right panel shows the super massive black hole mass versus  bulge  stellar  mass  relation  at $z=0$ compared  to  observational  data  from~\citet{mcconnell2013}. The lines in the right panel show the median of the predicted super massive black hole mass in bins of bulge mass and the shading the 10-90 percentiles of the model distributions. The blue dashed lines show the predictions from the GP18 model, while the red solid lines show the predictions from the model presented here. These data sets were used to calibrate the free parameters of the model.
} 
\end{figure*}
\begin{table}
\caption{
Differences between the model presented here and the \citet{gp18}, GP18, \gl implementation. 
The ram-pressure stripping parameter $\epsilon_{\rm strip}$ is described in Eq.~6 of \citet{font08}, and controls the stripping efficiency for reheated gas after the initial stripping event at the first pericentre \citep[see also][]{bensonbower10}. 
In the model used here, the ram-pressure stripping happens on longer timescales than in GP18. Here we use the updated model for the evolution of the super massive black holes (SMBH), as described in \citet{griffin}. This model assumes a mass of $10\,h^{-1}\,{\rm M}_{\odot}$ for the black hole seeds. The last three parameters in the table, f$_{\rm Edd}$, $\epsilon _{\rm heat}$ and $\alpha _{\rm cool}$, control the AGN feedback efficiency and their definitions can be found in \S~3.5.3 of \citet{lacey16}.
}
\hspace*{-0.7cm}
\label{tbl:galform}
\begin{center}
\begin{tabular}{|r|c|c|}
\hline
\gl parameter & GP18 & This work \\
\hline
   $\epsilon _{\rm strip}$ & 0.1 & 0.01 \\ 
   M$_{\rm BH seed} (h^{-1}{\rm M}_{\odot})$ &    0. & 10. \\
   f$_{\rm Edd}$   &  0.039 & 0.01 \\
   $\epsilon _{\rm heat}$  &  0.016 & 0.02 \\
   $\alpha _{\rm cool}$ &    0.9 & 0.8 \\
\hline
\end{tabular}
\end{center}
\end{table}


Semi-analytical models use simple, physically motivated rules to follow the fate of baryons in a universe in which structure grows hierarchically through gravitational instability \citep[see][for an overview of hierarchical galaxy formation models]{baugh06,benson10,somerville15}. 

\gl was introduced by \citet{cole00} and since then it has been enhanced and improved \citep[e.g.][]{baugh05,bower06,lagos11,lacey16,griffin}. \gl follows the physical processes that shape the formation and evolution of galaxies, including: (i) the collapse and merging of dark matter haloes; (ii) the shock-heating and radiative cooling of gas inside dark matter haloes, leading to the formation of galaxy discs; (iii) quiescent star formation in galaxy discs which takes into account both the atomic and molecular components of the gas \citep{lagos11}; (iv) feedback from supernovae, from active galactic nuclei \citep{bower06} and from photo-ionization of the intergalactic medium; (v) chemical enrichment of the stars and gas (assuming instantaneous recycling); (vi) galaxy mergers driven by dynamical friction within dark matter haloes. \gl predicts the number and properties of galaxies that reside within dark matter haloes of different masses and assembly histories. This information can be described in terms of a  non-parametric halo occupation distribution function, i.e. the mean number of galaxies as a function of halo mass. 

Currently there are two main branches of \gl: one with a universal stellar initial mass function \citep[IMF,][and this work]{gp14,gp18} and one that assumes different IMFs for quiescent and burst episodes of star formation \citep{lacey16}. These two models have also been re-calibrated to run on a dark matter simulation with a Planck cosmology \citep{planck14,cmb19}. 

Here we have modified the semi-analytical model described in \citet{gp18}, hereafter GP18, to better match the observed passive fraction of galaxies at $z=0$ and to include the updated treatment of the evolution of supermassive black holes (SMBHs) introduced by \citet{griffin}. These two aspects are described in more detail in \S~\ref{sec:passive} and \S~\ref{sec:bh}, respectively. The parameters that have been modified are summarised in Table~\ref{tbl:galform}, other free parameters have been inherited without change from the model described in GP18.

The free parameters presented in Table~\ref{tbl:galform} have been calibrated against observations at $z=0$: the luminosity function in the b$_J$ and K-bands (focusing on the region around the knee), the observed black hole-bulge mass relation (Fig.~\ref{fig:cal}) and the local passive fraction (Fig.~\ref{fig:pf_cal}). When calibrating the model presented here, our aim was to make the smallest number of changes to the GP18 model parameters, despite the introduction of the updated scheme for black hole growth.

\subsubsection{The treatment of gas in satellite galaxies}\label{sec:passive}
\begin{figure} 
%
\includegraphics[width=0.45\textwidth]{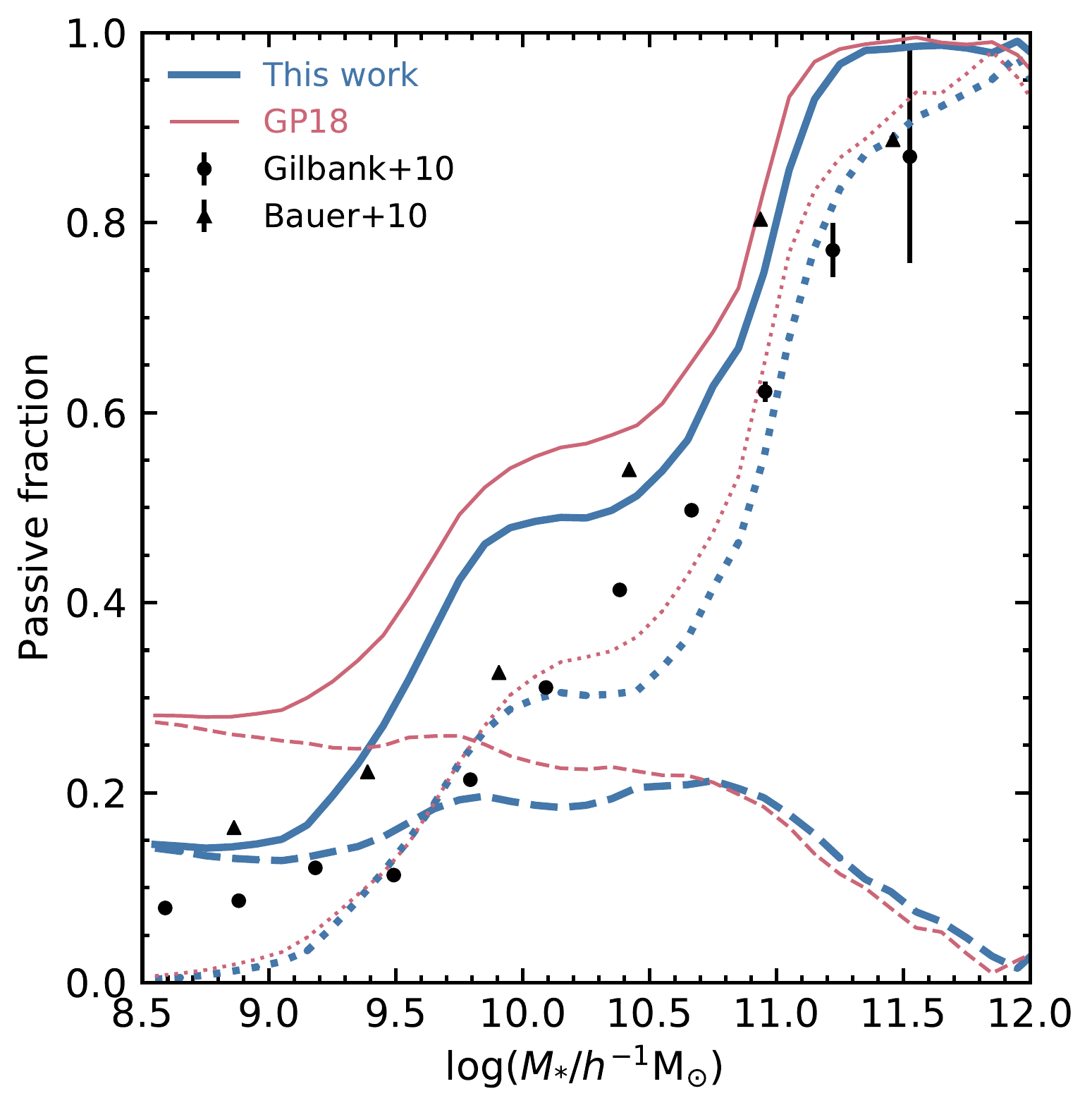}
\caption{\label{fig:pf_cal} The fraction of passive galaxies at $z=0$, i.e.\ those with sSFR $<0.3/{\rm t}_{\rm Hubble}(z=0)$, for this work (thick lines) and the GP18 model (thin lines), compared to the observational results from \citet{gilbank10} (circles) and  \citet{bauer13} (triangles) as extracted and presented in \citet{furlong15}. The solid lines show the total passive fraction, the contribution from satellite galaxies is shown by dashed lines and that from centrals by dotted lines.
} 
\end{figure}


The model we present here has been modified to achieve a local passive fraction of galaxies that is closer to that inferred from observations. We use the model of \citet{font08} for the gradual ram-pressure stripping of hot halo gas from satellite galaxies. In this model, the reservoirs of hot halo gas and reheated gas associated with the galaxy when it becomes a satellite are partially stripped according to the results of hydrodynamical simulations \citep{mccarthy08}. This mainly happens at the first pericentre of the satellite orbit, in the denser central parts of the hot halo where the ram pressure is highest. Ram-pressure stripping of the gas ejected from the galaxy by supernova feedback after it becomes a satellite is assumed to occur at an efficiency that is reduced by a factor $\epsilon_{\rm strip}$ compared to the initial stripping (as described in Eq.~6 from \citealt{font08}; see also \citealt{bensonbower10}). This extra suppression is invoked as the reheated gas is being ejected throughout the orbit of the satellite, not just at the pericenter. The satellite will spend more of its time in the outer parts of the hot halo where the ram pressure is lower, so less of the reheated gas should be stripped. In the model used here, we set $\epsilon_{\rm strip}=0.01$, which is ten times lower than the value of  $\epsilon_{\rm strip}=0.1$ assumed in \citet{font08} and in GP18, so that ram-pressure stripping after the first pericentre takes place ten times more slowly than in GP18. The main effect of this change, as seen in Fig.~\ref{fig:pf_cal}, is to reduce the passive fraction for low mass satellite galaxies in the model. In fact, the passive fraction at $z=0$ is practically insensitive to the precise value of $\epsilon_{\rm strip}$ once it is around $0.01$ or lower. Quiescent star formation in a satellite is followed by supernovae feedback that heats up the cold gas. This gas needs to cool down again before being available for further star formation and so a minimum passive fraction is reached for small satellite galaxies.

The separation of galaxies into passive and star-forming is done using the specific star formation rate (sSFR, i.e. the ratio between SFR and stellar mass) boundary proposed by \citet{franx08}: $sSFR = 0.3/t_{\rm Hubble}(z)$. 
Fig.~\ref{fig:pf_cal} shows that a slower removal of hot gas in satellites reduces the number of passive satellite galaxies. These dominate the low mass end. Around masses $M_* \sim 10^{10} \, h^{-1}{\rm M}_{\odot}$ at $z=0$, the model passive fraction displays a plateau dominated by the contribution of disks. This region of model masses is sensitive to the efficiency of the model AGN feedback. A more detailed exploration of the impact of the treatment of gas in models of galaxy formation will be developed in a future paper.

\subsubsection{The growth of model Supermassive Black Holes}\label{sec:bh}
\citet{griffin} presented a new version of \gl with an updated model of the growth and spin evolution for supermassive black holes (SMBHs). This model assumes that SMBHs can grow in three different ways: i) during starbursts triggered either by galaxy mergers or disc instabilities; ii) by accreting gas from the hot atmosphere of massive haloes; and iii) by SMBH-SMBH mergers after a galaxy merger. The model takes into account how the angular momentum of both the SMBH and the accretion disk affects the consumption of gas. 


In this updated model, SMBHs grow from seeds with mass $M_{\rm seed}$. When a galaxy is formed in the model, it is assigned a black hole of mass $M_{\rm seed}$. The value of this $M_{\rm seed}$ is a free parameter set, which we set to 10 $h^{-1}$M$_{\odot}$ (see also Table~\ref{tbl:galform}). For $M_{\rm seed}>0$, the black hole properties for SMBH in the observed mass range converge rapidly \citep[for further discussion see][]{griffin}. 

The angular momentum of the gas in the inner accretion disk is assumed to be periodically randomised with respect to the angular momentum of the SMBH, i.e. we assume a 'chaotic accretion' mode \citep{king2008,fanidakis11}. 


This updated model includes the evolution of SMBH spins. This evolution affects the growth of SMBHs and therefore the AGN activity. \citet{griffin} showed the new predicted AGN luminosity functions for a range of wavelengths. The SMBH mass versus bulge mass relation at $z=0$ is shown in the right panel of Fig.~\ref{fig:cal}. The updated model has a distribution consistent with that in  the GP18 model. 

\subsubsection{The emission line model}\label{sec:sams_lines}

Nebular emission lines are produced by gas heated by newly formed stars or  nuclear activity. Here we model the star-forming contribution. In \gl, the ratio between the \ox luminosity and the number of Lyman continuum photons is calculated using \HII region models of \citet{sta90}. The \gl model uses \HII region models tabulated for a range of gas metallicities but with a uniform density of 10 hydrogen particles per cm$^{-3}$ and one ionising star in the center of the region with an effective temperature of 45000 K. The ionisation parameter of the \HII region models is around $10^{-3}$, with the exact value  depending on the metallicity in a non-trivial way. These ionisation parameters are averages over the  grid of \HII regions provided by \citet{sta90}. Further details on the emission line model can be found in GP18 \citep[see also][]{Orsi2008}.

Nebular emission lines are assumed to be attenuated by dust in the same way as the stellar continuum~\citep{gp13,lacey16}.

The properties of model emission line galaxies derived using the \citet{sta90} default models are consistent with those obtained using the  \citet{anders03} model for typical \HII regions. The model emission line luminosity functions are also in reasonable agreement with the results derived from a model that assumes a large range of \HII regions~\citep{comparat15o2}. The nebular emission luminosity functions at different redshifts derived from this emission line model were found to be in agreement with observations~\citep{lagos14,gp18}. 

\subsection{The Cosmic Web}\label{sec:LSE}
Here, we apply two algorithms, \vweb and \pweb, to
classify the \LSE of the whole simulation box into knots, filaments, sheets and voids~\citep{Cui2018,Cui2019}. 

The \vweb method uses a dimensionless velocity shear tensor as the tracer to classify the \LSE. Following \cite{Hoffman2012}, at a given redshift, $z$, the velocity, $\vec{v}(\vec{r})$, shear tensor is defined as:
\begin{equation}
 \Sigma_{\alpha\beta} = - \frac{1}{2H(z)} \left( \frac{\partial v_\alpha}{\partial r_\beta} + \frac{\partial v_\beta}{\partial r_\alpha} \right),
 \label{eq:vweb}
\end{equation}
where, $H(z)$ is the Hubble constant at redshift $z$. The eigenvalues of $\Sigma_{\alpha\beta}$ are denoted as $\lambda^V_i$ ($i$ = 1, 2 and 3).

The \pweb method classifies the \LSE based on the tidal tensor, which is measured with the Hessian matrix of the gravitational potential field, $\phi(\vec{r})$. The gravitational potential is calculated from the matter density distribution via the Poisson equation, $\nabla^2 \phi = 4\pi G\rho \delta$, where $\rho$ is the mean density and $\delta$ is the density fluctuation. The tidal tensor, with units $s^{-2}$, is defined as follows~\citep{hahn07}:
\begin{equation}
    P_{\alpha\beta} = \frac{\partial^2\phi}{\partial r_\alpha \partial r_\beta} .
 \label{eq:pweb}
\end{equation}

The computation of the eigenvalues for both matrices is performed on the particles of the MS-W7 dark matter only simulation, split in regular $512^3$ cells grids. The typical side of a grid cell is $\sim 1\ h^{-1}$Mpc. We use a triangular-shaped cloud in cell prescription, for obtaining a smoothed density and velocity distribution at each point on the grid. These are smoothed further over a scale of $\sim 5\ h^{-1}$Mpc. Then, for every grid cell, the eigenvalues of the velocity shear tensor, tidal tensor for \pweb, are computed according to Eq.~\ref{eq:vweb} for \vweb and Eq.~\ref{eq:pweb} for \pweb. Note that although neither \vweb nor \pweb directly use the dark matter particles for their calculations, the larger the number of particles per cell, the more accurate the velocity or potential fields will be. The large smoothing scale, $\sim 5\ h^{-1}$Mpc, provides a robust velocity and density field in real space, and thereby reliable tensors.

Each individual cell is then classified as either `void', `sheet', `filament', or `knot' according to the eigenvalues $\lambda_1 > \lambda_2 > \lambda_3$:
\begin{itemize}
 \item[1.] Void, if $\lambda_1 < \lambda_{th}$,
 \item[2.] Sheet, if $\lambda_1 \geq \lambda_{th} > \lambda_2$,
 \item[3.] Filament, if $\lambda_2 \geq \lambda_{th} > \lambda_3$,
 \item[4.] Knot, if $\lambda_3 \geq \lambda_{th}$,
\end{itemize}
where $\lambda_{th}$ is a free threshold parameter \citep{Hoffman2012, Libeskind2012, Libeskind2013}. Following~\citet{Carlesi2014} and~\citet{Cui2018}, we find that the threshold adopted for \vweb at $z=0$, $\lambda^V_{th} = 0.1$, is also suitable for the simulation results at higher redshift \citep[see][for the redshift evolution of the mass and volume fractions of these large-scale structures with this fixed threshold]{Cui2019}. This value gives very convincing structures, which are visually comparable with the density field (see Appendix~\ref{App:check} for further details). 
However, we find that the threshold $\lambda^P_{th} = 0.01s^{-2}$ at $z$ = 0 for the \pweb method results in slightly larger voids and sheet structures at the two redshifts investigated in this paper, $z=0.83,0.99$. Therefore, we lower the threshold to $0.005s^{-2}$ for \pweb to provide consistent structures to \vweb. 

The effects of resolution and the choice of different thresholds for the cosmic web classification are discussed in detail in the Appendix \ref{App:check}.

\section{Model ELGs}\label{sec:modelelgs}
\begin{table*}
\caption{The cuts applied to the model galaxies in order to mimic the selection of \oem in the corresponding observational survey are the same as those summarised in Table 2 from \citet{gp18}, except for the eBOSS-SGC survey. We apply here the colour cuts described in \citet{raichoor17} for the eBOSS-SGC selection (further details can be found in Appendix~\ref{App:colours}) plus a cut in \ox flux to mimic the instrumentation limitation of the eBOSS-SGC survey. The magnitudes are on the AB system. The particular filter response used for the different cuts is indicated by a superscript on the magnitude column. 
}
\begin{center}
\begin{tabular}{|c|c|c|c|}
\hline
Cuts to & Apparent & \ox flux & Colour\\
mimic  & magnitude & (${\rm erg\, s^{-1}cm^{-2}}$) & selection\\
\hline \vspace{0.1cm}

DEEP2  & $R_{\rm AB}^{\rm DEIMOS}<24.1$ & $2.7\times 10^{-17}$ & None\\ \vspace{0.1cm}

VVDS-Deep &  $i_{\rm AB}^{\rm CFHT} \leq 24$ & $1.9\times 10^{-17}$ & None \\ \vspace{0.1cm}

VVDS-Wide &  $i_{\rm AB}^{\rm CFHT} \leq 22.5$ & $3.5\times 10^{-17}$ & None \\ \vspace{0.1cm}

eBOSS-SGC  &  $21.825< g_{\rm AB}^{\rm DECam} < 22.825$ & $1\times 10^{-16}$ & $-0.068(r-z) + 0.457 < (g-r) < 0.112(r-z) + 0.773$  \& \\ 
     & & & $0.218(g-r) + 0.571 < (r-z) < -0.555(g-r) + 1.901$ \\  

DESI &  $r_{\rm AB}^{\rm DECam} < 23.4$  & $8\times 10^{-17}$ & $(r-z) > 0.3$  \& $(g-r) > -0.3$ \& \\
     & & & $(g-r) < 1.1\cdot(r-z) - 0.13$ \& $(g-r) < -1.18\cdot(r-z) + 1.6$ \\

\hline
\end{tabular}
\end{center}
\label{tbl:obs}
\end{table*}
\begin{figure}  \includegraphics[width=0.5\textwidth]{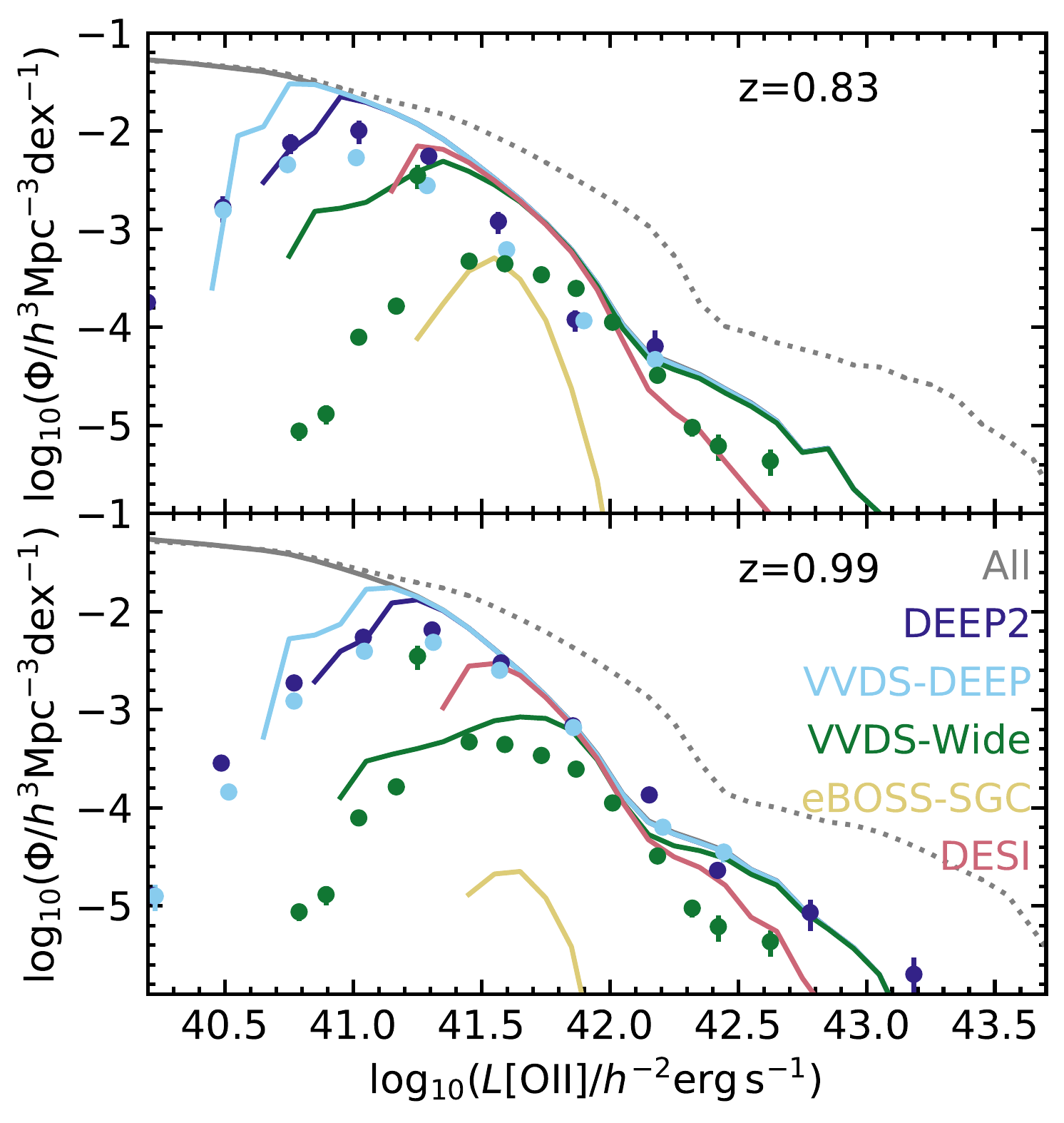}
\caption{\label{fig:lf} The luminosity function of \oem at $z=0.83$ and $z=0.99$ for model galaxies selected with the DEEP2 (dark blue lines, mostly over-plotted), VVDS-DEEP (light blue lines), VVDS-Wide (green lines), eBOSS-SGC (yellow lines) and DESI (red lines) cuts given in Table~\ref{tbl:obs}. The grey solid lines the model total dust attenuated L\ox luminosity function, the intrinsic one is shown by the dotted grey lines. The data from DEEP2 and VVDS are colour coded like the model galaxies selected to mimic both surveys. The observational errors come from jackknife re-sampling \citep{comparat16} and in some cases are smaller than the corresponding symbol. 
}
\end{figure}

A range of cosmological surveys, such as SDSS/eBOSS-SGC \citep{dawson16} and DESI \citep{desi1}, are or will be targeting star-forming ELG galaxies to probe the nature of dark energy using spectroscopic redshifts. ELGs selected with optical instruments at $z\sim 1$ are dominated by \oem \citep{comparat15o2}. As we previously did in GP18, we select \oem from the semi-analytical model described in \S\ref{sec:model} mimicking the samples from different surveys, \S~\ref{sec:elgsel}. We then explore how these model galaxies trace the \LSE in \S~\ref{sec:elgcw}. 

In this work we use a model that strips the gas in satellite galaxies slower than in GP18. This modification has a strong impact on decreasing the passive fraction of galaxies with masses below $10^{10}h^{-1}$M$_{\odot}$, however, this merely changes the fraction of model \oem by up to 5 per cent. Model ELGs at $z\sim 1$ are mostly centrals, with a satellite fraction between 4 and 9 per cent, dominated by star-forming galaxies with $sSFR>0.3/{\rm t}_{\rm Hubble}(z)$. 

\subsection{Model ELGs sample selection}\label{sec:elgsel}
We select model ELGs using the cuts specified in Table~\ref{tbl:obs} in apparent magnitude, \ox flux and colour. The magnitude and flux cuts reproduce the limits in the DEEP2 \citep{newman13} and VVDS \citep{lefevre13} surveys, applied to select the corresponding model \oem. No further colour cuts are applied to the model DEEP2 and VVDS selections, as the observational colour cuts were applied to restrict the redshift range and here we are limiting our study to two single simulation outputs at $z=0.83$ and $z=0.99$. We have additional colour cuts to select model DESI~\citep{desi1} and SDSS-IV/eBOSS~\citep{raichoor17} ELGs. These colour cuts were set observationally to target an spectroscopic galaxy sample with colours that minimally overlap with those from stars (further details can be found in appendix~\ref{App:colours}). 

Here we focus on two simulation outputs at $z=0.83$ and $z=0.99$, which are separated by 717 Myr. The lowest of these two redshifts is close to the effective redshift of the SDSS/eBOSS-SGC sample \citep{raichoor17}, $z=0.84$, which in turn is close to the average redshift of VVDS-DEEP \citep{comparat15o2}. The VVDS-Wide sample has a lower average redshift and will not be included in the clustering and environment analysis presented later in this work. The DESI ELG sample is designed to have a redshift baseline between 0.6 and 1.7 and an anticipated effective redshift of $z\sim 1$ \citep{desi1}. The clean sample of DEEP2 ELGs has a mean redshift of 0.97 \citep{comparat19vac}. Both values are close to the redshift $z=0.99$ of the simulation output. 

\subsubsection{\ox luminosity function}
Model \oem are selected in numbers that are in reasonable agreement with observational selections, as shown in Fig.~\ref{fig:lf}. Note that the dust attenuation in this model is such, that it mostly affects the most luminous and massive \oem. As reported in GP18, the change in slope of the luminosity functions shown in Fig.~\ref{fig:lf} is due to galaxies with an ongoing star-burst that dominate the bright end, $L {\rm [OII]}>10^{42} h^{-2}{\rm erg\,s^{-1}}$. This bright end is also dominated by galaxies with a bulge to total mass above $0.5$ (spheroids) and compact, with half mass radii smaller than $0.5 h^{-1} {\rm kpc}$. The luminosity functions shown in Fig.~\ref{fig:lf} are similar to those in GP18 and accompany Errata\footnote{Due to a problem with filter naming, the selection for VVDS was effectively done with the r-band, instead of the indicated i-band. This discrepancy has been corrected in this work.}.

The number density of model SDSS/eBOSS-SGC ELGs at $z\sim0.83$, $\sim 1.58\cdot10^{-4}h^3$Mpc$^{-3}$, is below the current observational estimations $\sim 2.67\cdot10^{-4}h^3$Mpc$^{-3}$ \citep{raichoor17}. In \cite{hongguo19} it was presented an empirical model directly calibrated with SDSS/eBOSS-SGC data and they compared their results with the model galaxies presented here. From this comparison it appeared that besides lacking satellite ELGs, as it was concluded in GP18, there might be a lack of massive central galaxies. This is also suggested by the results presented in \cite{comparat19vac} for DEEP2. Although dust attenuation affects the most luminous and massive galaxies, there might be other physical processes contributing to the discrepancies found, from the simplicity of our emission line modelling to a more fundamental aspect of the growth of massive galaxies \citep{mitchell18}.

\subsection{Model ELGs in the cosmic web}\label{sec:elgcw}

\begin{table}
\caption{Fraction of ELGs in the different \LSE structures as classified by the \vweb algorithm. The percentage of satellite galaxies for each selection is shown in brackets.
}
\label{tbl:elg_env}
\begin{center}
\begin{tabular}{|c|c|c|} 
\hline
z=0.83 & VVDS-DEEP (11 per cent) & eBOSS-SGC (5 per cent) \\
voids & 0.05 (3 per cent) & 0.04 (1 per cent) \\
sheets & 0.34 (6 per cent) & 0.32 (3 per cent) \\
filaments & 0.48 (11 per cent) & 0.51 (5 per cent) \\
knots & 0.13 (24 per cent) &  0.12 (13 per cent)\\
\hline
z=0.99 & DEEP2 (7 per cent) & DESI (4 per cent) \\
voids & 0.04 (2 per cent) & 0.04 (2 per cent) \\
sheets & 0.32 (4 per cent) & 0.34 (2 per cent) \\
filaments & 0.51 (7 per cent) & 0.51 (4 per cent) \\
knots & 0.13 (15 per cent) & 0.12 (8 per cent) \\
\hline
\end{tabular}
\end{center}
\end{table}

The \LSE of the dark matter in the N-body simulation has been classified using the algorithms described in~\S\ref{sec:methods} into: voids, sheets, filaments and knots. Table~\ref{tbl:elg_env} summarise how model ELGs are distributed within the different structures of the cosmic web, as classified by the \vweb algorithm, although similar results are found when using \pweb. We find that about 80 per cent of model ELGs live in either filaments or sheets, with half of them in filaments.

%
\begin{figure} 
\includegraphics[width=0.45\textwidth]{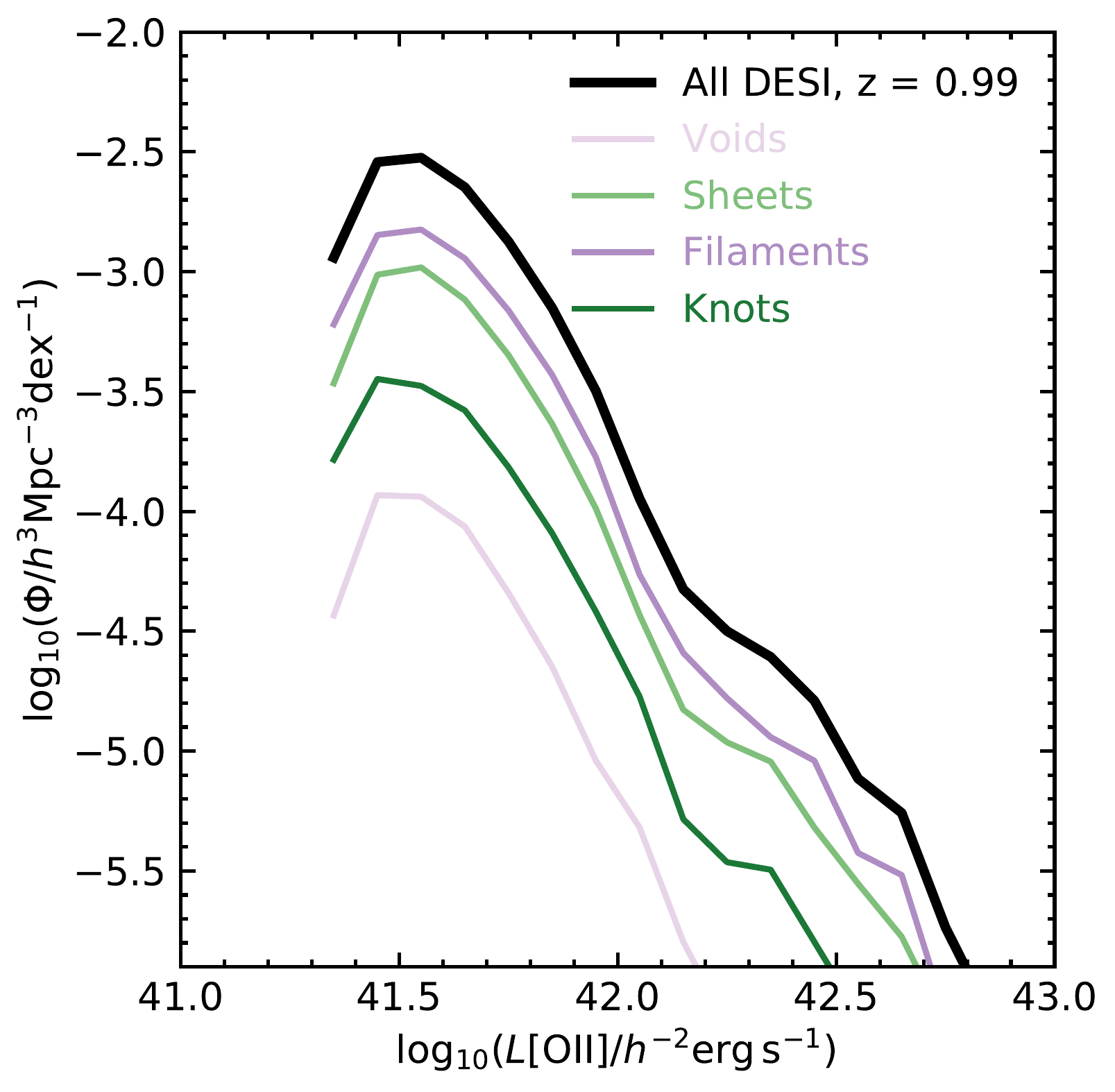}
\caption{\label{fig:lf_env} The \ox luminosity function for the model DESI model galaxies at $z=0.99$, thick line, and the contribution of the different \LSE structures, thin lines, as classified by the \vweb algorithm (see the legend). 
} 
\end{figure}

The distribution of ELGs in the cosmic web, summarised in Table~\ref{tbl:elg_env}, is also reflected in the split of the \ox luminosity function. This is shown in Fig.~\ref{fig:lf_env}, for DESI model galaxies at $z=0.99$, classified using the \vweb algorithm (similar results are found for \pweb). The \ox luminosity function varies in normalisation for the different \LSE structures, but the shape changes minimally. The brightest model \oem are found in the structures where they are most dominant: filaments and sheets.

\begin{figure} 
\includegraphics[width=0.45\textwidth]{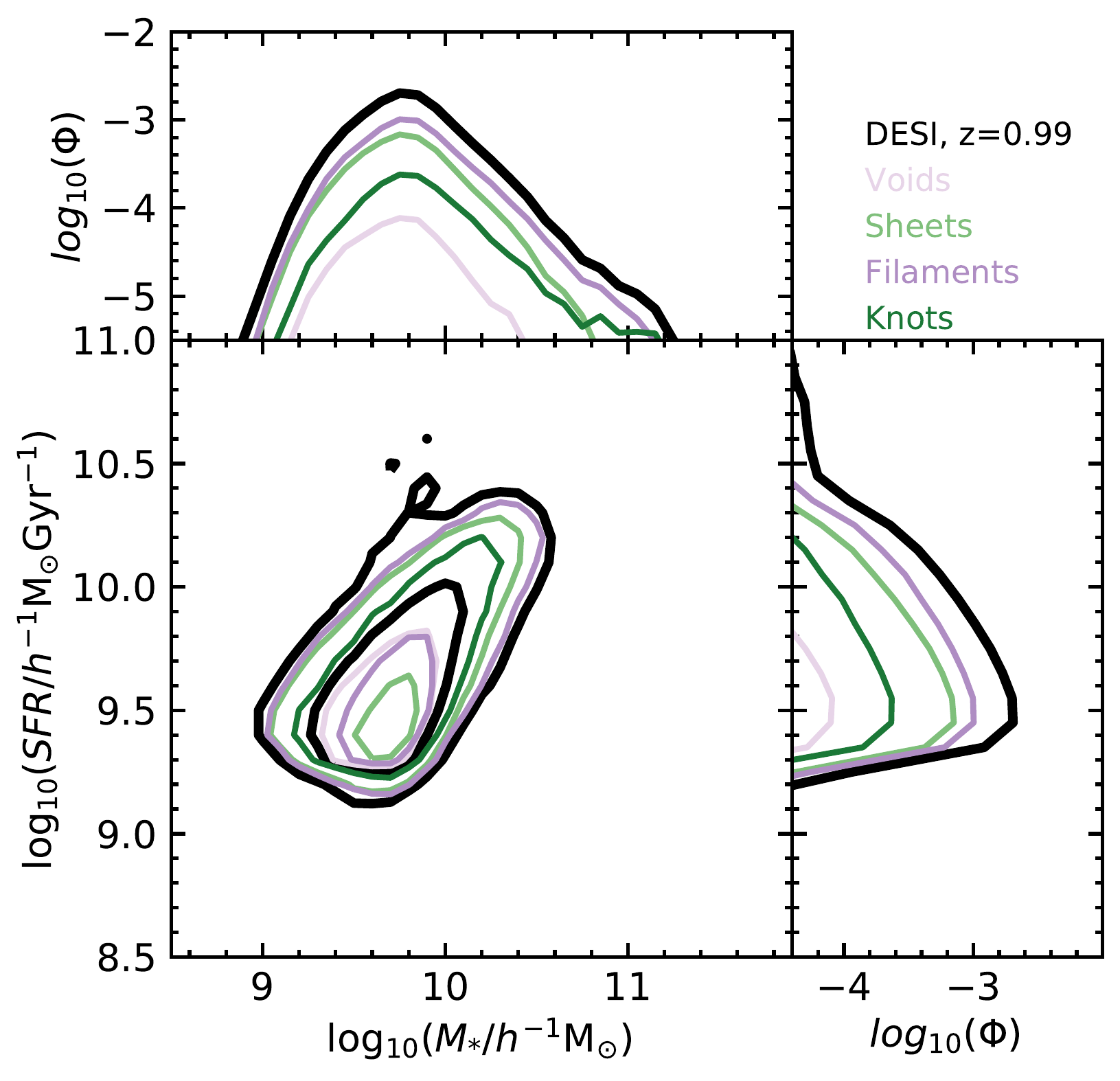}
\caption{\label{fig:sfr_env} The $z=0.99$ distribution of galaxies in the SFR-stellar mass plane for all DESI model galaxies, thick lines, and those living in different \LSE structures, thin lines as classified by the \vweb algorithm. The sSFR-stellar mass plane has been collapsed into the galaxy stellar mass function, top sub-panel, and the SFR function, right sub-panel. The corresponding densities shown are $\Phi (h^{3}{\rm Mpc}^{-3}{\rm dex}^{-1})$.
} 
\end{figure}

As the \ox luminosity function, the SFR function also shows different normalisations but similar shapes for galaxies in different cosmic web structures. Fig.~\ref{fig:sfr_env} shows the case for DESI model galaxies at $z=0.99$ classified with \vweb. Note that in \gl all galaxies have a SFR above zero, even if very small in some cases.

Fig.~\ref{fig:sfr_env} also shows the distribution of model galaxies in the SFR-stellar mass plane. It is clear from here, that model ELGs are not directly equivalent to imposing a cut in SFR. This was also reported in GP18 and is common to all the studied ELG selections.

The galaxy stellar mass function for DESI model galaxies is also shown in Fig.~\ref{fig:sfr_env}. In this case, there is a clear change in the shape for galaxies in knots, at high masses. Model ELGs in knots tend to be more massive. This is also found for the other ELG selections. This might be related with the larger fraction of satellite galaxies found in knots, as summarised in Table~\ref{tbl:elg_env}. 

As knots appear in denser regions, haloes are expected to be more massive and, thus, able to host several galaxies. Given an ELG selection, knots tend to host more satellites, with the differences being the largest for satellite galaxies with stellar masses around $10^{10}h^{-1}$M$_{\odot}$. This is the value above which the galaxy stellar mass function in knots starts to differ from the the other cosmic web structures, as it can be seen in Fig.~\ref{fig:sfr_env}.

We also find that the percentage of satellite galaxies as a function of stellar mass varies significantly between the different ELG selections (not shown in figure).

In knots, the gas fuelling star formation in satellite galaxies will be removed after some time and little new gas will be fuel to those galaxies. This gas will feed the central galaxy. We defer to the future studying the evolution of the star formation in model galaxies populating different cosmic web structures.

\subsubsection{The mean halo occupation distribution of ELGs}\label{sec:hodelg}
\begin{figure} 
\includegraphics[width=0.45\textwidth]{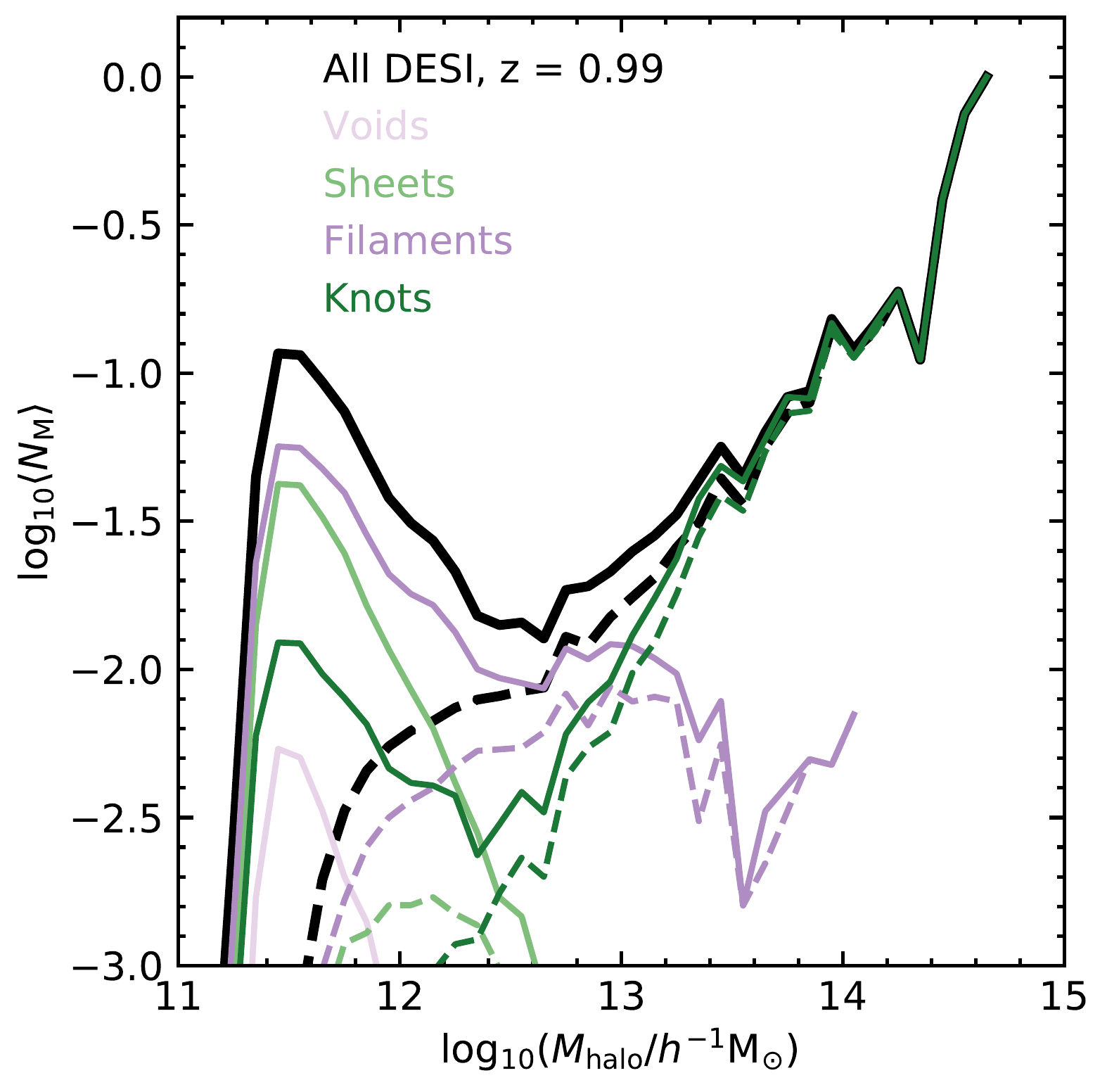}
\caption{\label{fig:hod_env} The mean halo occupation distribution (HOD) as a function of host halo mass for the model DESI galaxies at $z=0.99$, thick line, and the contribution of the different \LSE structures, thin lines, as classified by the \vweb algorithm (see the legend). Solid lines show the total mean HOD, the contribution from satellite galaxies is shown as dashed lines.
} 
\end{figure}

Fig.~\ref{fig:hod_env} shows the mean halo occupation distribution (HOD) for the DESI model galaxies at $z=0.99$. This HOD is well below having one galaxy per halo. The HOD of model central ELGs is close to an asymmetric Gaussian with maybe a plateau (see also GP18). Galaxy mock catalogues from HOD models usually assume a very different shape from that seen in Fig.~\ref{fig:hod_env}. The shape usually assumed for HOD models is that characteristic for stellar mass-selected samples, this will be further explored in \S~\ref{sec:nd}.

Fig.~\ref{fig:hod_env} shows that the normalization of the central galaxies peak decreases for different \LSE structures, following the trend in density reported in Table~\ref{tbl:elg_env}. The minimum halo mass to host an ELG remains practically independent of the cosmic web, except for voids, for which there is a slight increase in mass. The number of ELGs in voids is quite low, and those are mostly central. Note that the minimum halo mass in the model HOD shown in Fig.~\ref{fig:hod_env} is not affected by resolution effects.

In voids and sheets there are almost no satellite ELGs. This can be seen in Fig.~\ref{fig:hod_env} for model DESI galaxies. The contribution of satellite galaxies is so small in voids and sheets, that the global shape of the HOD for these environments can be described as an asymmetric Gaussian.

The shape of the mean HOD does change with environment. The HOD for central galaxies has a plateau in filaments and knots. There is a clear increase in the power law followed by model satellite ELGs in knots. The differences among the cosmic web structures highlight the importance that environmental processes have in shaping the evolution of galaxies. Environmental processes will therefore impact the small scale clustering derived for galaxies populating different large scale environments.

\subsubsection{The clustering of ELGs}\label{sec:xielg}
\begin{figure}
\includegraphics[width=0.45\textwidth]{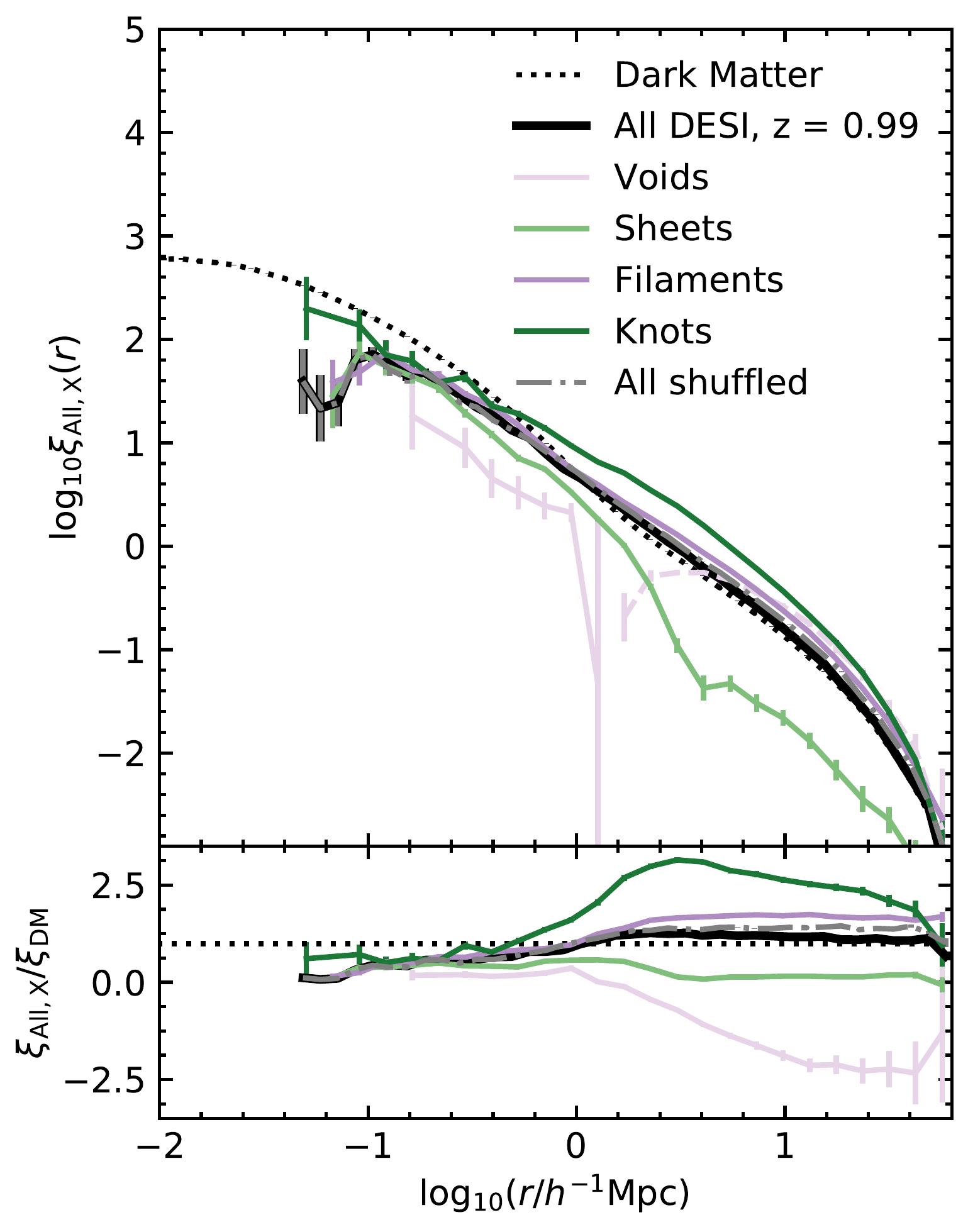}
\caption{\label{fig:xielg} The real-space two point correlation function, top panel, and ratio between the correlation functions of galaxies and of dark matter, $\xi_{\rm All,X}/\xi_{\rm DM}$. The dark matter auto-correlation function is shown as a dotted line. The auto-correlation function for all DESI model galaxies at $z=0.99$ is shown as thick solid lines and that for a shuffled sample as dot-dashed lines. Thin solid lines show the cross-correlation, $\xi_{\rm All,X}$, between the whole ELG sample and ELGs in different \LSE structures as classified by the \vweb algorithm (see the legend). The logarithm of the absolute value of negative $\xi_{\rm All,X}$ is shown with dashed thin lines. Poisson error bars are shown in both panels.
} 
\end{figure}
\begin{table}
\caption{Large-scale bias for the ELG samples presented in Table~\ref{tbl:obs}. The bias, $b$, and associated error have been obtained by minimizing $\chi^2$ for the model galaxy auto-correlation function in real space: $\xi_{\rm gg}=b^2\xi_{\rm Dark\,matter}$, in the range $8\leq r (h^{-1}{\rm Mpc}) \leq 50$. This range comprises the large scales available for the underlying dark matter simulation, for which we can consider to be measuring the linear scale bias.
}
\label{tbl:elg_bias}
\begin{center}
\begin{tabular}{|c|c|c|} 
\hline
z=0.83    & VVDS-DEEP  & eBOSS-SGC  \\
         & 1.12$\pm$0.01 & 1.09$\pm$0.01 \\
\hline
z=0.99    & DEEP2 & DESI \\
        & 1.17$\pm$0.01 &  1.07$\pm$0.01 \\
\hline
\end{tabular}
\end{center}
\end{table}

Here we study the clustering of ELGs living in different \LSE. Model galaxies come from a periodic simulated box and therefore the configuration space two-point auto-correlation function, $\xi$, is calculated using a simple estimator: $1+\xi(r)=2DD/(n^2VdV)$, where DD is the number of distinct model galaxy pairs with separation between $r$ and $r + $d$r$ and the denominator is the average number of neighbours found in the volume d$V$ of a spherical shell of radius r and thickness dr~\citep[see also][]{gp11}. The dark matter two-point correlation function is calculated using the particles from the MS-W7 dark matter only simulation. The calculation of the two-point auto-correlation functions has been done using the publicly available code {\sc cute}\footnote{\url{https://github.com/damonge/CUTE}~\citep{cute}.}. 

The 3-D pair-counts, $DD_{\rm ccf}$, needed to estimate the cross-correlation, $\xi_{\rm All,X}$, between the whole ELG sample and the sub-samples populating different \LSE structures is obtained with the publicly available Python package \texttt{Corrfunc.theory.DD}\footnote{\url{https://github.com/manodeep/Corrfunc}~\citep{corrfunc}}. The two point cross-correlation is then estimated as $1+\xi_{\rm All,X}(r)=DD_{\rm ccf}/(N_{\rm All}N_XdV/V)$, where $N_{\rm All}$ is the number of all ELGs and $N_X$ those ELGs in a given \LSE structure, within the simulation volume, $V$. The Poisson errors for the cross-correlation are estimated as $(1+\xi_{\rm All,X})/\sqrt(DD_X)$, where $DD_X$ is the number of unique pairs of the sub-sample of ELGs, living in either voids, sheets, filaments or knots.

Fig.~\ref{fig:xielg} shows the real-space two point auto-correlation function for all DESI model galaxies at $z=0.99$ compared with that for the dark matter. At large scales, $r>1h^{-1}$Mpc, DESI ELGs trace the dark matter clustering, with a linear bias close to 1. Table~\ref{tbl:elg_bias} presents the large-scale bias for each of the ELG samples studied here (see Table~\ref{tbl:obs}), which are all close to 1.

The properties of the model galaxies explored here are naturally affected by assembly bias, i.e., the dependence on halo assembly history as well as on halo mass \citep[e.g.][]{zenter2014}. Therefore the bias that we measure is the combination of the cosmological halo bias and that resulting from the assembly bias. The latter is expected to have a small or negative effect for star-forming samples, such as the ELGs we study here \citep{contreras2019}. This implies that the bias measured from a catalogue of galaxies constructed with a halo occupation distribution model without considering assembly bias, might be larger than the values reported here. To quantify the assembly bias, we present in Fig.~\ref{fig:xielg} the clustering for the model DESI sample shuffled within haloes of similar mass~\citep[see e.g.][]{jimenez2019}. In this case we found a small but negative signal. For the other ELG selections we find almost no galaxy assembly bias signal.

Fig.~\ref{fig:xielg} also shows the cross-correlation between all DESI ELGs at $z=0.99$ and those living in \LSE as classified by the \vweb algorithm. At small scales, $r\leq0.5h^{-1}$Mpc, the clustering of ELGs in follows closely the auto-correlation function in all \LSE, except  for knots from the DESI sample. This is clearer for the DEEP2 and VVDS-DEEP samples (not shown here), for which pairs of galaxies are found at separations smaller than $0.03h^{-1}$Mpc, extending the one halo term clustering to smaller scales than the DESI one. At larger scales, $r\geq0.5h^{-1}$Mpc, differences are found for the clustering of ELGs living in different \LSE. ELGs in filaments are the ones clustered most similarly to the two point auto-correlation function.

ELGs living in knots are the most clustered for separations $1\leq r(h^{-1}$Mpc$)\leq 10$. This can be seen in Fig.~\ref{fig:xielg} for DESI model galaxies, but it is also the case for the other ELG selections. The fraction of satellite ELGs in knots is the largest found for the explored cosmic web structures (see Table~\ref{tbl:elg_env}). The large number of satellites might explain the reported boost in the clustering at intermediate scales, corresponding to the transition between the 1-halo and 2-halo terms.

For all the studied ELG samples, at large scales, $r>1h^{-1}$Mpc, galaxies living in voids are less clustered than in any other environment, with cross-correlations having negative values and thus, negative bias values. There are few model ELGs found in voids (see Table~\ref{tbl:elg_env}), which are also the least dense regions, producing a low clustering at large scales. We find the differences to be dominated by the number density of each environment.

Similar results to those described above for the \vweb classification are found when using the \pweb algorithm to classify the \LSE.

\section{ELGs in context}\label{sec:nd}
\begin{figure*} 
\includegraphics[width=0.33\textwidth]{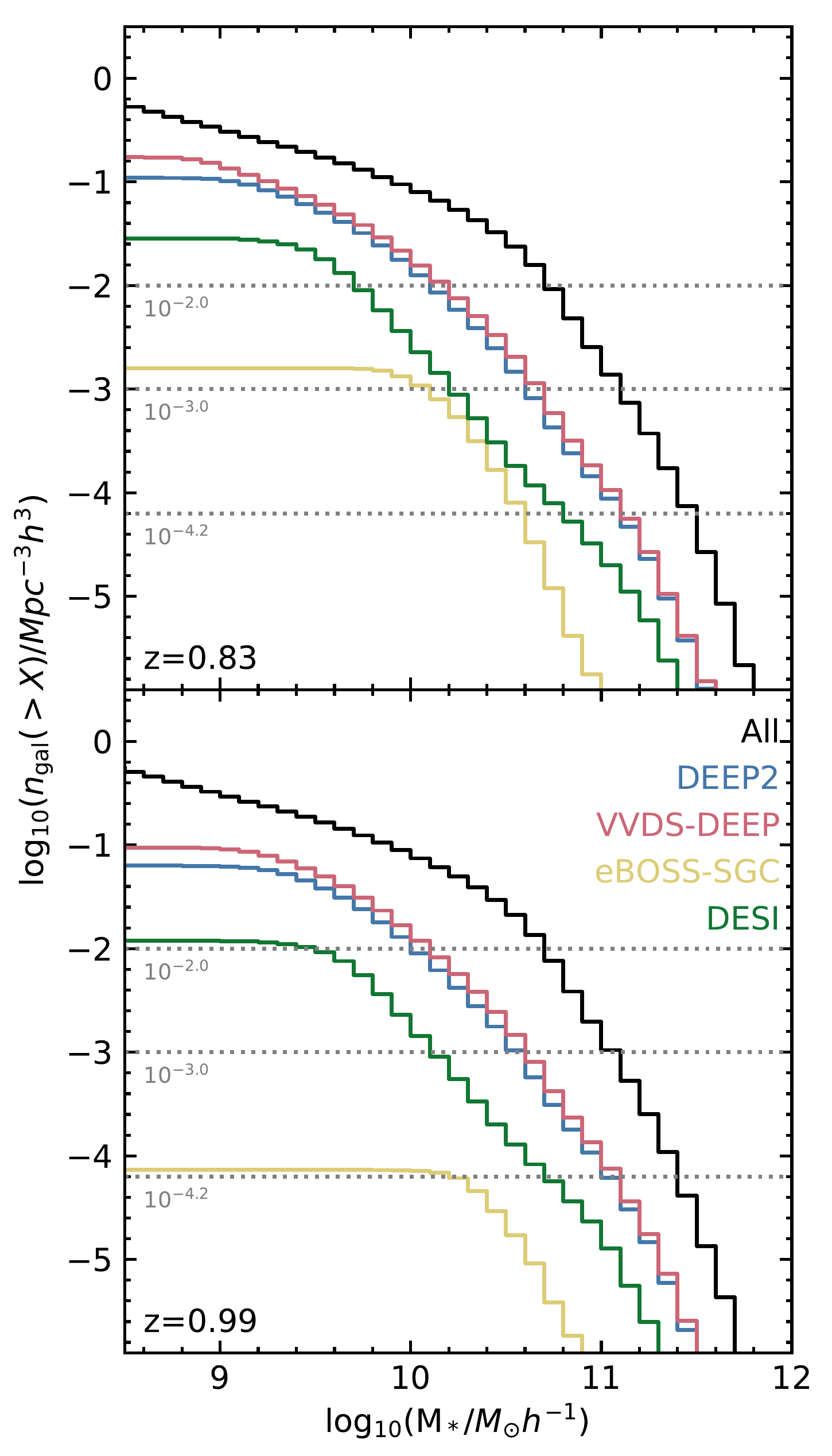}
\includegraphics[width=0.33\textwidth]{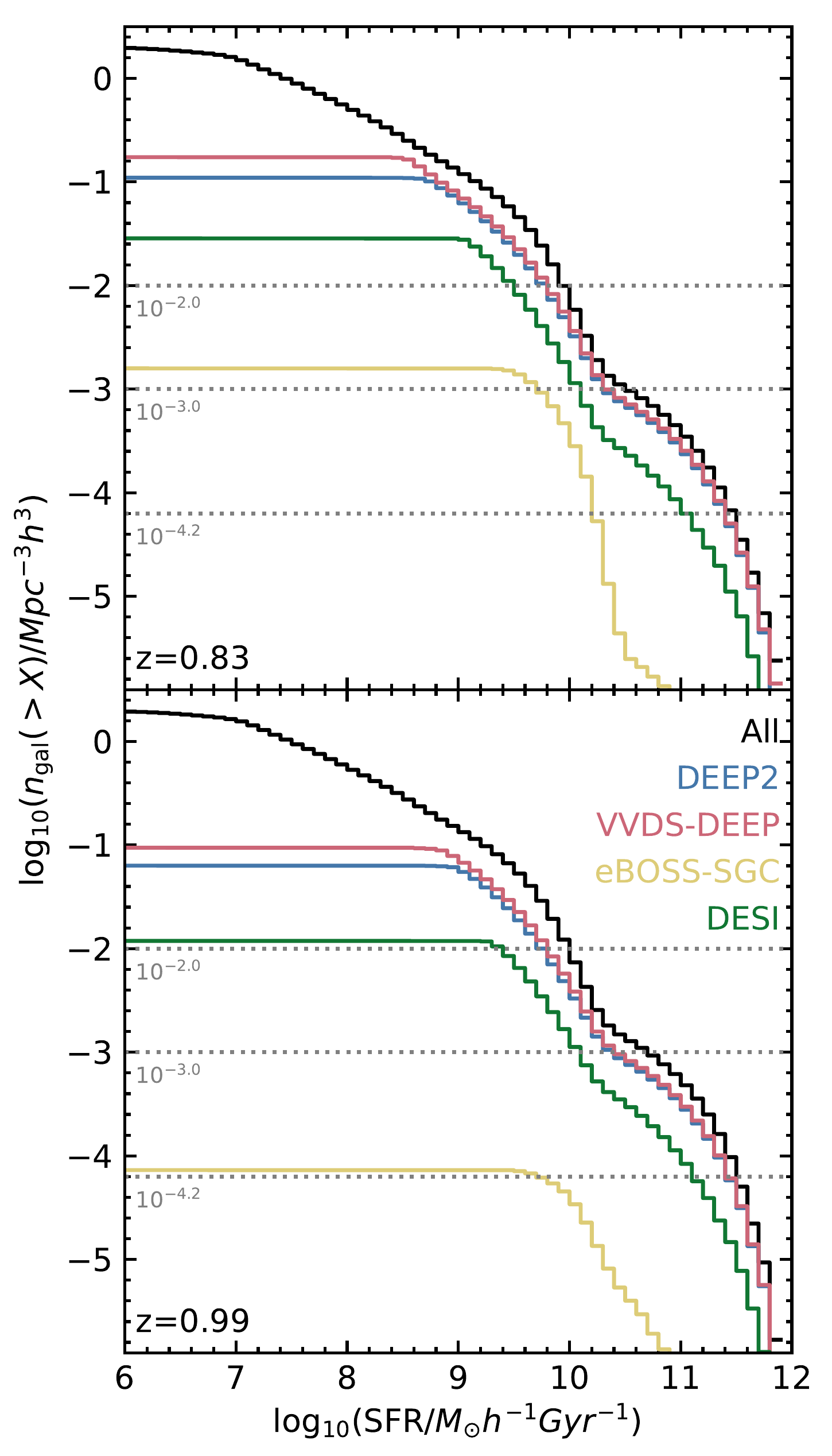}
\includegraphics[width=0.33\textwidth]{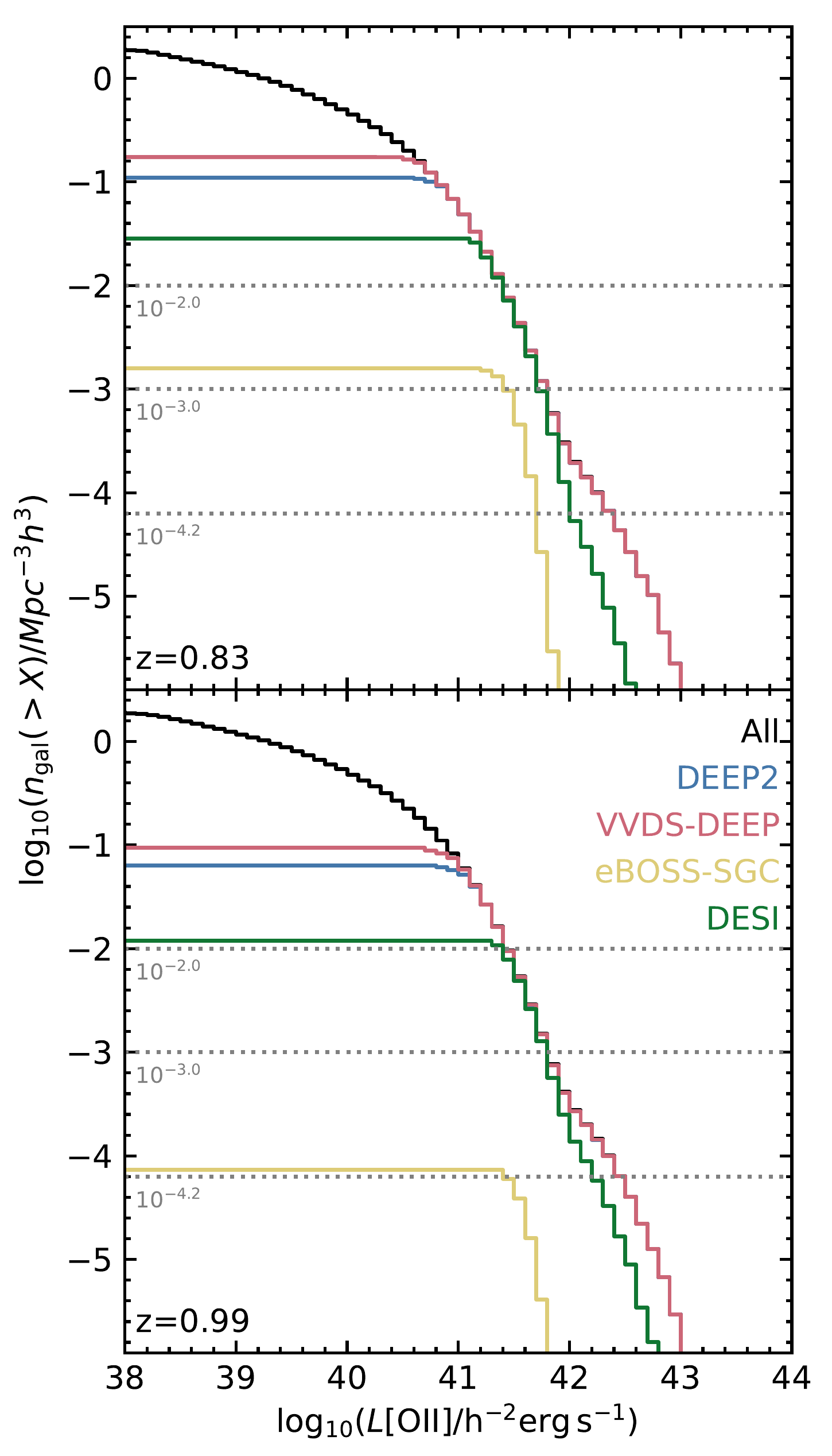}
\caption{\label{fig:cum} The cumulative abundance of all galaxies, black lines, and ELGs selected as summarised in Table~\ref{tbl:obs} and indicated in the legend, ranked by stellar L\ox, left, SFR, middle, and stellar mass, right, at $z=0.83$, top panel, and $z=0.99$, bottom panel. The three number density cuts used to define similar samples, $n_{\rm gal}=10^{-2},10^{-3},10^{-4.2}h^{3}{\rm Mpc}^{-3}$, are indicated by the dashed horizontal lines.
} 
\end{figure*}
\begin{figure*}
    \centering
    \includegraphics[width=.45\textwidth]{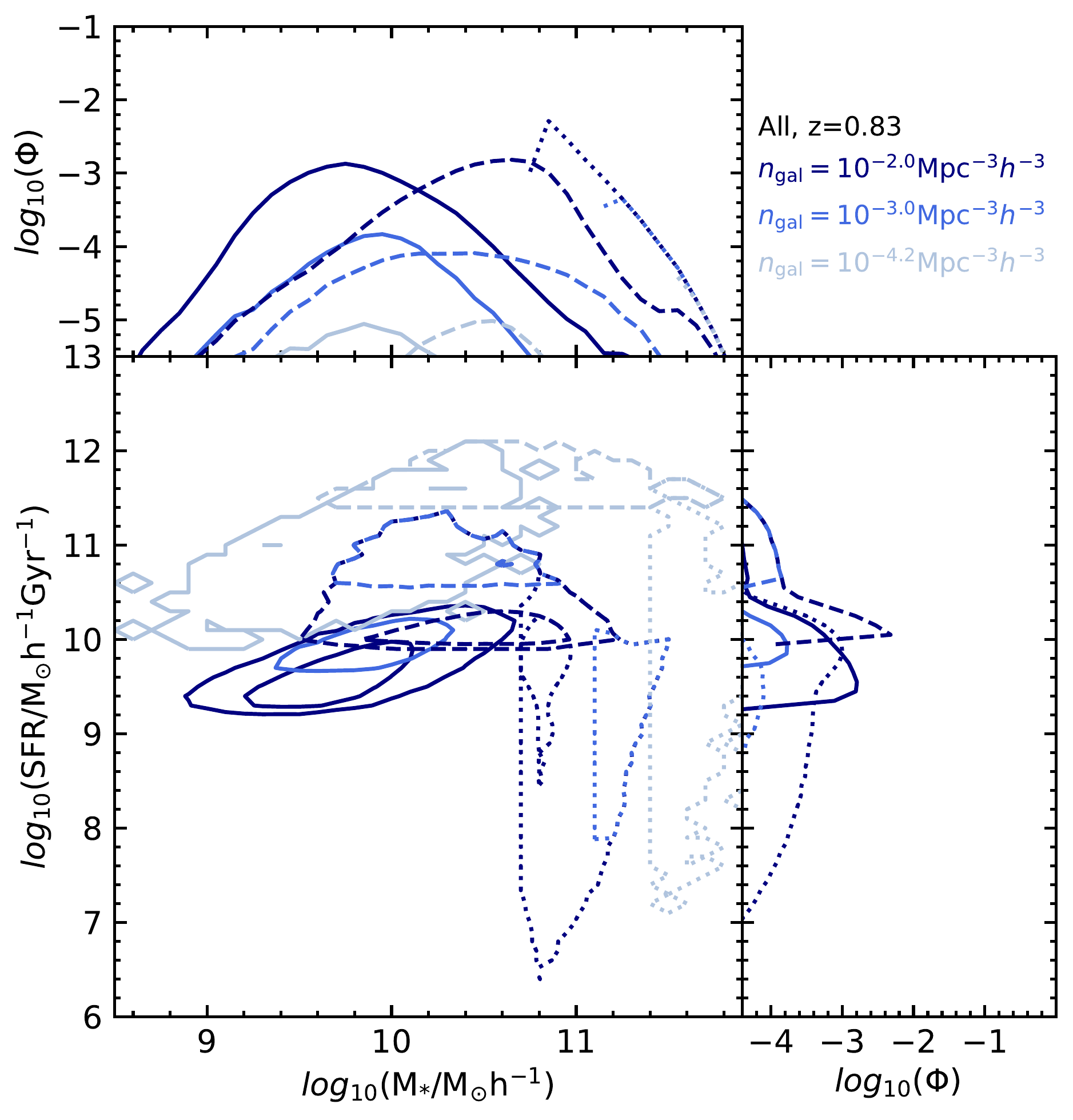}
    \includegraphics[width=.45\textwidth]{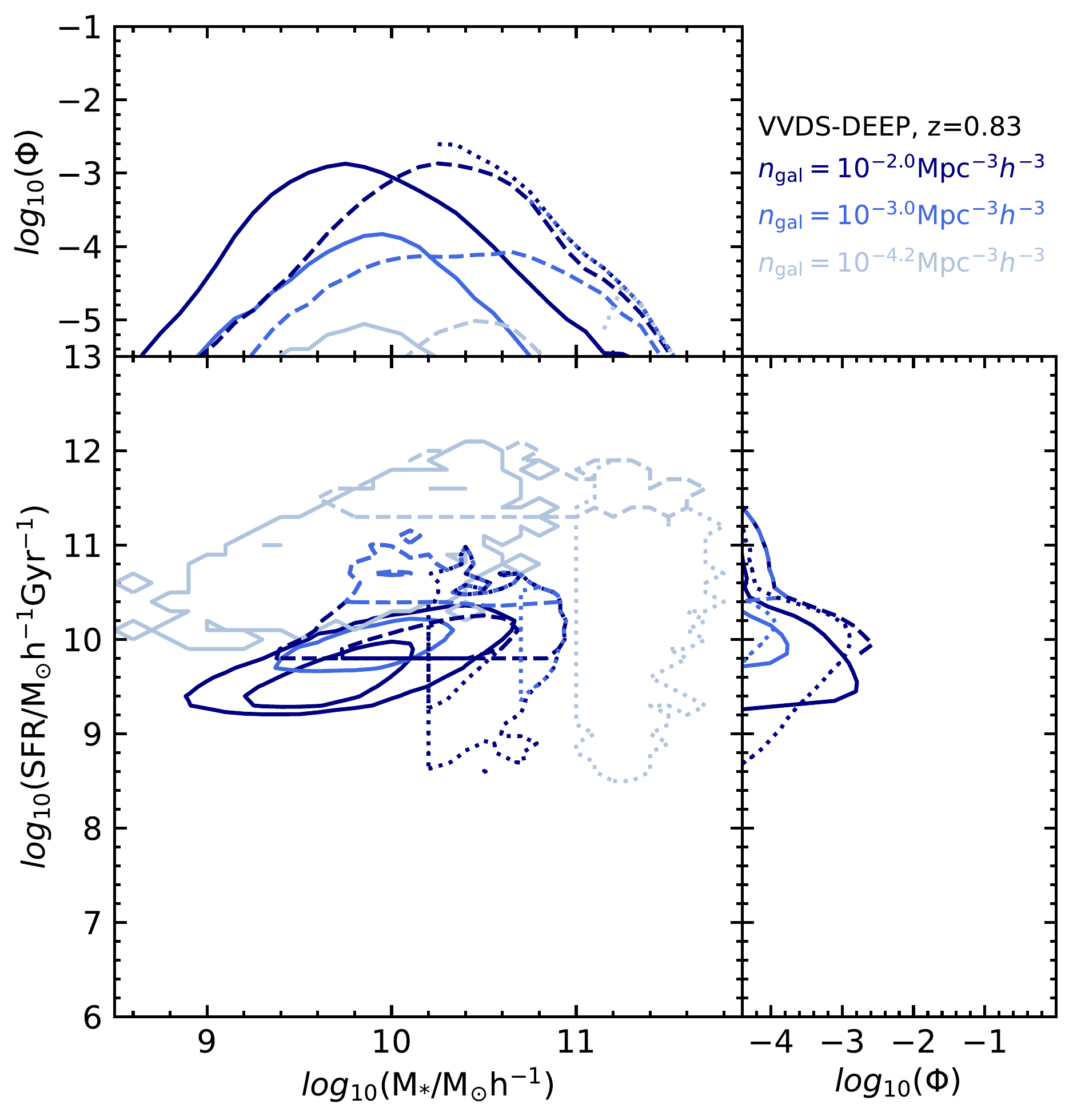}
\caption{\label{fig:compare_sfrm} The $z=0.83$ distribution of galaxies in the SFR-stellar mass plane for galaxies with the number densities indicated in the legend, selected by 
either imposing a single cut to the whole population of model galaxies (left panel) or an extra cut to the ELGs samples summarised in Table~\ref{tbl:obs} (the right panel shows VVDS-DEEP galaxies at $z=0.83$). The properties used for selecting the fixed number density samples are: stellar mass (dotted lines), SFR (dashed lines) and L\ox (solid lines). The sSFR-stellar mass plane has been collapsed into the galaxy stellar mass function, top subpanels, and the SFR function, right subpanels. The corresponding densities shown are $\Phi (h^{3}{\rm Mpc}^{-3}{\rm dex}^{-1})$.
} 
\end{figure*}

We have previously quantified how model ELGs trace the \LSE. ELGs are expected to typically trace less dense regions than mass selected galaxies, such as Luminous Red Galaxies~\citep[e.g.][]{alam2019}. Here we aim to contrast the properties of ELG samples with those selected using simpler criteria, such as stellar mass or star formation rate. This comparison will allow us to gain insight into  which aspects of the ELG populations are unique and might be a source of additional systematic errors when they are used as cosmological tracers~\citep[see e.g.][]{avila2020}.

To make a fair comparison between galaxy samples selected in different ways, we generate fixed number density samples with $n_{\rm gal}=10^{-2},10^{-3},10^{-4.2}\ h^{3}{\rm Mpc^{-3}}$. These fixed number density samples are generated by either imposing a single cut in stellar mass, SFR or L\ox or starting with the ELG samples described in \S~\ref{sec:elgsel} and then imposing an extra cut in one of the three mentioned properties.

Given either the effective or mean redshifts of the different surveys considered in this study, as described in \S~\ref{sec:modelelgs}, the analysis is done at $z=0.83$ for the eBOSS-SGC and VVDS-DEEP samples and at $z=0.99$ for the DEEP2 and DESI ones.

\subsection{Fixed number density samples}

Fig.~\ref{fig:cum} presents the cumulative abundance of the whole galaxy population and the ELGs subsamples ranked by their stellar mass, star formation rate and L\ox. From here, making cuts in these three properties, fixed number density samples are constructed with $n_{\rm gal}=10^{-2},10^{-3},10^{-4.2}\ h^{3}\ {\rm Mpc^{-3}}$, for the all model galaxies and the four ELG selections.

Fig.~\ref{fig:compare_sfrm} shows the SFR-stellar mass plane, galaxy stellar mass function and SFR function for samples with 3 different number densities selected either imposing cuts on the stellar mass, the SFR or L\ox on either the global population or the ELGs. Fig.~\ref{fig:compare_sfrm} only shows the results for the model VVDS-DEEP sample, but similar trends are found for the other ELG selections studied here, which are summarised in Table~\ref{tbl:obs}. Galaxies selected by their SFR have stellar masses spreading a large dynamical range. This is also the case for galaxies selected with a cut in L\ox. However, in this case, galaxies tend to have lower masses and SFR than the fixed number density sample selected with cuts in the SFR. Fig.~\ref{fig:compare_sfrm} shows that, as reported in GP18, the model ELG selection is not equivalent to imposing a cut in SFR.

The median mass of the host haloes increases with stellar mass limit (not shown), which decreases the number density of the galaxy sample. This trend is not found as clearly when either the SFR or L\ox are used to select the fixed number density samples. In this case, the selected galaxies spread a large range of stellar masses even for low number densities, as can be seen in Fig.~\ref{fig:compare_sfrm}. 

The median L\ox for fixed number density galaxy samples selected with a single cut in their stellar mass is below $10^{39.5}h^{-2}{\rm erg\, s}^{-1}$, while all the ELG fixed number density selections and those made with a single cut in SFR and L\ox have median L\ox above $10^{40.5}h^{-2}{\rm erg\, s}^{-1}$. Fixed number density ELGs have SFR$>10^8 h^{-1}{\rm M}_{\odot}{\rm Gyr}^{-1}$ at the studied redshifts. The median SFR increases with decreasing number density when galaxies are selected by their SFR and their L\ox.

\subsubsection{Mean HOD of fixed number density samples}
\begin{figure} \includegraphics[width=0.45\textwidth]{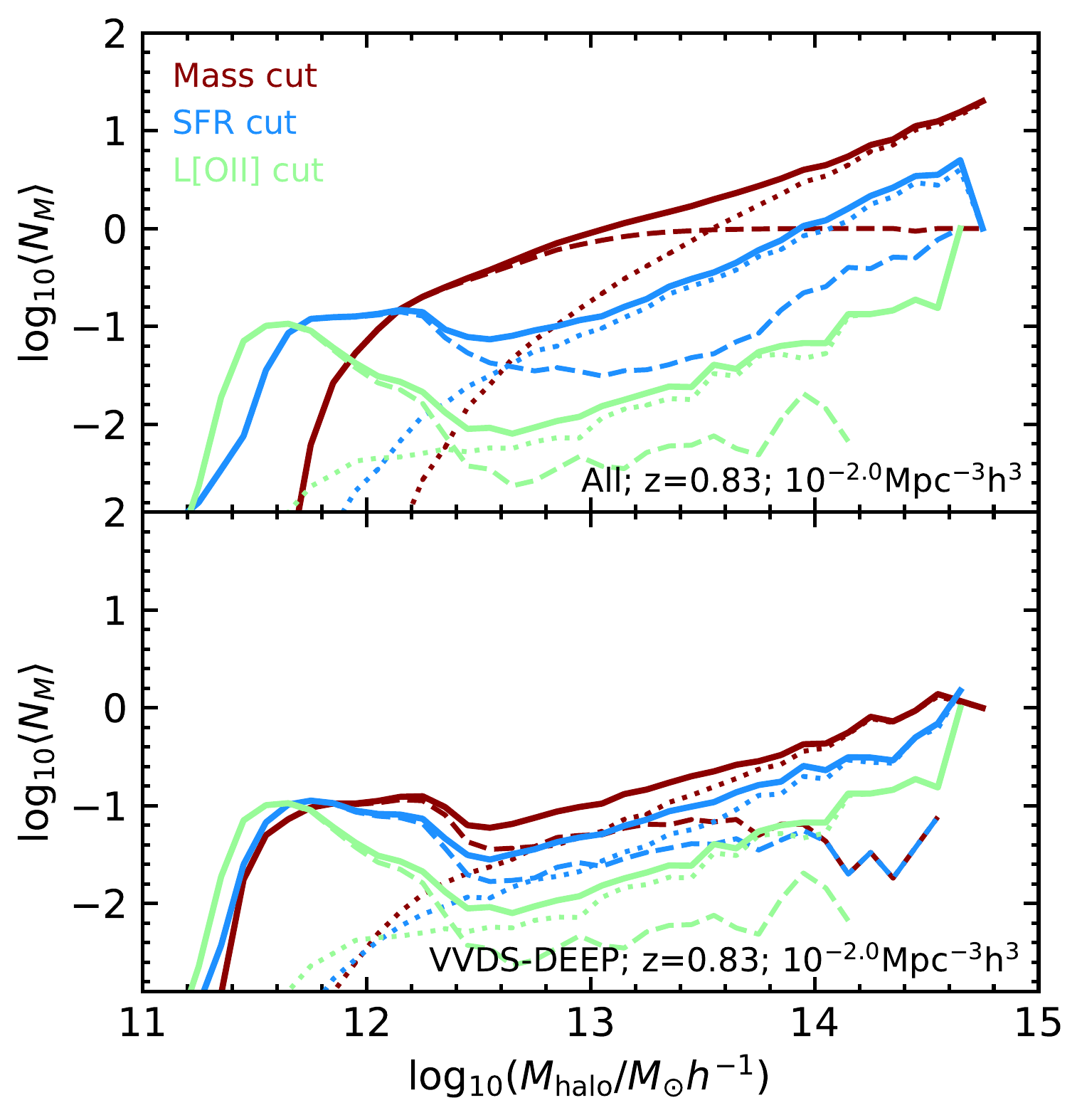}
\caption{\label{fig:compare_hod} The $z=0.83$ mean halo occupation distribution (HOD) for fixed number density samples, $10^{-2}h^{-3}{\rm Mpc}^{-3}$, of galaxies selected with either a single cut in one of the properties specified in the legend (top panel) or  applying a cut in one of those properties to the VVDS-DEEP (bottom panel). The dark red lines show the HOD when the number density selection is done using a cut in stellar mass, the blue lines when this is done using a cut in SFR and the green lines when a cut in L\ox is applied. The solid lines correspond to the total HOD, while the contribution from centrals is shown with dashed lines and that of satellites by dotted lines.
} 
\end{figure}

The mean halo occupation distribution (HOD) for fixed number density central galaxies selected with a cut in their stellar mass follows a soft step function, reaching unity: at least one galaxy of a given mass will be found in large enough haloes (see the top panel in Fig.~\ref{fig:compare_hod}). This is very different from the behaviour of fixed number density SFR selected central galaxies~\citep{zheng05,geach12,contreras13,cochrane2017,cochrane18}. As shown in Fig.~\ref{fig:compare_hod}, these follow a shape closer to an asymmetric Gaussian plus a shallow power law~\citep{cochrane2018eagle,gp18}. A star-forming galaxy is not found in all haloes above a certain mass. The top panel in Fig.~\ref{fig:compare_hod} shows that the same is true for fixed number density galaxies selectd with a single cut in L\ox. In this case, the suppression in the number of central galaxies found in massive haloes is even larger than for fixed number density samples selected with a SFR cut.

Fig.~\ref{fig:compare_hod} shows that the HOD of fixed number density samples of ELGs is very similar to that of SFR or L\ox selected samples, independently of the extra selection in either stellar mass, SFR or L\ox. This is because the ELGs we are studying are a sub-sample of star-forming galaxies. It is interesting to note that brighter \oem, have a reduced number of central galaxies in massive haloes, compared to fixed number density ELGs with an extra cut in SFR or stellar mass.

As it is shown in Fig.~\ref{fig:compare_hod}, the HOD of samples selected by their SFR or L\ox have a larger number of central galaxies with low masses and a lower number of satellites at larger masses, when compared with stellar mass selected samples. This difference gets larger for decreasing number densities. Such difference is reduced for the ELG samples with fixed number densities.

Fig.~\ref{fig:compare_hod} also shows that fixed number density L\ox selected samples populate slightly less massive haloes than the SFR selected samples. This difference is reduced for the fixed number density ELG samples, in particular for lower number densities.

Although Fig.~\ref{fig:compare_hod} only shows the results for the fixed number density VVDS-DEEP samples, similar trends are found for the other fixed number density ELG selections explored here, except for the DESI one. In this case, applying the different cuts result in minimal variations. 

Compared to the global population, the minimum halo masses needed to host an ELG selected with a fixed number density are closer among the the cuts using the three properties studied here, stellar mass, SFR and L\ox. Despite the similarities, the effective bias of these samples are different, as it is described in the next section, ~\S\ref{sec:ndxi}.

For the fixed number density stellar mass selections, the minimum halo mass needed to find a galaxy increases for smaller number densities. This trend is also reflected in the increase of the effective bias from fixed number density stellar mass selected samples, as it can be seen in Table~\ref{tbl:bias}. This is not as clear for fixed number density SFR and L\ox selections. 


\subsubsection{Clustering of fixed number density samples}\label{sec:ndxi}
\begin{figure}
\includegraphics[width=0.45\textwidth]{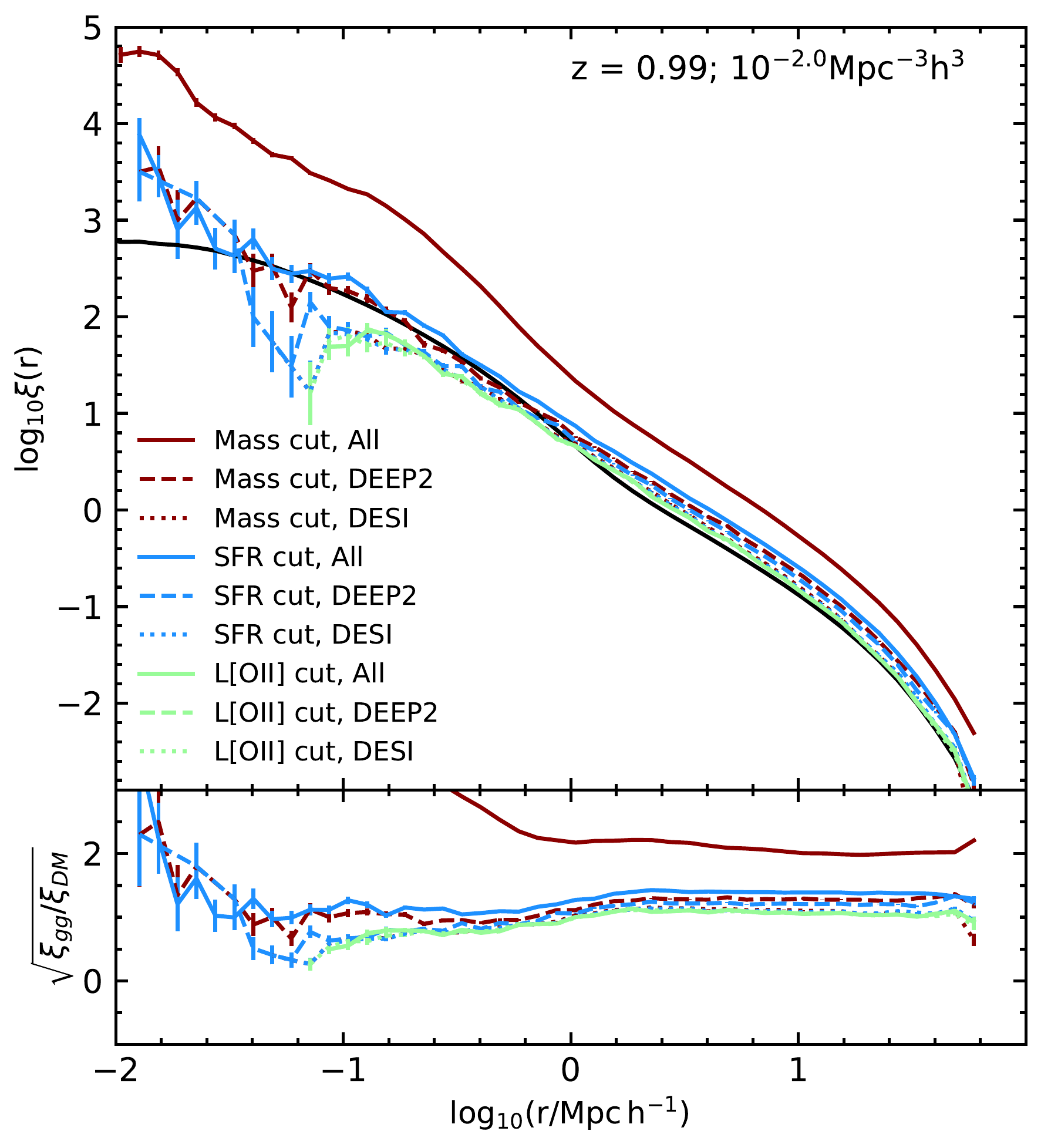}
\caption{\label{fig:xi} The real-space two point correlation function for model galaxies with a number density of $10^{-2}h^{3}$Mpc$^{-3}$ at redshift $z=0.99$, selected using different criteria, as indicated in the legend. The black solid line shows the dark matter correlation function. {\it Bottom panel:} The real space ratio $\sqrt{\xi_{\rm gg}/\xi_{\rm DM}}$. Poisson error bars are shown in both panels.
} 
\end{figure}
\begin{table*}
\caption{Bias for the fixed number density samples described in~\S\ref{sec:nd}, at $z=0.83$ and $z=0.99$. The bias is tabulated for the stellar mass, SFR and L\ox selected samples separated by a comma. The bias and associated error have been obtained as described in~\S\ref{sec:xielg} in the range $8\leq r (h^{-1}{\rm Mpc}) \leq 50$.
}
\label{tbl:bias}
\begin{center}
\begin{tabular}{ |c|c|c|c|c| } 
\hline
z & Survey & $10^{-2}h^{3}$Mpc$^{-3}$ & $10^{-3}h^{3}$Mpc$^{-3}$ & $10^{-4.2}h^{3}$Mpc$^{-3}$\\
\hline
0.83 & All & 1.90$\pm$0.01, 1.27$\pm$0.01, 0.99 $\pm$0.01 & 2.56$\pm$0.03, 1.21$\pm$0.03, 0.97 $\pm$0.05 & 4.36$\pm$0.06, 1.48$\pm$0.68, 1.76 $\pm$0.96 \\
 & VVDS-DEEP & 1.26$\pm$0.01, 1.13$\pm$0.01, 0.99 $\pm$0.01 & 1.50$\pm$0.04, 1.18$\pm$0.02, 0.96 $\pm$0.09 & 2.35$\pm$0.40, 1.11$\pm$0.71, 1.83 $\pm$0.91 \\
 & eBOSS-SGC & -, -, - & 1.23$\pm$0.01, 1.12$\pm$0.01, 1.05 $\pm$0.01 & 1.42$\pm$0.02, 1.20$\pm$0.01, 1.22 $\pm$0.69 \\ 
\hline
0.99 & All & 2.00$\pm$0.01, 1.39$\pm$0.01, 1.06 $\pm$0.01 & 2.70$\pm$0.03, 1.29$\pm$0.06, 1.15 $\pm$0.01 & 4.28$\pm$0.45, 1.97$\pm$0.95, 1.80$\pm$0.29 \\ 
 & DEEP2 & 1.28$\pm$0.01, 1.21$\pm$0.01, 1.06 $\pm$0.01  & 1.57$\pm$0.06, 1.29$\pm$0.01, 1.16$\pm$0.04 & 2.95$\pm$0.63, 1.36$\pm$0.76, 1.52 $\pm$0.68 \\ 
 & DESI &  1.09$\pm$0.01, 1.08$\pm$0.01, 1.05 $\pm$0.01  & 1.28$\pm$0.04, 1.17$\pm$0.06, 1.08 $\pm$0.06  & 1.27$\pm$0.80, 1.70$\pm$0.96, 1.85 $\pm$1.09 \\ 
\hline
\end{tabular}
\end{center}
\end{table*}
The real-space two point correlation function for galaxies with a fixed number density of $10^{-2}h^{3}$Mpc$^{-3}$ at redshift $z=0.99$, is shown in Fig.~\ref{fig:xi}. The calculation of the two point correlation function has been done following the description in~\S\ref{sec:xielg}.

Fig.~\ref{fig:xi} shows that at large scales, $r>8h^{-1}$Mpc, the two point correlation function of fixed number dentisy SFR and L\ox selected galaxies remain close, independently of starting with the whole galaxy population or ELGs. In fact, at a given number density, the bias of SFR and L\ox cut samples for all galaxies and ELGs at $z=0.83$ and $z=0.99$ remain within a 0.6 range (0.3 if only number densities above $10^{-4.2}h^{3}$Mpc$^{-3}$  are considered). The bias of all studied samples can be seen in Table~\ref{tbl:bias}. The bias has been calculated in the range $8\leq r (h^{-1}{\rm Mpc}) \leq 50$, as $\sqrt{\xi_{\rm gg}/\xi_{\rm DM}}$. 

Both Fig.~\ref{fig:xi} and Table~\ref{tbl:bias} show that galaxies selected with a single cut in stellar mass are more clustered than the rest of the samples (solid red line versus the rest in Fig.~\ref{fig:xi}). Although this is also true at large scales for fixed number density ELGs selected with an extra cut in stellar mass, in these cases the differences are much smaller (see the dashed and dotted red lines versus the blue ones in Fig.~\ref{fig:xi}, for the case of DEEP2 selected galaxies). There is one exception to this: the least dense DESI sample, for which all selections are consistent at large scales. 

Table~\ref{tbl:bias} shows that, except for the DESI sample, the bias of mass selected galaxies grow with lower number densities. Such a trend does not seem to exist for the other galaxy selections.

At large scales, the SFR  and L\ox cut samples trace closely the dark matter clustering, with biases, between 0.95 and 1.4 for samples with number densities above $10^{-4.2}h^{3}$Mpc$^{-3}$ (see Table~\ref{tbl:bias}). For these number densities, the L\ox sample has bias slightly lower than the SFR one, being closer to 1. The clustering in the lowest studied number density bin becomes very noisy and despite the biases reaching values close to 2, their corresponding error bars are close to 1. 

The bias of galaxies selected with a single cut in either SFR or L\ox are comparable to that of ELGs with a fixed number density.

As shown in Fig.~\ref{fig:xi}, at small scales, the clustering of ELGs and galaxies selected with a single cut in SFR are different, except for the DESI-like sample. The clustering of this sample is consistent within the Poisson error bars for both mass and SFR selected samples. 

Fig.~\ref{fig:xi} shows that pairs of galaxies selected by their L\ox are not found at the shortest separations found for the stellar mass or SFR samples. This is the case for all the studied selections. At $z=0.99$ and $10^{-2}h^{3}$Mpc$^{-3}$ (shown in Fig.~\ref{fig:xi}), no pairs of L\ox selected galaxies are found with separations $r\lesssim0.04h^{-1}$Mpc. At $z=0.83$ for the same number density, no L\ox selected galaxies are found with $r\lesssim0.03h^{-1}$Mpc. This values increase for decreasing number densities, a trend also seen for the other selections. The difference seen in Fig.~\ref{fig:xi} is striking. This difference is smaller for VVDS-DEEP at $z=0.83$ and for lower number density samples. Nevertheless, this difference is worth exploring as its origin is unclear. In Fig.~\ref{fig:compare_sfrm}, different galaxy selections are compared in the SFR-stellar mass plane. From here it is clear that very different ranges of stellar mass and SFR are covered by the different samples. These will cause differences in the clustering. We defer to an other study the intra-halo analysis of these samples, needed to better understand the differences in the clustering of galaxies within the same halo, the 1-halo term.

\subsection{Fixed number density samples in the Cosmic Web}\label{sec:cosmicweb}

\begin{figure}
\includegraphics[clip, trim=2.2cm 0.2cm 2.5cm 0.5cm,width=0.48\textwidth]{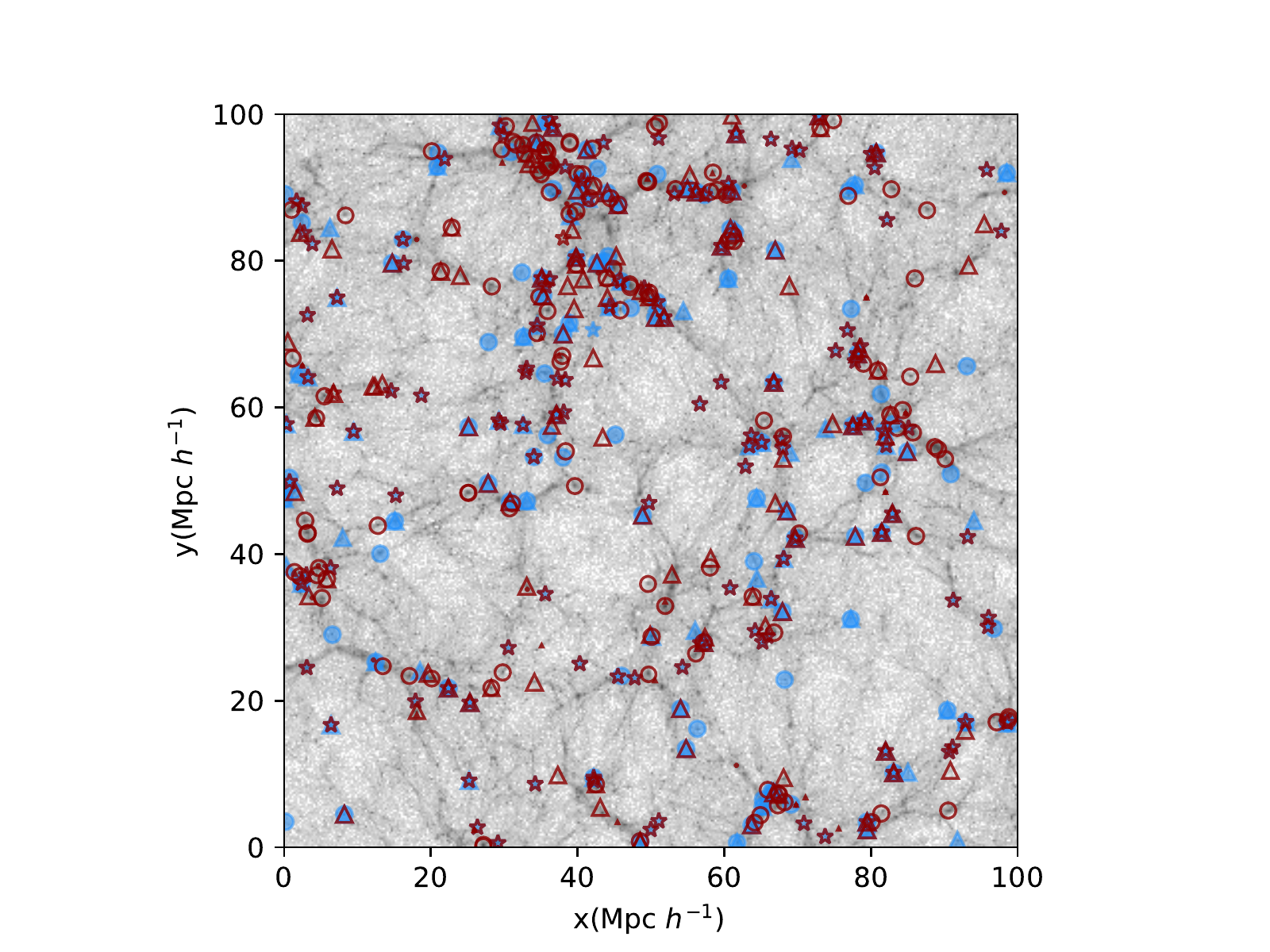}
\caption{\label{fig:pretty} Distribution of galaxies with a number density of $10^{-2}h^{3}$Mpc$^{-3}$, on top of the smooth underlying dark matter distribution (grey). This slice of 10$h^{-1}$Mpc thickness is taken from the MS-W7 simulation at $z=0.99$. Red symbols show fixed number density galaxies selected with a single cut in stellar mass (circles), SFR (triangles) or L\ox (stars), while blue ones show those fixed number density DEEP2 galaxies selected with an extra cut on the mentioned properties. The area of the symbols is proportional to the log$_{10}(L_{\rm [OII]})$.
} 
\end{figure}

Here we study the fixed-number-density samples constructed in Section~\S\ref{sec:nd}, to understand ELGs compared to mass and SFR selected samples within the cosmic web. Fig.~\ref{fig:pretty} presents a $100\times 100\times 10h^{-3}{\rm Mpc}^3$ slice of the whole simulation box at redshift $z=0.99$, highlighting in grey the cosmic web of the dark matter, together with the location of galaxies with a fixed number density, $10^{-2}h^{3}$Mpc$^{-3}$, selected with single cuts on their stellar mass, SFR or L\ox and ELGs with and extra cut on these properties, as described previously. Fig.~\ref{fig:pretty} shows that, at least qualitatively, even when the number density is fixed, star-forming galaxies tend to trace less dense environments than mass selected samples. In Fig.~\ref{fig:pretty} it is unclear if there are significant differences between the \LSE traced by ELGs and galaxies selected by their SFR to have the same number density. In this section we attempt to quantify the \LSE of ELGs and galaxies selected by their stellar mass and SFR.

\subsubsection{Large scale environment distribution}
\begin{figure*}
    \centering
    \begin{subfigure}[b]{\textwidth}
    \includegraphics[width=.33\textwidth]{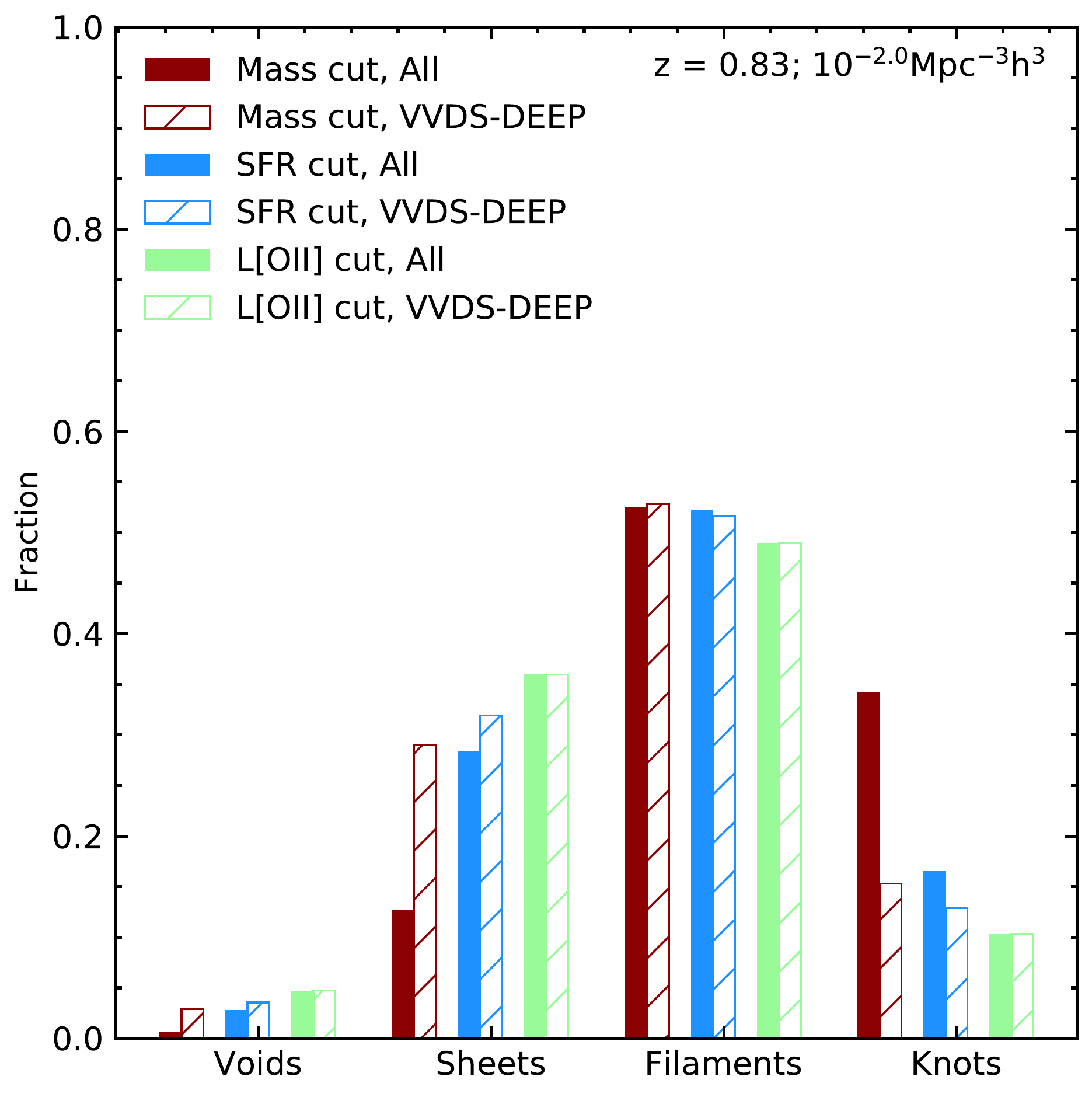}
    \includegraphics[width=.33\textwidth]{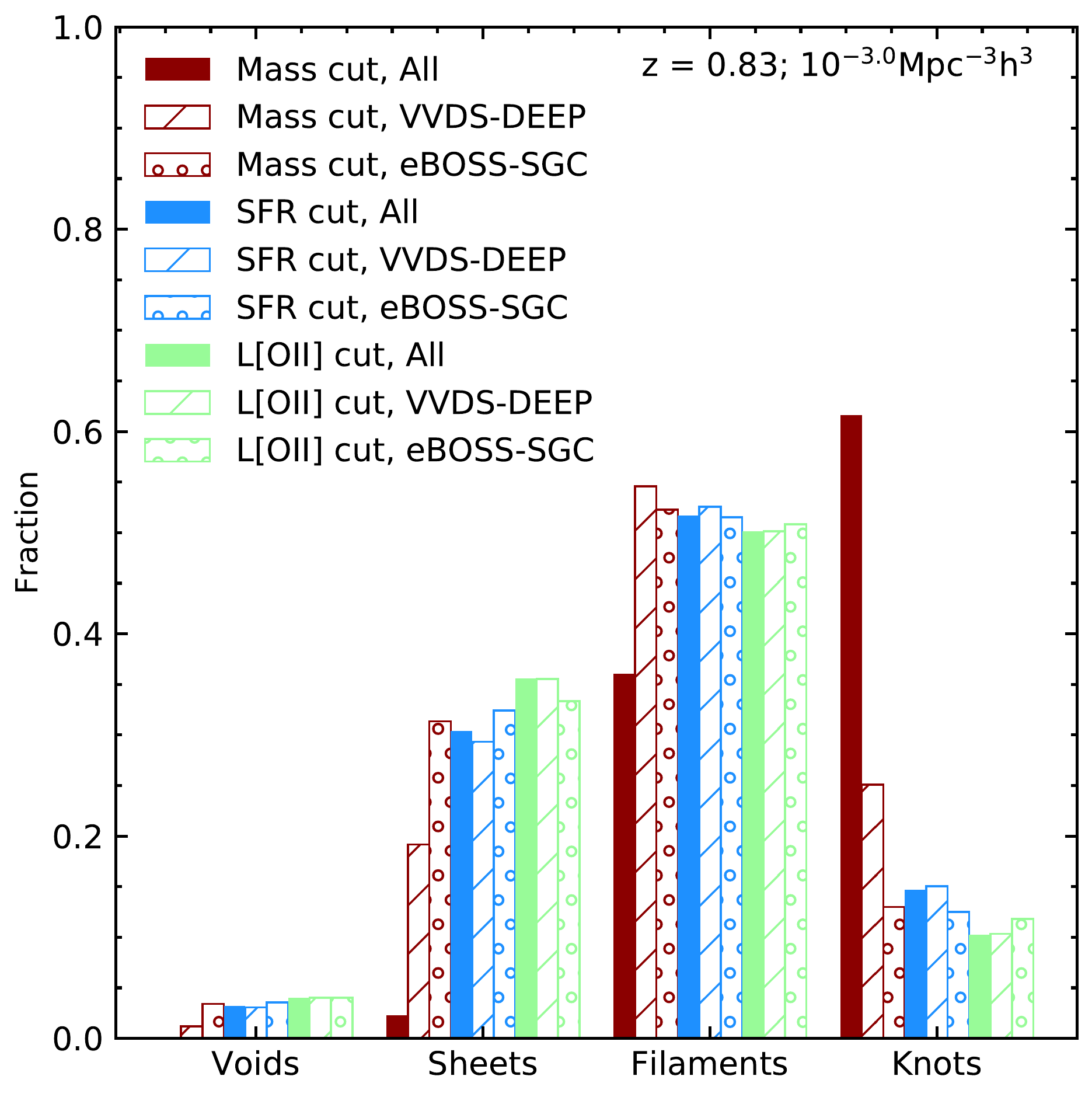}
    \includegraphics[width=.33\textwidth]{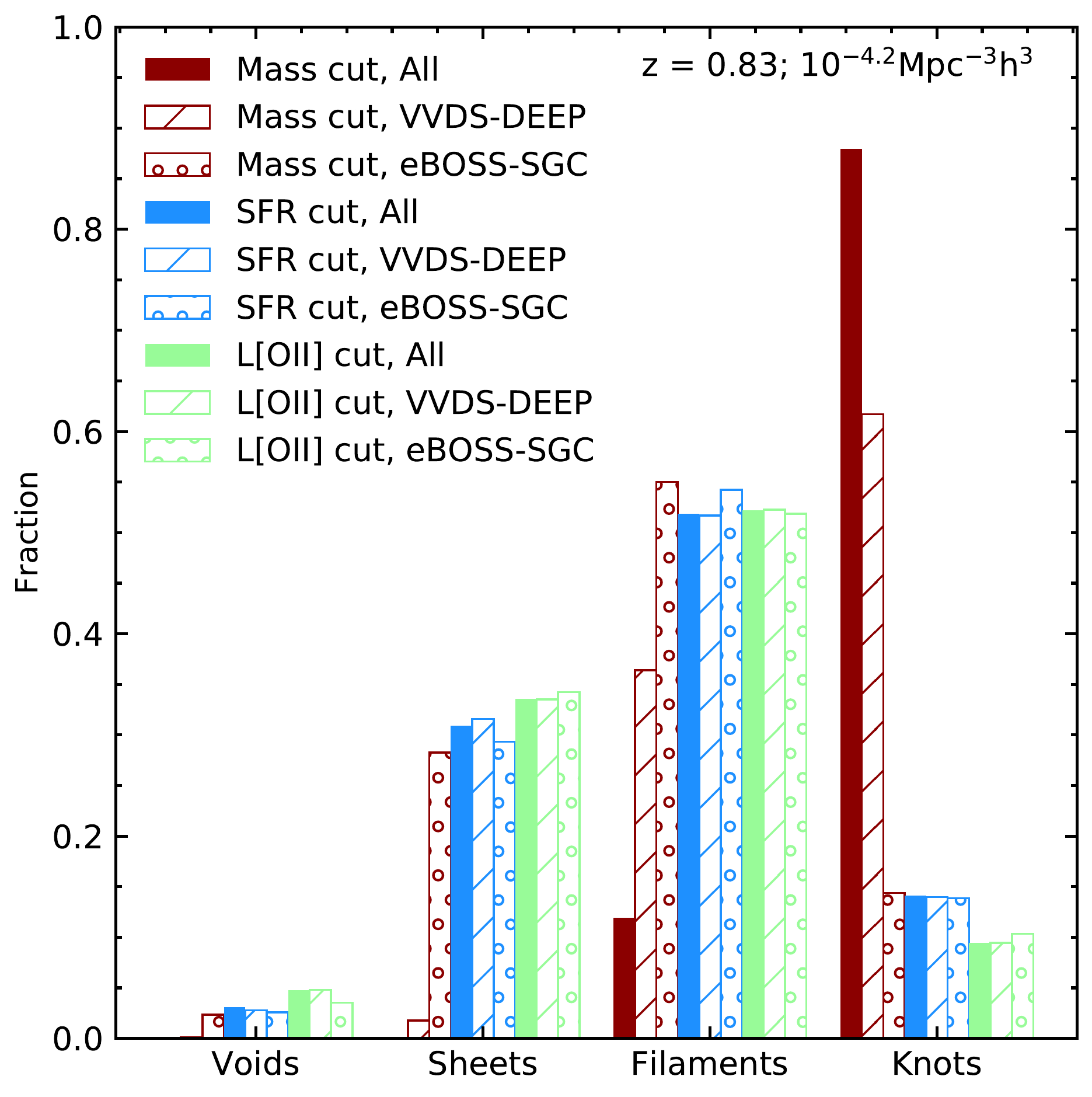}
    \end{subfigure}
    \begin{subfigure}[b]{\textwidth}
    \includegraphics[width=.33\textwidth]{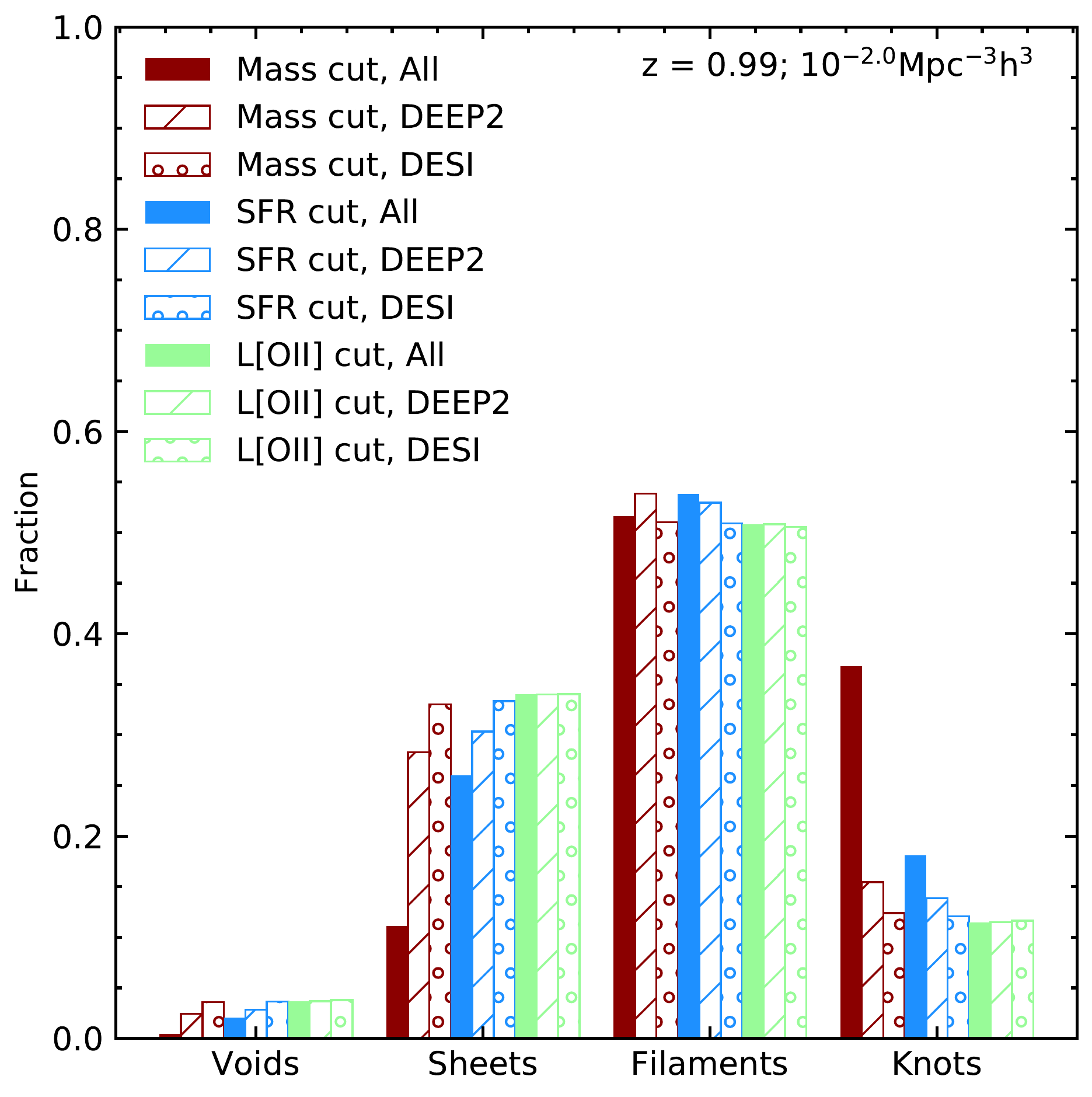}
    \includegraphics[width=.33\textwidth]{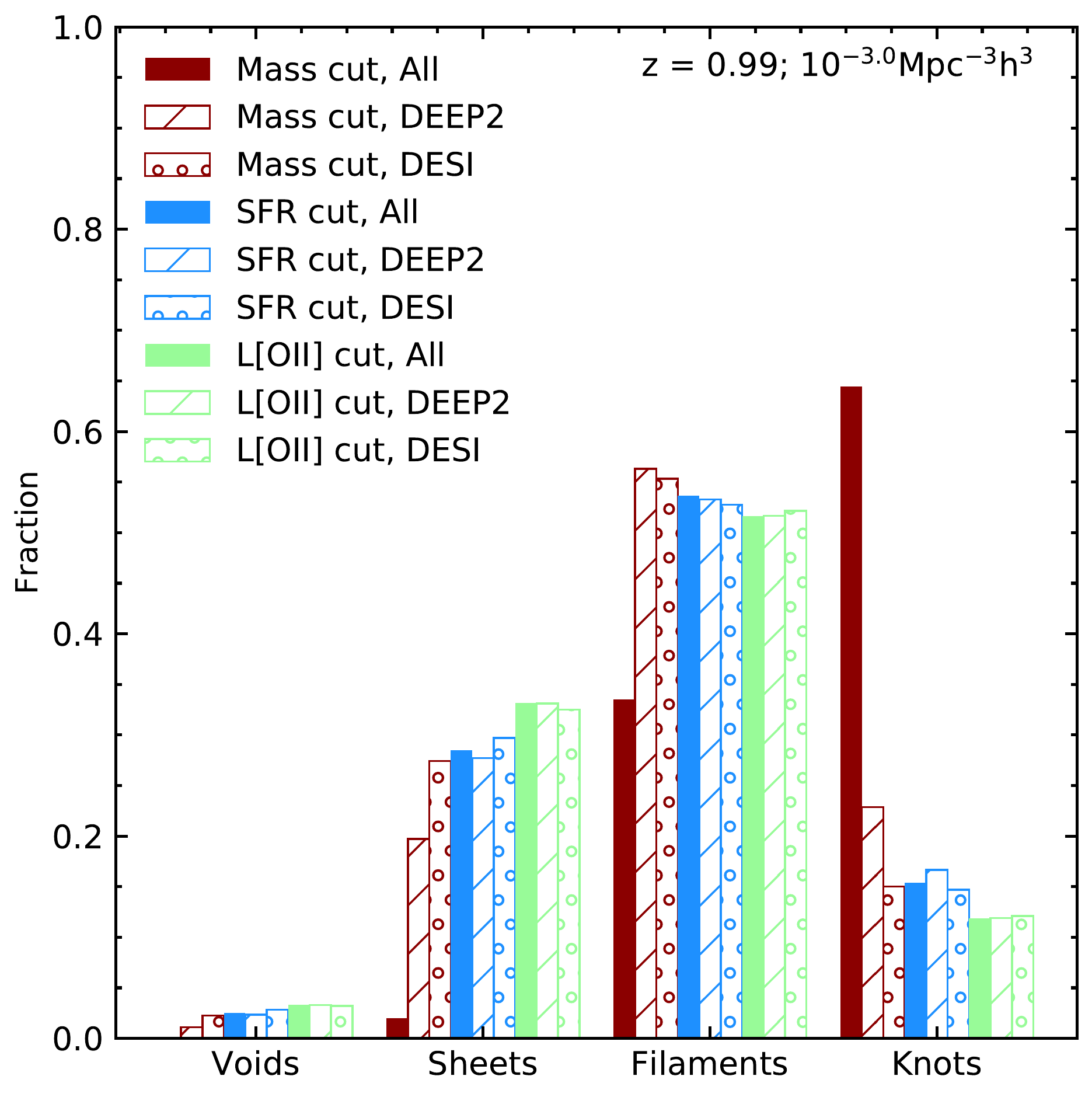}
    \includegraphics[width=.33\textwidth]{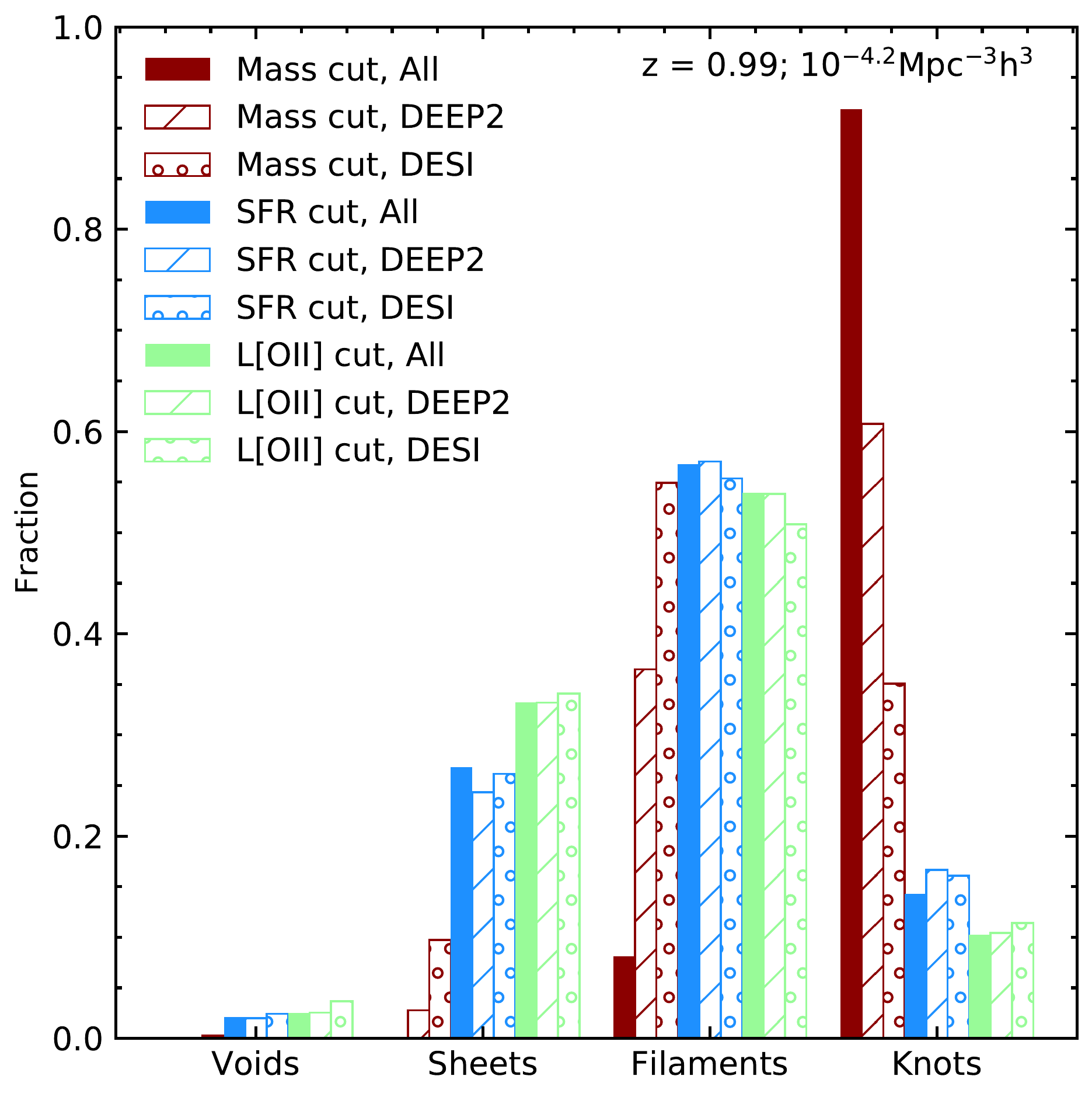}
    \end{subfigure}
\caption{\label{fig:environ} Histograms with the fraction of fixed-number-density samples of galaxies in voids, sheets, filaments and knots. The \LSE has been classified using \vweb (see \S~\ref{sec:LSE} for details). The top row show galaxies at $z=0.83$ and the bottom row at $z=0.99$. Selections with number densities of $10^{-2}h^3{\rm Mpc}^{-3}$ are shown in the left column, with $10^{-3}h^3{\rm Mpc}^{-3}$ in the middle column and with $10^{-4.2}h^3{\rm Mpc}^{-3}$ in the right column. Each panel shows galaxies selected with a single stellar mass, SFR or L\ox cut and those adding these cuts to the ELG selections, as indicated in the legend. Note that there are no eBOSS-SGC galaxies with number densities of $10^{-2}h^3{\rm Mpc}^{-3}$ at $z=0.83$, as it can be seen in Fig.~\ref{fig:cum}.
} 
\end{figure*}

Following the methods described in \S~\ref{sec:LSE}, we classify the \LSE into voids, sheets, filaments and knots using a velocity-shear-tensor algorithm, \vweb, with a 0.1 threshold, for the samples of galaxies with fixed number density constructed in~\S\ref{sec:nd}. Fig.~\ref{fig:environ} compares selections in stellar mass, SFR and L\ox with the same number density. For most of the galaxy selections considered, about half of these galaxies populate filaments. This is not the case for the mass selected sample with number densities below $10^{-2}h^3{\rm Mpc}^{-3}$ and for DEEP2 and VVDS-DEEP galaxies with an extra cut in stellar mass to achieve the lowest number density studied here, $10^{-4.2}h^3{\rm Mpc}^{-3}$.

As expected, samples  based on a single stellar mass cuts have a higher presence in knots than the rest of the selections, which are star-forming galaxies. For all samples, the presence in knots increases for lower number densities.

L\ox selected samples trace the same \LSE structures, independently of being selected with just a single cut in L\ox or not. This stresses that, at least for the studied number densities, the particular magnitude and colour cuts applied to select ELGs are secondary to the L\ox limits. 

L\ox selected samples are, in general, more present in sheets, $\sim30$ per cent, than galaxies selected with a single cut in SFR. For number densities below $10^{-2}h^3{\rm Mpc}^{-3}$, L\ox galaxies are about 5 per cent more present in sheets, and less present in knots, than ELGs selected in other ways. This is accordant with  the difference found for the clustering of their the 1-halo term, reported in \S~\ref{sec:ndxi}.

All the studied ELG selections are distributed in the cosmic web close to that of samples with the same number density based on a single SFR cut for number densities above $10^{-4.2}h^3{\rm Mpc}^{-3}$, with differences below a 0.11 ratio. ELGs selected with a number density of $10^{-2}h^3{\rm Mpc}^{-3}$ are about 5 per cent more present in sheets than the sample selected only by a cut in SFR.

The ELG sample with an extra cut in SFR to fix the number density, closely follow the distribution of the SFR sample, with differences in fractions up to 0.07. The differences between the ELG sample with an extra cut in stellar mass and the SFR one increase with decreasing number densities. At the lowest studied number density, the ELGs with an extra stellar mass cut have a much larger presence in knots than the SFR sample, with difference in fractions up to 0.48.

Above 70 per cent of the model ELG samples with number densities $10^{-2}h^3{\rm Mpc}^{-3}$ and $10^{-3}h^3{\rm Mpc}^{-3}$ are found in either filaments or sheets and about half of them are indeed in filaments. For the samples with a number density of $10^{-4.2}h^3{\rm Mpc}^{-3}$, this is only true for ELG samples with an extra cut in SFR or L\ox; for stellar mass selected ELG samples, the percentage drops for all the ELGs, except for the eBOSS-SGC.

The environmental split does agree with the differences in the clustering amplitudes reported in \S~\ref{sec:ndxi}. Fig.~\ref{fig:xi} shows that when more galaxies are found in knots the 1-halo term of the two point correlation function is much higher than that of the dark matter.

We have done a similar analysis but classifying the \LSE with a tidal-tensor algorithm, \pweb, with a $0.005s^{-2}$ threshold. 
The results with \pweb are quantitatively similar to those described above and can be seen in the Appendix~\ref{App:pweb}. 

ELGs and L\ox selected galaxies tend to occupy either filaments or sheets. ELGs and L\ox selected galaxies roughly populate the same \LSE as galaxies selected based on their SFR, for number densities above or equal to $10^{-3}h^3{\rm Mpc}^{-3}$. Below this number density, the differences can be large for stellar mass selected ELGs, in particular in knots and voids.

\subsubsection{Comparison of global properties}\label{sec:props}

\begin{figure*}
\includegraphics[height=0.25\textheight]{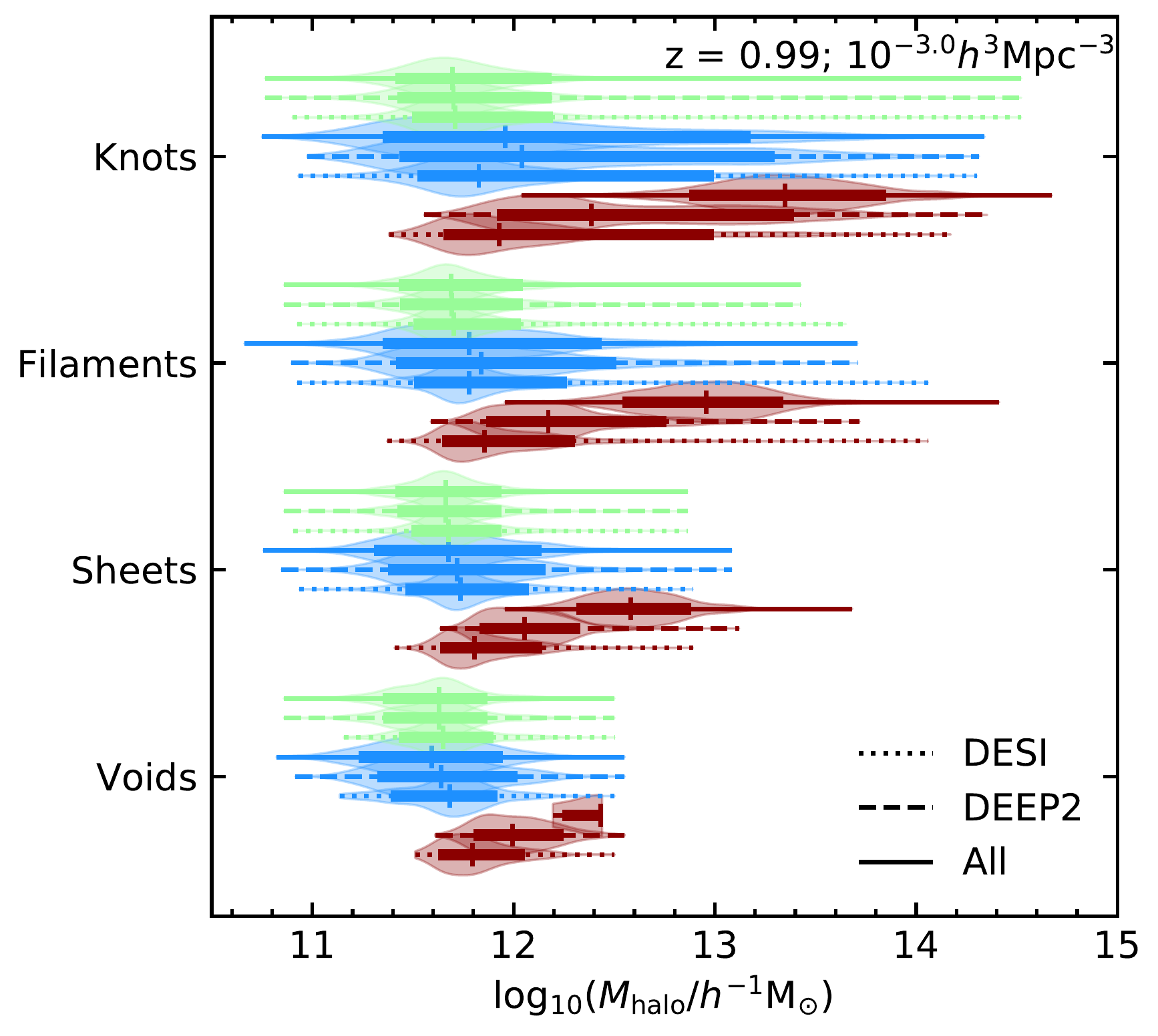} 
\includegraphics[height=0.25\textheight]{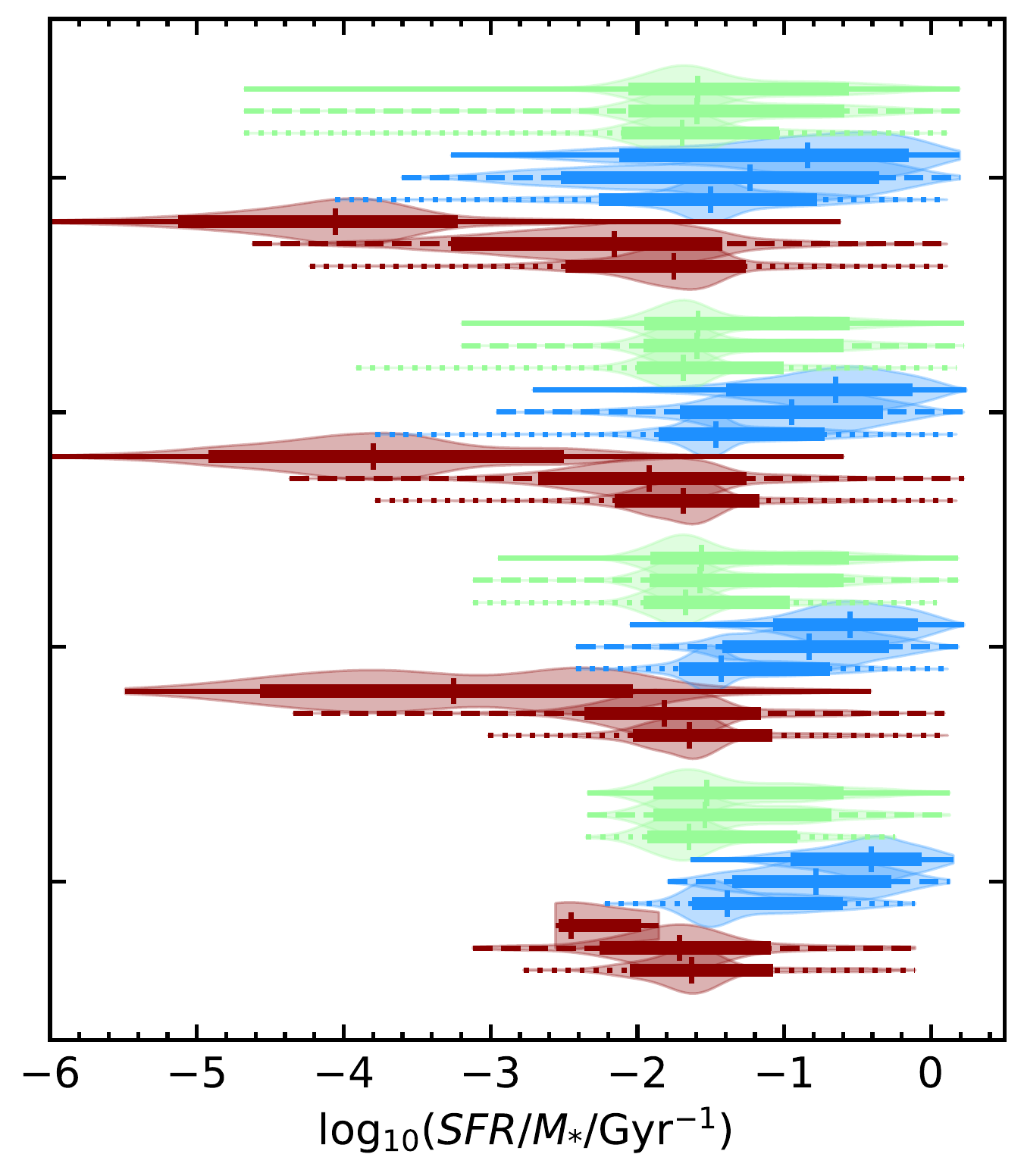}  
\includegraphics[height=0.25\textheight]{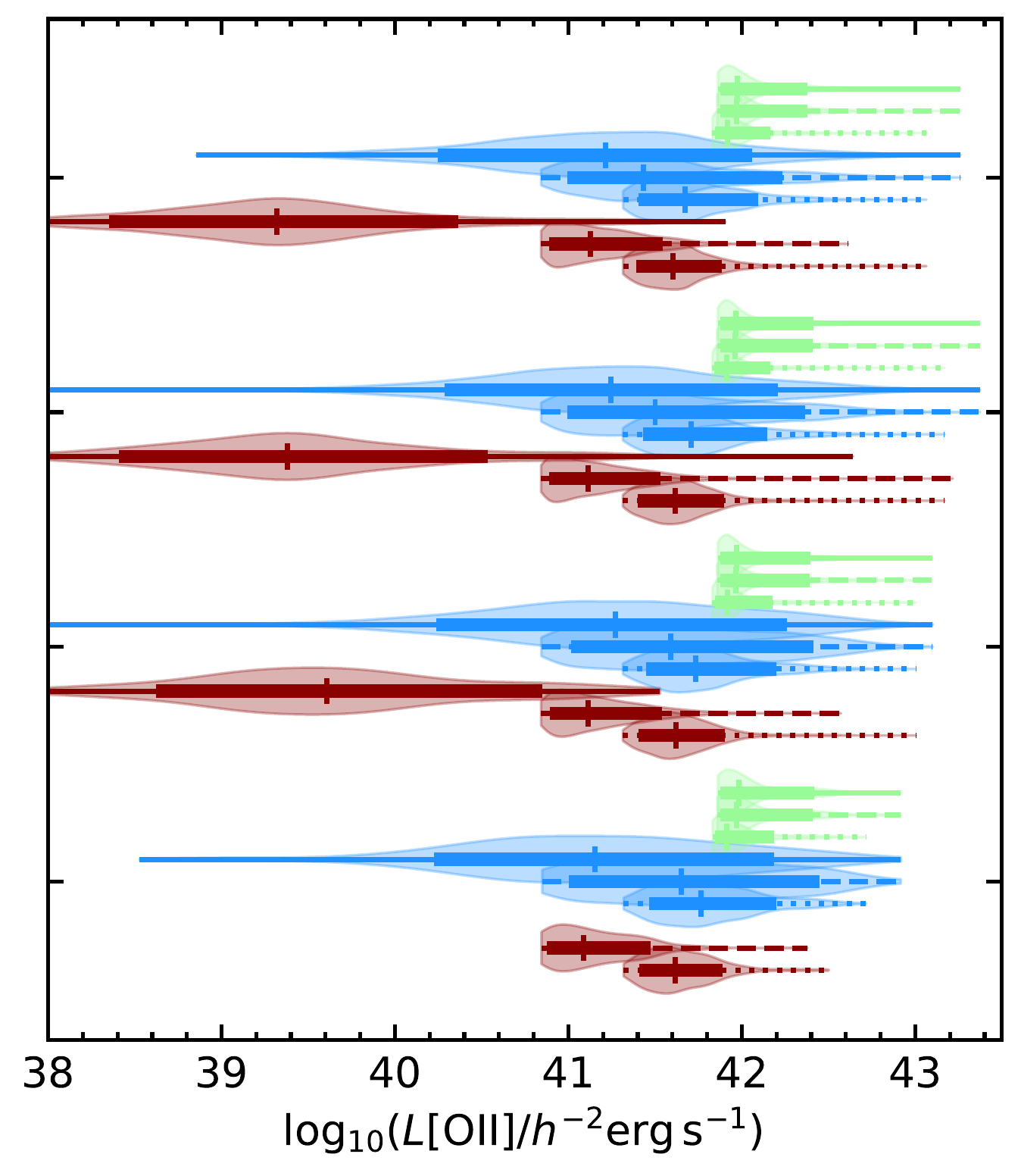}
\caption{\label{fig:violin} Distribution of the host halo mass, left panel, specific SFR, middle panel, and $L$\ox, right panel, of galaxies with a fixed number density of $10^{-3}h^3Mpc^{-3}$ at $z=0.99$, in the knots, filaments, sheets and voids classifications of the \LSE done with the \vweb algorithm. The distributions of L\ox selected samples is shown in green, SFR selected samples in blue and those for stellar mass selected samples in red. Horizontal lines indicate the extent of the variables for each galaxy selection. Solid lines correspond to selections made with a single cut in either stellar mass or SFR, while those on top of the ELG selections summarised in Table~\ref{tbl:obs} are shown by dashed or dotted lines as indicated in the legend. For each selection, the vertical line indicates its median, the thick horizontal line the range of the 10 to 90 per cent of the distribution and the violin type areas show the data distribution using a Gaussian kernel density estimation. Note that galaxies selected with a single mass cut have no \ox emission in voids.
} 
\end{figure*}
For a given galaxy sample, the median stellar mass is comparable for galaxies in knots, filaments, sheets and voids. As expected, the maximum stellar mass of galaxies decreases towards less dense environments, i.e. it decreases from knots to voids. A similar trend is seen for the host halo mass of the galaxies, shown in the left panel in Fig.~\ref{fig:violin} for galaxies selected at $z=0.99$ with $10^{-3}h^3{\rm Mpc}^{-3}$. This trend does affect the distribution of host halo masses, such that median halo masses can decrease from knots to voids.

The median SFR is comparable for a given sample of galaxies in knots, filaments and sheets. In voids, stellar mass selected galaxies have minimum SFR above those for knots, filaments and Sheet. For example, galaxies with $10^{-3}h^3{\rm Mpc}^{-3}$ at $z=0.99$ in voids have SFR$>10^9h^{-1}{\rm M}_\odot{\rm Gyr}^{-1}$, while in the other large-scale structures galaxies with
SFR$<10^7h^{-1}{\rm M}_\odot{\rm Gyr}^{-1}$ can be found. This suggests that galaxies in voids at $z\sim1$ are less affected by the quenching of the star formation than in other \LSE. A similar difference between voids and the other large-scale structures is found for the distribution of mass selected samples as a function of specific SFR, as shown in the middle panel of shown in Fig.~\ref{fig:violin} for galaxies selected with $10^{-3}h^3{\rm Mpc}^{-3}$ at $z=0.99$. However, the distribution as a function of specific SFR of galaxies selected with a single stellar mass cut has a larger variation with the large-scale structure. In Fig.~\ref{fig:violin} the differences between filaments, sheets and voids are clear for galaxies selected with a single cut in stellar mass. This trend is in agreement with star formation being quenched more effectively in the densest large-scale structures for a mass selected sample. This is not as clear for star-forming galaxies, for which a minimum SFR or L\ox has already been imposed.

The right panel in Fig.~\ref{fig:violin} shows as a function of $L$\ox the distribution of galaxies with $10^{-3}h^3{\rm Mpc}^{-3}$ at $z=0.99$. The distributions are comparable for galaxies in different large-scale structures. 

The trends discussed above are found for the classifications of the \LSE done with both the \vweb and \pweb algorithms.

\begin{table}
\caption{Percentage of satellite galaxies for the fixed number density samples with $10^{-3}h^{3}$Mpc$^{-3}$ (see~\S\ref{sec:nd}), at $z=0.83$ and $z=0.99$. The percentage of satellites is tabulated for the stellar mass, SFR and L\ox selected samples separated by a comma, for each \LSE environment classified using the \vweb algorithm (see \S~\ref{sec:cosmicweb}). 
}
\label{tbl:fsat}
\begin{center}
\begin{tabular}{ |c|c|c|c| } 
\hline
z=0.83 & All & VVDS-DEEP & eBOSS-SGC \\
voids & 0.0, 1.8, 1.6 &  0.7, 1.8, 1.6 &  1.6, 1.3, 0.8 \\
sheets & 0.3, 1.4, 1.7 & 1.3, 1.3, 1.7 & 2.6, 2.2, 2.4 \\
filaments & 2.5, 2.3, 3.4 & 3.7, 2.2, 3.4 & 5.2, 4.5, 4.4 \\
knots & 8.7, 4.3, 6.1 & 8.8, 4.5, 6.0 & 14.3, 12.4, 10.2 \\ 
\hline
z=0.99 & All & DEEP2 &  DESI  \\
voids & 0.0, 1.6, 1.7 &  0.7, 0.7, 1.7 & 0.7, 0.3, 1.5 \\
sheets & 0.6, 1.4, 1.6 & 0.9, 1.5, 1.6 & 0.9, 1.3, 1.7 \\
filaments & 2.1, 2.6, 2.8 & 2.6, 2.2, 2.8 & 2.7, 2.4, 3.0 \\
knots &  7.9, 5.4, 7.3 & 6.2, 4.4, 7.3 & 5.4, 5.1, 7.5 \\
\hline
\end{tabular}
\end{center}
\end{table}

The percentages of satellites in different \LSE with $10^{-3}h^{3}$Mpc$^{-3}$ are summarised in Table~\ref{tbl:fsat}. The percentage of satellite galaxies decreases with number density, as rarer objects are more likely to be central galaxies. For all the studied number densities, less than 10 per cent of model galaxies in either voids, sheets or filaments are satellites. In knots, this percentage increases and it can go up to 30 per cent for stellar mass and SFR selected samples with a fixed number density of $10^{-2}h^{3}$Mpc$^{-3}$. The percentage of satellites within the L\ox selected sample remains below 15 per cent in all studied cases. Groups and clusters of galaxies tend to occur in knots and thus, the fraction of satellite galaxies is expected to be larger there~\citep[e.g][]{guo2015}. We find here that this trend is maintained also for star-forming galaxies, although the variation is slightly smaller, as can be seen in Table~\ref{tbl:fsat}. The percentage of satellite galaxies in different \LSE are comparable between the \vweb and \pweb classifications. 


\section{Summary and conclusions}\label{sec:conclusions}

Star-forming emission line galaxies (ELGs) are being targeted by current and future cosmological redshift surveys. Here we have studied how they populate the cosmic web structure according to a semi-analytical model (SAM) of galaxy formation and evolution. In the future, we expect observational studies to characterise how ELGs trace the cosmic web. Here we have also contrasted ELGs with samples selected in simpler, and perhaps more generic, ways, by ranking galaxies in properties such as stellar mass or star formation rate and applying a cut.

We have used a new flavour of the semi-analytical model \gl run on the MS-W7 dark matter only simulation, with a WMAP7 cosmology and a simulation box of side $500h^{-1}$Mpc. This new version improves on the model presented in~\citet{gp18} by (i) stripping the gas in satellites more slowly, such that the observed passive fraction at $z=0$ is better matched (see Fig.~\ref{fig:pf_cal} in \S~\ref{sec:model}); and (ii) including the updated treatment of the evolution of supermassive black holes introduced in~\citet{griffin}. The last point is relevant for this work as this improvement results in a different evolution of the AGN feedback. The model has been calibrated against local observations.

Model ELGs (driven by star-formation rather than nuclear activity) are selected by imposing cuts on apparent magnitude and \ox flux to mimic five observational surveys: DEEP2, VVDS-Deep, VVDS-Wide, eBOSS-SGC and DESI (see Table~\ref{tbl:obs}). Further colour cuts are imposed in the latter two cases to mimic the spectroscopic selection that avoids targetting objects with colours that can be confused with stars. 

The large scale bias of model ELGs is close to unity (see Fig.~\ref{fig:xielg} in \S~\ref{sec:xielg}). These model galaxies are naturally affected by assembly bias, as the semi-analytical model of galaxy formation and evolution includes the effect of different halo assembly histories. Thus, the bias measured from galaxy catalogues constructed with a halo occupation distribution model without considering assembly bias, might be slightly larger than the values reported here.

The \LSE at $z=0.83,\,0.99$ has been classified into voids, sheets, filaments and knots using: (i) a velocity-shear-tensor algorithm, \vweb, with a threshold of 0.1; and (ii) a tidal-tensor algorithm, \pweb, with a threshold of $0.005s^{-2}$. 
Similar conclusions are reached with both algorithms.

Half of the model ELGs live in filaments and a third in sheets (see \S~\ref{sec:elgcw}). Model ELGs in knots have the largest percentage of satellite galaxies and a tail of massive galaxies that sets them apart when comparing the galaxy stellar mass function in each environment. We find that the shape of the mean halo occupation distribution (HOD) of model ELGs varies widely with \LSE (see Fig~\ref{fig:hod_env}), partly due to the different presence of satellite ELGs in different cosmic web structures. The mean HOD of voids and sheets, where almost all galaxies are centrals, has a shape close to an asymmetric Gaussian. The mean HOD of central galaxies in filaments and knots has a plateau. The presence of satellite galaxies is most important among ELGs in knots, for which the mean satellite HOD follows a typical power law. 

We have explored the cross-correlation between the whole ELG sample and those living in voids, sheets, filaments or knots (see Fig.~\ref{fig:xielg}). We find that, for all the studied ELG samples, the clustering of ELGs in knots is boosted at $1\leq r (h^{-1}{\rm Mpc})\leq10$, while ELGs in voids are less  clustered on large scales, $r>1h^{-1}$Mpc.

To put in context the results obtained for model ELGs, we have defined samples with three fixed number densities, $10^{-2},10^{-3},10^{-4.2}h^3$Mpc$^{-3}$. These samples have been selected by imposing an extra cut in either stellar mass, SFR or L\ox for ELGs or by imposing a single cut in one of these three properties to the whole sample of model galaxies, as shown in Fig.~\ref{fig:cum}. The median L\ox for galaxies selected only by their stellar mass is below $10^{39.5}h^{-2}{\rm erg}\,{\rm s}^{-1}$, while all the other selections are much brighter, with median L\ox above $10^{40.5}h^{-2}{\rm erg}\,{\rm s}^{-1}$.

The mean HODs of model ELGs with fixed number densities have shapes close to those of star-forming samples, selected either based on a SFR or L\ox cut (see Fig.~\ref{fig:compare_hod}). The studied ELGs are indeed a subsample of the star-forming population.

For a fixed number density, we find that, in general, star-forming galaxies are less clustered than stellar mass selected ones (see Fig.~\ref{fig:xi}). Fixed number density ELG, SFR and L\ox selected samples have very similar large scale bias. However, their clustering differs below separations of $1h^{-1}$Mpc. For instance, no pairs of L\ox selected samples are found at the smallest separations considered. This might have implications for the expectations of redshift-space distortions derived assuming that ELGs are equivalent to galaxies selected by a single cut in SFR \citep[e.g.][]{orsi2018,jimenez2019}. 

As expected, fixed number density samples selected with a single stellar mass cut have a higher presence in knots than either ELGs or galaxies selected by their SFR or L\ox.

For a fixed number density, the distribution of star-forming ELGs in the cosmic web follows closely that of  samples selected with a single cut in eitheir SFR or L\ox~(see Fig.~\ref{fig:environ}). The differences are more significant for low number density samples, at least with respect to SFR selected samples.

Over 70 per cent of the model ELG samples with number densities $10^{-2}h^{3}$Mpc$^{-3}$ and $10^{-3}h^{3}$Mpc$^{-3}$ are found in either filaments or sheets. About half of them are in filaments. For samples with lower number densities, this percentage drops, except for the eBOSS-SGC model sample. 

The maximum stellar mass and host halo mass decreases from knots to voids for both star-forming and stellar mass selected samples with fixed number densities (see Fig.~\ref{fig:violin}). The specific star formation of fixed number density model samples is largely independent of the \LSE for star-forming galaxies, but increases moving from knots to voids for galaxies selected with a single cut in stellar mass. For a fixed number density model sample, the $L$\ox appears to be independent of the large-scale environment.

The agreement between the properties of the ELGs, SFR and L\ox selected samples, at least for number densities above $10^{-4.2}h^{3}$Mpc$^{-3}$, shows the robustness of our results. For large scales, one could use the dispersion in the two point correlation function among these 'star-forming' samples, as a reasonable estimate of the systematic error when producing mock catalogues. For small scales, variations are found among star-forming galaxies selected in different ways.

\section{Data availability}

The programs and a small fraction of the data used to generate the plots presented in this paper can be found in \url{https://github.com/viogp/plots4papers/tree/master/elg_cw_plots}. Other sets of data can be shared upon request.

\section*{Acknowledgements}
The authors would like to thank the help provided by Nuala McCullagh, Lee Stothert and Alex Smith to run {\sc cute} on hdf5 files. VGP acknowledges past support from the University of Portsmouth through the Dennis Sciama Fellowship award. This project has received funding from the European Research Council (ERC) under the European Union's Horizon 2020 research and innovation programme (grant agreement No 769130).
WC was supported by the {\it Ministerio de Econom\'ia y Competitividad} and the {\it Fondo Europeo de Desarrollo Regional} (MINECO/FEDER, UE) in Spain through grant AYA2015-63810-P. WC further acknowledges the support from the European Research Council under grant number 670193.
SC acknowledges the support from the Juan de la Cierva Formaci\'on Fellowship (FJCI-2017-33816).
AG acknowledges support from the STCF studentship ST/N50404X/1.
CGL and CMB acknowledge support from the STFC grant ST/P000541/1.
AK is supported by MINECO/FEDER (Spain) under research grants AYA2015-63819-P and PGC2018-094975-C2. He further acknowledges support from the Spanish Red Consolider MultiDark FPA2017-90566-REDC and thanks Tocotronic for the red album.
PN acknowledges the support of the Royal Society through the award of a University Research Fellowship, and the European Research Council, through receipt of a Starting Grant (DEGAS-259586).
This work used the DiRAC@Durham facility managed by the Institute for Computational Cosmology on behalf of the STFC DiRAC HPC Facility (www.dirac.ac.uk). The equipment was funded by BEIS capital funding via STFC capital grants ST/P002293/1, ST/R002371/1 and ST/S002502/1, Durham University and STFC operations grant ST/R000832/1. DiRAC is part of the National e-Infrastructure. This work has benefited from the publicly available programming language {\sc python} (\url{https://www.python.org/}) and the package {\sc matplotlib} (\url{https://matplotlib.org/}).




\bibliographystyle{mnras}
\bibliography{cosmicweb_elg.bib} 

\begin{thebibliography}{}
\makeatletter
\relax
\def\mn@urlcharsother{\let\do\@makeother \do\$\do\&\do\#\do\^\do\_\do\%\do\~}
\def\mn@doi{\begingroup\mn@urlcharsother \@ifnextchar [ {\mn@doi@}
  {\mn@doi@[]}}
\def\mn@doi@[#1]#2{\def\@tempa{#1}\ifx\@tempa\@empty \href
  {http://dx.doi.org/#2} {doi:#2}\else \href {http://dx.doi.org/#2} {#1}\fi
  \endgroup}
\def\mn@eprint#1#2{\mn@eprint@#1:#2::\@nil}
\def\mn@eprint@arXiv#1{\href {http://arxiv.org/abs/#1} {{\tt arXiv:#1}}}
\def\mn@eprint@dblp#1{\href {http://dblp.uni-trier.de/rec/bibtex/#1.xml}
  {dblp:#1}}
\def\mn@eprint@#1:#2:#3:#4\@nil{\def\@tempa {#1}\def\@tempb {#2}\def\@tempc
  {#3}\ifx \@tempc \@empty \let \@tempc \@tempb \let \@tempb \@tempa \fi \ifx
  \@tempb \@empty \def\@tempb {arXiv}\fi \@ifundefined
  {mn@eprint@\@tempb}{\@tempb:\@tempc}{\expandafter \expandafter \csname
  mn@eprint@\@tempb\endcsname \expandafter{\@tempc}}}

\bibitem[\protect\citeauthoryear{{Ahumada} et~al.,}{{Ahumada}
  et~al.}{2019}]{dr16}
{Ahumada} R.,  et~al., 2019, arXiv e-prints, \href
  {https://ui.adsabs.harvard.edu/abs/2019arXiv191202905A} {p. arXiv:1912.02905}

\bibitem[\protect\citeauthoryear{Alam, Peacock, Kraljic, Ross  \&
  Comparat}{Alam et~al.}{2019}]{alam2019}
Alam S.,  Peacock J.~A.,  Kraljic K.,  Ross A.~J.,   Comparat J.,  2019,
  arXiv:1910.05095 [astro-ph]

\bibitem[\protect\citeauthoryear{{Alonso}}{{Alonso}}{2012}]{cute}
{Alonso} D.,  2012, arXiv e-prints, \href
  {https://ui.adsabs.harvard.edu/abs/2012arXiv1210.1833A} {p. arXiv:1210.1833}

\bibitem[\protect\citeauthoryear{{Alpaslan} et~al.,}{{Alpaslan}
  et~al.}{2016}]{alpaslan2016}
{Alpaslan} M.,  et~al., 2016, \mn@doi [\mnras] {10.1093/mnras/stw134}, \href
  {https://ui.adsabs.harvard.edu/abs/2016MNRAS.457.2287A} {457, 2287}

\bibitem[\protect\citeauthoryear{{Anders} \& {Fritze-v.~Alvensleben}}{{Anders}
  \& {Fritze-v.~Alvensleben}}{2003}]{anders03}
{Anders} P.,  {Fritze-v.~Alvensleben} U.,  2003, \mn@doi [\aap]
  {10.1051/0004-6361:20030151}, 401, 1063

\bibitem[\protect\citeauthoryear{{Avila} et~al.,}{{Avila}
  et~al.}{2020}]{avila2020}
{Avila} S.,  et~al., 2020, arXiv e-prints, \href
  {https://ui.adsabs.harvard.edu/abs/2020arXiv200709012A} {p. arXiv:2007.09012}

\bibitem[\protect\citeauthoryear{{Bauer} et~al.,}{{Bauer}
  et~al.}{2013}]{bauer13}
{Bauer} A.~E.,  et~al., 2013, \mn@doi [\mnras] {10.1093/mnras/stt1011}, 434,
  209

\bibitem[\protect\citeauthoryear{{Baugh}}{{Baugh}}{2006}]{baugh06}
{Baugh} C.~M.,  2006, \mn@doi [Reports of Progress in Physics]
  {10.1088/0034-4885/69/12/R02}, 69, 3101

\bibitem[\protect\citeauthoryear{{Baugh}, {Lacey}, {Frenk}, {Granato}, {Silva},
  {Bressan}, {Benson}  \& {Cole}}{{Baugh} et~al.}{2005}]{baugh05}
{Baugh} C.~M.,  {Lacey} C.~G.,  {Frenk} C.~S.,  {Granato} G.~L.,  {Silva} L.,
  {Bressan} A.,  {Benson} A.~J.,   {Cole} S.,  2005, \mn@doi [\mnras]
  {10.1111/j.1365-2966.2004.08553.x}, \href
  {http://adsabs.harvard.edu/abs/2005MNRAS.356.1191B} {356, 1191}

\bibitem[\protect\citeauthoryear{Baugh et~al.,}{Baugh et~al.}{2018}]{cmb19}
Baugh C.~M.,  et~al., 2018, \mn@doi [Monthly Notices of the Royal Astronomical
  Society] {10.1093/mnras/sty3427}, 483, 4922

\bibitem[\protect\citeauthoryear{{Benson}}{{Benson}}{2010}]{benson10}
{Benson} A.~J.,  2010, \mn@doi [\physrep] {10.1016/j.physrep.2010.06.001}, 495,
  33

\bibitem[\protect\citeauthoryear{{Benson} \& {Bower}}{{Benson} \&
  {Bower}}{2010}]{bensonbower10}
{Benson} A.~J.,  {Bower} R.,  2010, \mn@doi [\mnras]
  {10.1111/j.1365-2966.2010.16592.x}, \href
  {http://adsabs.harvard.edu/abs/2010MNRAS.405.1573B} {405, 1573}

\bibitem[\protect\citeauthoryear{{Bond}, {Kofman}  \& {Pogosyan}}{{Bond}
  et~al.}{1996}]{Bond1996}
{Bond} J.~R.,  {Kofman} L.,   {Pogosyan} D.,  1996, \mn@doi [\nat]
  {10.1038/380603a0}, \href
  {https://ui.adsabs.harvard.edu/abs/1996Natur.380..603B} {380, 603}

\bibitem[\protect\citeauthoryear{{Bower}, {Benson}, {Malbon}, {Helly}, {Frenk},
  {Baugh}, {Cole}  \& {Lacey}}{{Bower} et~al.}{2006}]{bower06}
{Bower} R.~G.,  {Benson} A.~J.,  {Malbon} R.,  {Helly} J.~C.,  {Frenk} C.~S.,
  {Baugh} C.~M.,  {Cole} S.,   {Lacey} C.~G.,  2006, \mn@doi [\mnras]
  {10.1111/j.1365-2966.2006.10519.x}, 370, 645

\bibitem[\protect\citeauthoryear{Carlesi, Knebe, Lewis, Wales  \&
  Yepes}{Carlesi et~al.}{2014}]{Carlesi2014}
Carlesi E.,  Knebe A.,  Lewis G.~F.,  Wales S.,   Yepes G.,  2014, \mn@doi
  [Monthly Notices of the Royal Astronomical Society, Volume 439, Issue 3,
  p.2943-2957] {10.1093/mnras/stu150}, 439, 2943

\bibitem[\protect\citeauthoryear{{Cautun}, {van de Weygaert}, {Jones}  \&
  {Frenk}}{{Cautun} et~al.}{2014}]{Cautun2014}
{Cautun} M.,  {van de Weygaert} R.,  {Jones} B. J.~T.,   {Frenk} C.~S.,  2014,
  \mn@doi [\mnras] {10.1093/mnras/stu768}, \href
  {https://ui.adsabs.harvard.edu/abs/2014MNRAS.441.2923C} {441, 2923}

\bibitem[\protect\citeauthoryear{{Chen} et~al.,}{{Chen}
  et~al.}{2017}]{chen2017}
{Chen} Y.-C.,  et~al., 2017, \mn@doi [\mnras] {10.1093/mnras/stw3127}, \href
  {https://ui.adsabs.harvard.edu/abs/2017MNRAS.466.1880C} {466, 1880}

\bibitem[\protect\citeauthoryear{{Cochrane} \& {Best}}{{Cochrane} \&
  {Best}}{2018}]{cochrane2018eagle}
{Cochrane} R.~K.,  {Best} P.~N.,  2018, \mn@doi [\mnras]
  {10.1093/mnras/sty1708}, \href
  {https://ui.adsabs.harvard.edu/abs/2018MNRAS.480..864C} {480, 864}

\bibitem[\protect\citeauthoryear{{Cochrane}, {Best}, {Sobral}, {Smail}, {Wake},
  {Stott}  \& {Geach}}{{Cochrane} et~al.}{2017}]{cochrane2017}
{Cochrane} R.~K.,  {Best} P.~N.,  {Sobral} D.,  {Smail} I.,  {Wake} D.~A.,
  {Stott} J.~P.,   {Geach} J.~E.,  2017, \mn@doi [\mnras]
  {10.1093/mnras/stx957}, \href
  {https://ui.adsabs.harvard.edu/abs/2017MNRAS.469.2913C} {469, 2913}

\bibitem[\protect\citeauthoryear{{Cochrane}, {Best}, {Sobral}, {Smail},
  {Geach}, {Stott}  \& {Wake}}{{Cochrane} et~al.}{2018}]{cochrane18}
{Cochrane} R.~K.,  {Best} P.~N.,  {Sobral} D.,  {Smail} I.,  {Geach} J.~E.,
  {Stott} J.~P.,   {Wake} D.~A.,  2018, \mn@doi [\mnras]
  {10.1093/mnras/stx3345}, \href
  {https://ui.adsabs.harvard.edu/abs/2018MNRAS.475.3730C} {475, 3730}

\bibitem[\protect\citeauthoryear{{Cole}, {Lacey}, {Baugh}  \& {Frenk}}{{Cole}
  et~al.}{2000}]{cole00}
{Cole} S.,  {Lacey} C.~G.,  {Baugh} C.~M.,   {Frenk} C.~S.,  2000, \mnras, 319,
  168

\bibitem[\protect\citeauthoryear{{Colless} et~al.,}{{Colless}
  et~al.}{2001}]{colless2001}
{Colless} M.,  et~al., 2001, \mn@doi [\mnras]
  {10.1046/j.1365-8711.2001.04902.x}, \href
  {https://ui.adsabs.harvard.edu/abs/2001MNRAS.328.1039C} {328, 1039}

\bibitem[\protect\citeauthoryear{{Comparat} et~al.,}{{Comparat}
  et~al.}{2013a}]{comparat2013elgsdss}
{Comparat} J.,  et~al., 2013a, \mn@doi [\mnras] {10.1093/mnras/sts127}, \href
  {https://ui.adsabs.harvard.edu/abs/2013MNRAS.428.1498C} {428, 1498}

\bibitem[\protect\citeauthoryear{{Comparat} et~al.,}{{Comparat}
  et~al.}{2013b}]{comparat2013}
{Comparat} J.,  et~al., 2013b, \mn@doi [\mnras] {10.1093/mnras/sts127}, \href
  {https://ui.adsabs.harvard.edu/abs/2013MNRAS.428.1498C} {428, 1498}

\bibitem[\protect\citeauthoryear{{Comparat} et~al.,}{{Comparat}
  et~al.}{2013c}]{comparat2013elglensing}
{Comparat} J.,  et~al., 2013c, \mn@doi [\mnras] {10.1093/mnras/stt797}, \href
  {https://ui.adsabs.harvard.edu/abs/2013MNRAS.433.1146C} {433, 1146}

\bibitem[\protect\citeauthoryear{{Comparat} et~al.,}{{Comparat}
  et~al.}{2015}]{comparat15o2}
{Comparat} J.,  et~al., 2015, \mn@doi [\aap] {10.1051/0004-6361/201424767},
  \href {http://adsabs.harvard.edu/abs/2015A%26A...575A..40C} {575, A40}

\bibitem[\protect\citeauthoryear{{Comparat} et~al.,}{{Comparat}
  et~al.}{2016a}]{comparat16}
{Comparat} J.,  et~al., 2016a, \mn@doi [\mnras] {10.1093/mnras/stw1393}, \href
  {http://adsabs.harvard.edu/abs/2016MNRAS.461.1076C} {461, 1076}

\bibitem[\protect\citeauthoryear{{Comparat} et~al.,}{{Comparat}
  et~al.}{2016b}]{comparat15eboss}
{Comparat} J.,  et~al., 2016b, \mn@doi [\aap] {10.1051/0004-6361/201527377},
  \href {http://adsabs.harvard.edu/abs/2016A%26A...592A.121C} {592, A121}

\bibitem[\protect\citeauthoryear{{Comparat} et~al.,}{{Comparat}
  et~al.}{2017}]{comparat19vac}
{Comparat} J.,  et~al., 2017, arXiv e-prints, \href
  {http://adsabs.harvard.edu/abs/2017arXiv171106575C} {}

\bibitem[\protect\citeauthoryear{{Contreras}, {Baugh}, {Norberg}  \&
  {Padilla}}{{Contreras} et~al.}{2013}]{contreras13}
{Contreras} S.,  {Baugh} C.~M.,  {Norberg} P.,   {Padilla} N.,  2013, \mn@doi
  [\mnras] {10.1093/mnras/stt629}, 432, 2717

\bibitem[\protect\citeauthoryear{{Contreras}, {Zehavi}, {Padilla}, {Baugh},
  {Jim{\'e}nez}  \& {Lacerna}}{{Contreras} et~al.}{2019}]{contreras2019}
{Contreras} S.,  {Zehavi} I.,  {Padilla} N.,  {Baugh} C.~M.,  {Jim{\'e}nez} E.,
    {Lacerna} I.,  2019, \mn@doi [\mnras] {10.1093/mnras/stz018}, \href
  {https://ui.adsabs.harvard.edu/abs/2019MNRAS.484.1133C} {484, 1133}

\bibitem[\protect\citeauthoryear{{Cui}, {Borgani}, {Dolag}, {Murante}  \&
  {Tornatore}}{{Cui} et~al.}{2012}]{Cui2012}
{Cui} W.,  {Borgani} S.,  {Dolag} K.,  {Murante} G.,   {Tornatore} L.,  2012,
  \mn@doi [\mnras] {10.1111/j.1365-2966.2012.21037.x}, \href
  {https://ui.adsabs.harvard.edu/abs/2012MNRAS.423.2279C} {423, 2279}

\bibitem[\protect\citeauthoryear{{Cui}, {Borgani}  \& {Murante}}{{Cui}
  et~al.}{2014}]{Cui2014}
{Cui} W.,  {Borgani} S.,   {Murante} G.,  2014, \mn@doi [\mnras]
  {10.1093/mnras/stu673}, \href
  {https://ui.adsabs.harvard.edu/abs/2014MNRAS.441.1769C} {441, 1769}

\bibitem[\protect\citeauthoryear{{Cui}, {Knebe}, {Yepes}, {Yang}, {Borgani},
  {Kang}, {Power}  \& {Staveley-Smith}}{{Cui} et~al.}{2018}]{Cui2018}
{Cui} W.,  {Knebe} A.,  {Yepes} G.,  {Yang} X.,  {Borgani} S.,  {Kang} X.,
  {Power} C.,   {Staveley-Smith} L.,  2018, \mn@doi [\mnras]
  {10.1093/mnras/stx2323}, \href
  {https://ui.adsabs.harvard.edu/abs/2018MNRAS.473...68C} {473, 68}

\bibitem[\protect\citeauthoryear{{Cui} et~al.,}{{Cui} et~al.}{2019}]{Cui2019}
{Cui} W.,  et~al., 2019, \mn@doi [\mnras] {10.1093/mnras/stz565}, \href
  {https://ui.adsabs.harvard.edu/abs/2019MNRAS.485.2367C} {485, 2367}

\bibitem[\protect\citeauthoryear{{DESI Collaboration} et~al.,}{{DESI
  Collaboration} et~al.}{2016}]{desi1}
{DESI Collaboration} et~al., 2016, preprint (\mn@eprint {arXiv} {1611.00036})

\bibitem[\protect\citeauthoryear{{Darvish}, {Sobral}, {Mobasher}, {Scoville},
  {Best}, {Sales}  \& {Smail}}{{Darvish} et~al.}{2014}]{darvish2014}
{Darvish} B.,  {Sobral} D.,  {Mobasher} B.,  {Scoville} N.~Z.,  {Best} P.,
  {Sales} L.~V.,   {Smail} I.,  2014, \mn@doi [\apj]
  {10.1088/0004-637X/796/1/51}, \href
  {https://ui.adsabs.harvard.edu/abs/2014ApJ...796...51D} {796, 51}

\bibitem[\protect\citeauthoryear{{Darvish}, {Mobasher}, {Sobral}, {Hemmati},
  {Nayyeri}  \& {Shivaei}}{{Darvish} et~al.}{2015}]{darvish2015}
{Darvish} B.,  {Mobasher} B.,  {Sobral} D.,  {Hemmati} S.,  {Nayyeri} H.,
  {Shivaei} I.,  2015, \mn@doi [\apj] {10.1088/0004-637X/814/2/84}, \href
  {https://ui.adsabs.harvard.edu/abs/2015ApJ...814...84D} {814, 84}

\bibitem[\protect\citeauthoryear{{Dawson} et~al.,}{{Dawson}
  et~al.}{2016}]{dawson16}
{Dawson} K.~S.,  et~al., 2016, \mn@doi [\aj] {10.3847/0004-6256/151/2/44}, 151,
  44

\bibitem[\protect\citeauthoryear{{Delubac} et~al.,}{{Delubac}
  et~al.}{2017}]{delubac17}
{Delubac} T.,  et~al., 2017, \mn@doi [\mnras] {10.1093/mnras/stw2741}, \href
  {http://adsabs.harvard.edu/abs/2017MNRAS.465.1831D} {465, 1831}

\bibitem[\protect\citeauthoryear{{Drinkwater} et~al.,}{{Drinkwater}
  et~al.}{2010}]{drinkwater2010}
{Drinkwater} M.~J.,  et~al., 2010, \mn@doi [\mnras]
  {10.1111/j.1365-2966.2009.15754.x}, \href
  {https://ui.adsabs.harvard.edu/abs/2010MNRAS.401.1429D} {401, 1429}

\bibitem[\protect\citeauthoryear{{Drinkwater} et~al.,}{{Drinkwater}
  et~al.}{2018}]{drinkwater18}
{Drinkwater} M.~J.,  et~al., 2018, \mn@doi [\mnras] {10.1093/mnras/stx2963},
  \href {http://adsabs.harvard.edu/abs/2018MNRAS.474.4151D} {474, 4151}

\bibitem[\protect\citeauthoryear{{Driver} et~al.,}{{Driver}
  et~al.}{2012}]{driver12}
{Driver} S.~P.,  et~al., 2012, \mn@doi [\mnras]
  {10.1111/j.1365-2966.2012.22036.x}, 427, 3244

\bibitem[\protect\citeauthoryear{{Fanidakis}, {Baugh}, {Benson}, {Bower},
  {Cole}, {Done}  \& {Frenk}}{{Fanidakis} et~al.}{2011}]{fanidakis11}
{Fanidakis} N.,  {Baugh} C.~M.,  {Benson} A.~J.,  {Bower} R.~G.,  {Cole} S.,
  {Done} C.,   {Frenk} C.~S.,  2011, \mn@doi [\mnras]
  {10.1111/j.1365-2966.2010.17427.x}, \href
  {http://adsabs.harvard.edu/abs/2011MNRAS.410...53F} {410, 53}

\bibitem[\protect\citeauthoryear{{Favole} et~al.,}{{Favole}
  et~al.}{2016}]{favole16}
{Favole} G.,  et~al., 2016, \mn@doi [\mnras] {10.1093/mnras/stw1483}, \href
  {http://adsabs.harvard.edu/abs/2016MNRAS.461.3421F} {461, 3421}

\bibitem[\protect\citeauthoryear{{Font} et~al.,}{{Font} et~al.}{2008}]{font08}
{Font} A.~S.,  et~al., 2008, \mn@doi [\mnras]
  {10.1111/j.1365-2966.2008.13698.x}, 389, 1619

\bibitem[\protect\citeauthoryear{{Franx}, {van Dokkum}, {F{\"o}rster
  Schreiber}, {Wuyts}, {Labb{\'e}}  \& {Toft}}{{Franx} et~al.}{2008}]{franx08}
{Franx} M.,  {van Dokkum} P.~G.,  {F{\"o}rster Schreiber} N.~M.,  {Wuyts} S.,
  {Labb{\'e}} I.,   {Toft} S.,  2008, \mn@doi [\apj] {10.1086/592431}, 688, 770

\bibitem[\protect\citeauthoryear{{Furlong} et~al.,}{{Furlong}
  et~al.}{2015}]{furlong15}
{Furlong} M.,  et~al., 2015, \mn@doi [\mnras] {10.1093/mnras/stv852}, 450, 4486

\bibitem[\protect\citeauthoryear{{Geach}, {Sobral}, {Hickox}, {Wake}, {Smail},
  {Best}, {Baugh}  \& {Stott}}{{Geach} et~al.}{2012}]{geach12}
{Geach} J.~E.,  {Sobral} D.,  {Hickox} R.~C.,  {Wake} D.~A.,  {Smail} I.,
  {Best} P.~N.,  {Baugh} C.~M.,   {Stott} J.~P.,  2012, \mn@doi [\mnras]
  {10.1111/j.1365-2966.2012.21725.x}, \href
  {http://adsabs.harvard.edu/abs/2012MNRAS.426..679G} {426, 679}

\bibitem[\protect\citeauthoryear{{Geller} \& {Huchra}}{{Geller} \&
  {Huchra}}{1989}]{Geller1989}
{Geller} M.~J.,  {Huchra} J.~P.,  1989, \mn@doi [Science]
  {10.1126/science.246.4932.897}, \href
  {https://ui.adsabs.harvard.edu/abs/1989Sci...246..897G} {246, 897}

\bibitem[\protect\citeauthoryear{{Gilbank}, {Baldry}, {Balogh}, {Glazebrook}
  \& {Bower}}{{Gilbank} et~al.}{2010}]{gilbank10}
{Gilbank} D.~G.,  {Baldry} I.~K.,  {Balogh} M.~L.,  {Glazebrook} K.,   {Bower}
  R.~G.,  2010, \mn@doi [\mnras] {10.1111/j.1365-2966.2010.16640.x}, 405, 2594

\bibitem[\protect\citeauthoryear{{Goh} et~al.,}{{Goh} et~al.}{2019}]{goh2019}
{Goh} T.,  et~al., 2019, \mn@doi [\mnras] {10.1093/mnras/sty3153}, \href
  {https://ui.adsabs.harvard.edu/abs/2019MNRAS.483.2101G} {483, 2101}

\bibitem[\protect\citeauthoryear{{Gonzalez-Perez}, {Baugh}, {Lacey}  \&
  {Kim}}{{Gonzalez-Perez} et~al.}{2011}]{gp11}
{Gonzalez-Perez} V.,  {Baugh} C.~M.,  {Lacey} C.~G.,   {Kim} J.~W.,  2011,
  \mn@doi [\mnras] {10.1111/j.1365-2966.2011.19294.x}, \href
  {https://ui.adsabs.harvard.edu/abs/2011MNRAS.417..517G} {417, 517}

\bibitem[\protect\citeauthoryear{{Gonzalez-Perez}, {Lacey}, {Baugh}, {Frenk}
  \& {Wilkins}}{{Gonzalez-Perez} et~al.}{2013}]{gp13}
{Gonzalez-Perez} V.,  {Lacey} C.~G.,  {Baugh} C.~M.,  {Frenk} C.~S.,
  {Wilkins} S.~M.,  2013, \mn@doi [\mnras] {10.1093/mnras/sts446}, 429, 1609

\bibitem[\protect\citeauthoryear{{Gonzalez-Perez}, {Lacey}, {Baugh}, {Lagos},
  {Helly}, {Campbell}  \& {Mitchell}}{{Gonzalez-Perez} et~al.}{2014}]{gp14}
{Gonzalez-Perez} V.,  {Lacey} C.~G.,  {Baugh} C.~M.,  {Lagos} C.~D.~P.,
  {Helly} J.,  {Campbell} D.~J.~R.,   {Mitchell} P.~D.,  2014, \mn@doi [\mnras]
  {10.1093/mnras/stt2410}, 439, 264

\bibitem[\protect\citeauthoryear{{Gonzalez-Perez} et~al.,}{{Gonzalez-Perez}
  et~al.}{2018}]{gp18}
{Gonzalez-Perez} V.,  et~al., 2018, \mn@doi [\mnras] {10.1093/mnras/stx2807},
  \href {http://adsabs.harvard.edu/abs/2018MNRAS.474.4024G} {474, 4024}

\bibitem[\protect\citeauthoryear{{Gott}, {Juri{\'c}}, {Schlegel}, {Hoyle},
  {Vogeley}, {Tegmark}, {Bahcall}  \& {Brinkmann}}{{Gott}
  et~al.}{2005}]{gott2005}
{Gott} J.~Richard I.,  {Juri{\'c}} M.,  {Schlegel} D.,  {Hoyle} F.,  {Vogeley}
  M.,  {Tegmark} M.,  {Bahcall} N.,   {Brinkmann} J.,  2005, \mn@doi [\apj]
  {10.1086/428890}, \href
  {https://ui.adsabs.harvard.edu/abs/2005ApJ...624..463G} {624, 463}

\bibitem[\protect\citeauthoryear{{Griffin}, {Lacey}, {Gonzalez-Perez}, {Lagos},
  {Baugh}  \& {Fanidakis}}{{Griffin} et~al.}{2019}]{griffin}
{Griffin} A.~J.,  {Lacey} C.~G.,  {Gonzalez-Perez} V.,  {Lagos} C. d.~P.,
  {Baugh} C.~M.,   {Fanidakis} N.,  2019, \mn@doi [\mnras]
  {10.1093/mnras/stz1216}, \href
  {https://ui.adsabs.harvard.edu/abs/2019MNRAS.487..198G} {487, 198}

\bibitem[\protect\citeauthoryear{{Guo}, {White}, {Angulo}, {Henriques},
  {Lemson}, {Boylan-Kolchin}, {Thomas}  \& {Short}}{{Guo} et~al.}{2013}]{guo13}
{Guo} Q.,  {White} S.,  {Angulo} R.~E.,  {Henriques} B.,  {Lemson} G.,
  {Boylan-Kolchin} M.,  {Thomas} P.,   {Short} C.,  2013, \mn@doi [\mnras]
  {10.1093/mnras/sts115}, 428, 1351

\bibitem[\protect\citeauthoryear{{Guo}, {Tempel}  \& {Libeskind}}{{Guo}
  et~al.}{2015}]{guo2015}
{Guo} Q.,  {Tempel} E.,   {Libeskind} N.~I.,  2015, \mn@doi [\apj]
  {10.1088/0004-637X/800/2/112}, \href
  {https://ui.adsabs.harvard.edu/abs/2015ApJ...800..112G} {800, 112}

\bibitem[\protect\citeauthoryear{Guo et~al.,}{Guo et~al.}{2018a}]{guo2018}
Guo H.,  et~al., 2018a, \mn@doi [arXiv:1810.05318 [astro-ph]]
  {10.3847/1538-4357/aaf9ad}

\bibitem[\protect\citeauthoryear{{Guo} et~al.,}{{Guo}
  et~al.}{2018b}]{hongguo19}
{Guo} H.,  et~al., 2018b, arXiv e-prints, \href
  {http://adsabs.harvard.edu/abs/2018arXiv181005318G} {}

\bibitem[\protect\citeauthoryear{{Hahn}, {Porciani}, {Carollo}  \&
  {Dekel}}{{Hahn} et~al.}{2007}]{hahn07}
{Hahn} O.,  {Porciani} C.,  {Carollo} C.~M.,   {Dekel} A.,  2007, \mn@doi
  [\mnras] {10.1111/j.1365-2966.2006.11318.x}, \href
  {http://adsabs.harvard.edu/abs/2007MNRAS.375..489H} {375, 489}

\bibitem[\protect\citeauthoryear{{Hayashi} et~al.,}{{Hayashi}
  et~al.}{2020}]{hayashi2020}
{Hayashi} M.,  et~al., 2020, arXiv e-prints, \href
  {https://ui.adsabs.harvard.edu/abs/2020arXiv200707413H} {p. arXiv:2007.07413}

\bibitem[\protect\citeauthoryear{{Hill} et~al.,}{{Hill} et~al.}{2008}]{hetdex}
{Hill} G.~J.,  et~al., 2008, in {Kodama} T.,  {Yamada} T.,   {Aoki} K.,  eds,
  Astronomical Society of the Pacific Conference Series Vol. 399, Panoramic
  Views of Galaxy Formation and Evolution. p.~115 (\mn@eprint {arXiv}
  {0806.0183})

\bibitem[\protect\citeauthoryear{Hoffman, Metuki, Yepes, Gottl{\"{o}}ber,
  Forero-Romero, Libeskind  \& Knebe}{Hoffman et~al.}{2012}]{Hoffman2012}
Hoffman Y.,  Metuki O.,  Yepes G.,  Gottl{\"{o}}ber S.,  Forero-Romero J.~E.,
  Libeskind N.~I.,   Knebe A.,  2012, \mn@doi [Monthly Notices of the Royal
  Astronomical Society, Volume 425, Issue 3, pp. 2049-2057.]
  {10.1111/j.1365-2966.2012.21553.x}, 425, 2049

\bibitem[\protect\citeauthoryear{{Hounsell} et~al.,}{{Hounsell}
  et~al.}{2018}]{wfirst}
{Hounsell} R.,  et~al., 2018, \mn@doi [\apj] {10.3847/1538-4357/aac08b}, \href
  {http://adsabs.harvard.edu/abs/2018ApJ...867...23H} {867, 23}

\bibitem[\protect\citeauthoryear{{Jiang}, {Helly}, {Cole}  \& {Frenk}}{{Jiang}
  et~al.}{2014}]{jiang14}
{Jiang} L.,  {Helly} J.~C.,  {Cole} S.,   {Frenk} C.~S.,  2014, \mn@doi
  [\mnras] {10.1093/mnras/stu390}, 440, 2115

\bibitem[\protect\citeauthoryear{{Jim{\'e}nez}, {Contreras}, {Padilla},
  {Zehavi}, {Baugh}  \& {Gonzalez-Perez}}{{Jim{\'e}nez}
  et~al.}{2019}]{jimenez2019}
{Jim{\'e}nez} E.,  {Contreras} S.,  {Padilla} N.,  {Zehavi} I.,  {Baugh} C.~M.,
    {Gonzalez-Perez} V.,  2019, arXiv e-prints, \href
  {https://ui.adsabs.harvard.edu/abs/2019arXiv190604298J} {p. arXiv:1906.04298}

\bibitem[\protect\citeauthoryear{{Khostovan} et~al.,}{{Khostovan}
  et~al.}{2018}]{khostovan2018}
{Khostovan} A.~A.,  et~al., 2018, \mn@doi [\mnras] {10.1093/mnras/sty925},
  \href {https://ui.adsabs.harvard.edu/abs/2018MNRAS.478.2999K} {478, 2999}

\bibitem[\protect\citeauthoryear{{King}, {Pringle}  \& {Hofmann}}{{King}
  et~al.}{2008}]{king2008}
{King} A.~R.,  {Pringle} J.~E.,   {Hofmann} J.~A.,  2008, \mn@doi [\mnras]
  {10.1111/j.1365-2966.2008.12943.x}, \href
  {http://adsabs.harvard.edu/abs/2008MNRAS.385.1621K} {385, 1621}

\bibitem[\protect\citeauthoryear{{Klypin} \& {Shandarin}}{{Klypin} \&
  {Shandarin}}{1983}]{Klypin1983}
{Klypin} A.~A.,  {Shandarin} S.~F.,  1983, \mn@doi [\mnras]
  {10.1093/mnras/204.3.891}, \href
  {https://ui.adsabs.harvard.edu/abs/1983MNRAS.204..891K} {204, 891}

\bibitem[\protect\citeauthoryear{{Komatsu} et~al.,}{{Komatsu}
  et~al.}{2011}]{wmap7}
{Komatsu} E.,  et~al., 2011, \mn@doi [\apjs] {10.1088/0067-0049/192/2/18}, 192,
  18

\bibitem[\protect\citeauthoryear{{Kraljic} et~al.,}{{Kraljic}
  et~al.}{2018}]{kraljic2018}
{Kraljic} K.,  et~al., 2018, \mn@doi [\mnras] {10.1093/mnras/stx2638}, \href
  {https://ui.adsabs.harvard.edu/abs/2018MNRAS.474..547K} {474, 547}

\bibitem[\protect\citeauthoryear{{Kraljic} et~al.,}{{Kraljic}
  et~al.}{2019}]{kraljic2019}
{Kraljic} K.,  et~al., 2019, \mn@doi [\mnras] {10.1093/mnras/sty3216}, \href
  {https://ui.adsabs.harvard.edu/abs/2019MNRAS.483.3227K} {483, 3227}

\bibitem[\protect\citeauthoryear{{Lacey} et~al.,}{{Lacey}
  et~al.}{2016}]{lacey16}
{Lacey} C.~G.,  et~al., 2016, \mn@doi [\mnras] {10.1093/mnras/stw1888}, \href
  {http://adsabs.harvard.edu/abs/2016MNRAS.462.3854L} {462, 3854}

\bibitem[\protect\citeauthoryear{{Lagos}, {Lacey}, {Baugh}, {Bower}  \&
  {Benson}}{{Lagos} et~al.}{2011}]{lagos11}
{Lagos} C.~D.~P.,  {Lacey} C.~G.,  {Baugh} C.~M.,  {Bower} R.~G.,   {Benson}
  A.~J.,  2011, \mn@doi [\mnras] {10.1111/j.1365-2966.2011.19160.x}, 416, 1566

\bibitem[\protect\citeauthoryear{{Lagos}, {Davis}, {Lacey}, {Zwaan}, {Baugh},
  {Gonzalez-Perez}  \& {Padilla}}{{Lagos} et~al.}{2014}]{lagos14}
{Lagos} C.~d.~P.,  {Davis} T.~A.,  {Lacey} C.~G.,  {Zwaan} M.~A.,  {Baugh}
  C.~M.,  {Gonzalez-Perez} V.,   {Padilla} N.~D.,  2014, \mn@doi [\mnras]
  {10.1093/mnras/stu1209}, 443, 1002

\bibitem[\protect\citeauthoryear{{Laigle} et~al.,}{{Laigle}
  et~al.}{2015}]{laigle2015}
{Laigle} C.,  et~al., 2015, \mn@doi [\mnras] {10.1093/mnras/stu2289}, \href
  {https://ui.adsabs.harvard.edu/abs/2015MNRAS.446.2744L} {446, 2744}

\bibitem[\protect\citeauthoryear{{Laigle} et~al.,}{{Laigle}
  et~al.}{2018}]{laigle2018}
{Laigle} C.,  et~al., 2018, \mn@doi [\mnras] {10.1093/mnras/stx3055}, \href
  {https://ui.adsabs.harvard.edu/abs/2018MNRAS.474.5437L} {474, 5437}

\bibitem[\protect\citeauthoryear{{Laureijs} et~al.,}{{Laureijs}
  et~al.}{2011}]{laureijs11}
{Laureijs} R.,  et~al., 2011, preprint (\mn@eprint {arXiv} {1110.3193})

\bibitem[\protect\citeauthoryear{{Le F{\`e}vre} et~al.,}{{Le F{\`e}vre}
  et~al.}{2013}]{lefevre13}
{Le F{\`e}vre} O.,  et~al., 2013, \aap, 559, A14

\bibitem[\protect\citeauthoryear{{Leauthaud} et~al.,}{{Leauthaud}
  et~al.}{2007}]{stars}
{Leauthaud} A.,  et~al., 2007, \mn@doi [\apjs] {10.1086/516598}, 172, 219

\bibitem[\protect\citeauthoryear{{Levi} et~al.,}{{Levi} et~al.}{2013}]{levi13}
{Levi} M.,  et~al., 2013, preprint (\mn@eprint {arXiv} {1308.0847})

\bibitem[\protect\citeauthoryear{{Liao} \& {Gao}}{{Liao} \&
  {Gao}}{2019}]{liao2019}
{Liao} S.,  {Gao} L.,  2019, \mn@doi [\mnras] {10.1093/mnras/stz441}, \href
  {https://ui.adsabs.harvard.edu/abs/2019MNRAS.485..464L} {485, 464}

\bibitem[\protect\citeauthoryear{{Libeskind}, {Hoffman}, {Knebe}, {Steinmetz},
  {Gottl{\"o}ber}, {Metuki}  \& {Yepes}}{{Libeskind}
  et~al.}{2012}]{Libeskind2012}
{Libeskind} N.~I.,  {Hoffman} Y.,  {Knebe} A.,  {Steinmetz} M.,
  {Gottl{\"o}ber} S.,  {Metuki} O.,   {Yepes} G.,  2012, \mn@doi [\mnras]
  {10.1111/j.1745-3933.2012.01222.x}, \href
  {http://adsabs.harvard.edu/abs/2012MNRAS.421L.137L} {421, L137}

\bibitem[\protect\citeauthoryear{{Libeskind}, {Hoffman}, {Forero-Romero},
  {Gottl{\"o}ber}, {Knebe}, {Steinmetz}  \& {Klypin}}{{Libeskind}
  et~al.}{2013}]{Libeskind2013}
{Libeskind} N.~I.,  {Hoffman} Y.,  {Forero-Romero} J.,  {Gottl{\"o}ber} S.,
  {Knebe} A.,  {Steinmetz} M.,   {Klypin} A.,  2013, \mn@doi [\mnras]
  {10.1093/mnras/sts216}, \href
  {http://adsabs.harvard.edu/abs/2013MNRAS.428.2489L} {428, 2489}

\bibitem[\protect\citeauthoryear{{Libeskind} et~al.,}{{Libeskind}
  et~al.}{2018}]{Libeskind2018}
{Libeskind} N.~I.,  et~al., 2018, \mn@doi [\mnras] {10.1093/mnras/stx1976},
  \href {https://ui.adsabs.harvard.edu/abs/2018MNRAS.473.1195L} {473, 1195}

\bibitem[\protect\citeauthoryear{{Luber}, {van Gorkom}, {Hess}, {Pisano},
  {Fern{\'a}ndez}  \& {Momjian}}{{Luber} et~al.}{2019}]{luber2019}
{Luber} N.,  {van Gorkom} J.~H.,  {Hess} K.~M.,  {Pisano} D.~J.,
  {Fern{\'a}ndez} X.,   {Momjian} E.,  2019, \mn@doi [\aj]
  {10.3847/1538-3881/ab1b6e}, \href
  {https://ui.adsabs.harvard.edu/abs/2019AJ....157..254L} {157, 254}

\bibitem[\protect\citeauthoryear{{Malavasi} et~al.,}{{Malavasi}
  et~al.}{2017}]{malavasi2017}
{Malavasi} N.,  et~al., 2017, \mn@doi [\mnras] {10.1093/mnras/stw2864}, \href
  {https://ui.adsabs.harvard.edu/abs/2017MNRAS.465.3817M} {465, 3817}

\bibitem[\protect\citeauthoryear{{McCarthy}, {Frenk}, {Font}, {Lacey}, {Bower},
  {Mitchell}, {Balogh}  \& {Theuns}}{{McCarthy} et~al.}{2008}]{mccarthy08}
{McCarthy} I.~G.,  {Frenk} C.~S.,  {Font} A.~S.,  {Lacey} C.~G.,  {Bower}
  R.~G.,  {Mitchell} N.~L.,  {Balogh} M.~L.,   {Theuns} T.,  2008, \mn@doi
  [\mnras] {10.1111/j.1365-2966.2007.12577.x}, \href
  {http://adsabs.harvard.edu/abs/2008MNRAS.383..593M} {383, 593}

\bibitem[\protect\citeauthoryear{{McConnell} \& {Ma}}{{McConnell} \&
  {Ma}}{2013}]{mcconnell2013}
{McConnell} N.~J.,  {Ma} C.-P.,  2013, \mn@doi [\apj]
  {10.1088/0004-637X/764/2/184}, \href
  {https://ui.adsabs.harvard.edu/abs/2013ApJ...764..184M} {764, 184}

\bibitem[\protect\citeauthoryear{{Mitchell} et~al.,}{{Mitchell}
  et~al.}{2018}]{mitchell18}
{Mitchell} P.~D.,  et~al., 2018, \mn@doi [\mnras] {10.1093/mnras/stx2770},
  \href {http://adsabs.harvard.edu/abs/2018MNRAS.474..492M} {474, 492}

\bibitem[\protect\citeauthoryear{{Newman} et~al.,}{{Newman}
  et~al.}{2013}]{newman13}
{Newman} J.~A.,  et~al., 2013, \mn@doi [\apjs] {10.1088/0067-0049/208/1/5},
  208, 5

\bibitem[\protect\citeauthoryear{{Norberg} et~al.,}{{Norberg}
  et~al.}{2002}]{norberg02}
{Norberg} P.,  et~al., 2002, \mn@doi [\mnras]
  {10.1046/j.1365-8711.2002.05831.x}, 336, 907

\bibitem[\protect\citeauthoryear{{Orsi} \& {Angulo}}{{Orsi} \&
  {Angulo}}{2018}]{orsi2018}
{Orsi} {\'A}.~A.,  {Angulo} R.~E.,  2018, \mn@doi [\mnras]
  {10.1093/mnras/stx3349}, \href
  {https://ui.adsabs.harvard.edu/abs/2018MNRAS.475.2530O} {475, 2530}

\bibitem[\protect\citeauthoryear{{Orsi}, {Lacey}, {Baugh}  \& {Infante}}{{Orsi}
  et~al.}{2008}]{Orsi2008}
{Orsi} A.,  {Lacey} C.~G.,  {Baugh} C.~M.,   {Infante} L.,  2008, \mn@doi
  [\mnras] {10.1111/j.1365-2966.2008.14010.x}, \href
  {https://ui.adsabs.harvard.edu/abs/2008MNRAS.391.1589O} {391, 1589}

\bibitem[\protect\citeauthoryear{{Paulino-Afonso}, {Sobral}, {Darvish},
  {Ribeiro}, {Smail}, {Best}, {Stroe}  \& {Cairns}}{{Paulino-Afonso}
  et~al.}{2019}]{paulino-afonso2019}
{Paulino-Afonso} A.,  {Sobral} D.,  {Darvish} B.,  {Ribeiro} B.,  {Smail} I.,
  {Best} P.,  {Stroe} A.,   {Cairns} J.,  2019, arXiv e-prints, \href
  {https://ui.adsabs.harvard.edu/abs/2019arXiv191104517P} {p. arXiv:1911.04517}

\bibitem[\protect\citeauthoryear{{Percival} et~al.,}{{Percival}
  et~al.}{2019}]{percival2019}
{Percival} W.~J.,  et~al., 2019, arXiv e-prints, \href
  {https://ui.adsabs.harvard.edu/abs/2019arXiv190303158P} {p. arXiv:1903.03158}

\bibitem[\protect\citeauthoryear{{Planck Collaboration} et~al.,}{{Planck
  Collaboration} et~al.}{2014}]{planck14}
{Planck Collaboration} et~al., 2014, \mn@doi [\aap]
  {10.1051/0004-6361/201321591}, \href
  {http://adsabs.harvard.edu/abs/2014A%26A...571A..16P} {571, A16}

\bibitem[\protect\citeauthoryear{{Poudel}, {Hein{\"a}m{\"a}ki}, {Tempel},
  {Einasto}, {Lietzen}  \& {Nurmi}}{{Poudel} et~al.}{2017}]{poudel2017}
{Poudel} A.,  {Hein{\"a}m{\"a}ki} P.,  {Tempel} E.,  {Einasto} M.,  {Lietzen}
  H.,   {Nurmi} P.,  2017, \mn@doi [\aap] {10.1051/0004-6361/201629639}, \href
  {https://ui.adsabs.harvard.edu/abs/2017A&A...597A..86P} {597, A86}

\bibitem[\protect\citeauthoryear{{Raichoor} et~al.,}{{Raichoor}
  et~al.}{2017}]{raichoor17}
{Raichoor} A.,  et~al., 2017, preprint, \href
  {http://adsabs.harvard.edu/abs/2017arXiv170400338R} {} (\mn@eprint {arXiv}
  {1704.00338})

\bibitem[\protect\citeauthoryear{{Sinha} \& {Garrison}}{{Sinha} \&
  {Garrison}}{2020}]{corrfunc}
{Sinha} M.,  {Garrison} L.~H.,  2020, \mn@doi [\mnras] {10.1093/mnras/stz3157},
  \href {https://ui.adsabs.harvard.edu/abs/2020MNRAS.491.3022S} {491, 3022}

\bibitem[\protect\citeauthoryear{{Sobral}, {Best}, {Matsuda}, {Smail}, {Geach}
  \& {Cirasuolo}}{{Sobral} et~al.}{2012}]{sobral12}
{Sobral} D.,  {Best} P.~N.,  {Matsuda} Y.,  {Smail} I.,  {Geach} J.~E.,
  {Cirasuolo} M.,  2012, \mn@doi [\mnras] {10.1111/j.1365-2966.2011.19977.x},
  420, 1926

\bibitem[\protect\citeauthoryear{{Sobral}, {Stroe}, {Koyama}, {Darvish},
  {Calhau}, {Afonso}, {Kodama}  \& {Nakata}}{{Sobral} et~al.}{2016}]{sobral16}
{Sobral} D.,  {Stroe} A.,  {Koyama} Y.,  {Darvish} B.,  {Calhau} J.,  {Afonso}
  A.,  {Kodama} T.,   {Nakata} F.,  2016, \mn@doi [\mnras]
  {10.1093/mnras/stw534}, 458, 3443

\bibitem[\protect\citeauthoryear{{Somerville} \& {Dav{\'e}}}{{Somerville} \&
  {Dav{\'e}}}{2015}]{somerville15}
{Somerville} R.~S.,  {Dav{\'e}} R.,  2015, \mn@doi [\araa]
  {10.1146/annurev-astro-082812-140951}, 53, 51

\bibitem[\protect\citeauthoryear{{Stasi{\'n}ska}}{{Stasi{\'n}ska}}{1990}]{sta90}
{Stasi{\'n}ska} G.,  1990, \aaps, 83, 501

\bibitem[\protect\citeauthoryear{{Takada} et~al.,}{{Takada}
  et~al.}{2014}]{takada14}
{Takada} M.,  et~al., 2014, \mn@doi [\pasj] {10.1093/pasj/pst019}, 66, R1

\bibitem[\protect\citeauthoryear{Valentino et~al.,}{Valentino
  et~al.}{2017}]{valentino2017}
Valentino F.,  et~al., 2017, \mn@doi [Monthly Notices of the Royal Astronomical
  Society] {10.1093/mnras/stx2305}, 472, 4878

\bibitem[\protect\citeauthoryear{{Vulcani} et~al.,}{{Vulcani}
  et~al.}{2019}]{vulcani2019}
{Vulcani} B.,  et~al., 2019, \mn@doi [\mnras] {10.1093/mnras/stz1399}, \href
  {https://ui.adsabs.harvard.edu/abs/2019MNRAS.487.2278V} {487, 2278}

\bibitem[\protect\citeauthoryear{{Wang} et~al.,}{{Wang} et~al.}{2018}]{wang18}
{Wang} Y.,  et~al., 2018, arXiv e-prints, \href
  {https://ui.adsabs.harvard.edu/\#abs/2018arXiv180201539W} {p.
  arXiv:1802.01539}

\bibitem[\protect\citeauthoryear{{White} \& {Rees}}{{White} \&
  {Rees}}{1978}]{whiterees1978}
{White} S.~D.~M.,  {Rees} M.~J.,  1978, \mn@doi [\mnras]
  {10.1093/mnras/183.3.341}, \href
  {https://ui.adsabs.harvard.edu/abs/1978MNRAS.183..341W} {183, 341}

\bibitem[\protect\citeauthoryear{{Zentner}, {Hearin}  \& {van den
  Bosch}}{{Zentner} et~al.}{2014}]{zenter2014}
{Zentner} A.~R.,  {Hearin} A.~P.,   {van den Bosch} F.~C.,  2014, \mn@doi
  [\mnras] {10.1093/mnras/stu1383}, \href
  {https://ui.adsabs.harvard.edu/abs/2014MNRAS.443.3044Z} {443, 3044}

\bibitem[\protect\citeauthoryear{{Zheng} et~al.,}{{Zheng}
  et~al.}{2005}]{zheng05}
{Zheng} Z.,  et~al., 2005, \mn@doi [\apj] {10.1086/466510}, 633, 791

\bibitem[\protect\citeauthoryear{{de Jong} et~al.,}{{de Jong}
  et~al.}{2014}]{deJong14}
{de Jong} R.~S.,  et~al., 2014, in Ground-based and Airborne Instrumentation
  for Astronomy V. p. 91470M

\makeatother
\end{thebibliography}



\appendix

\section{eBOSS-SGC and DESI colour cuts}\label{App:colours}
\begin{figure}  \includegraphics[width=0.45\textwidth]{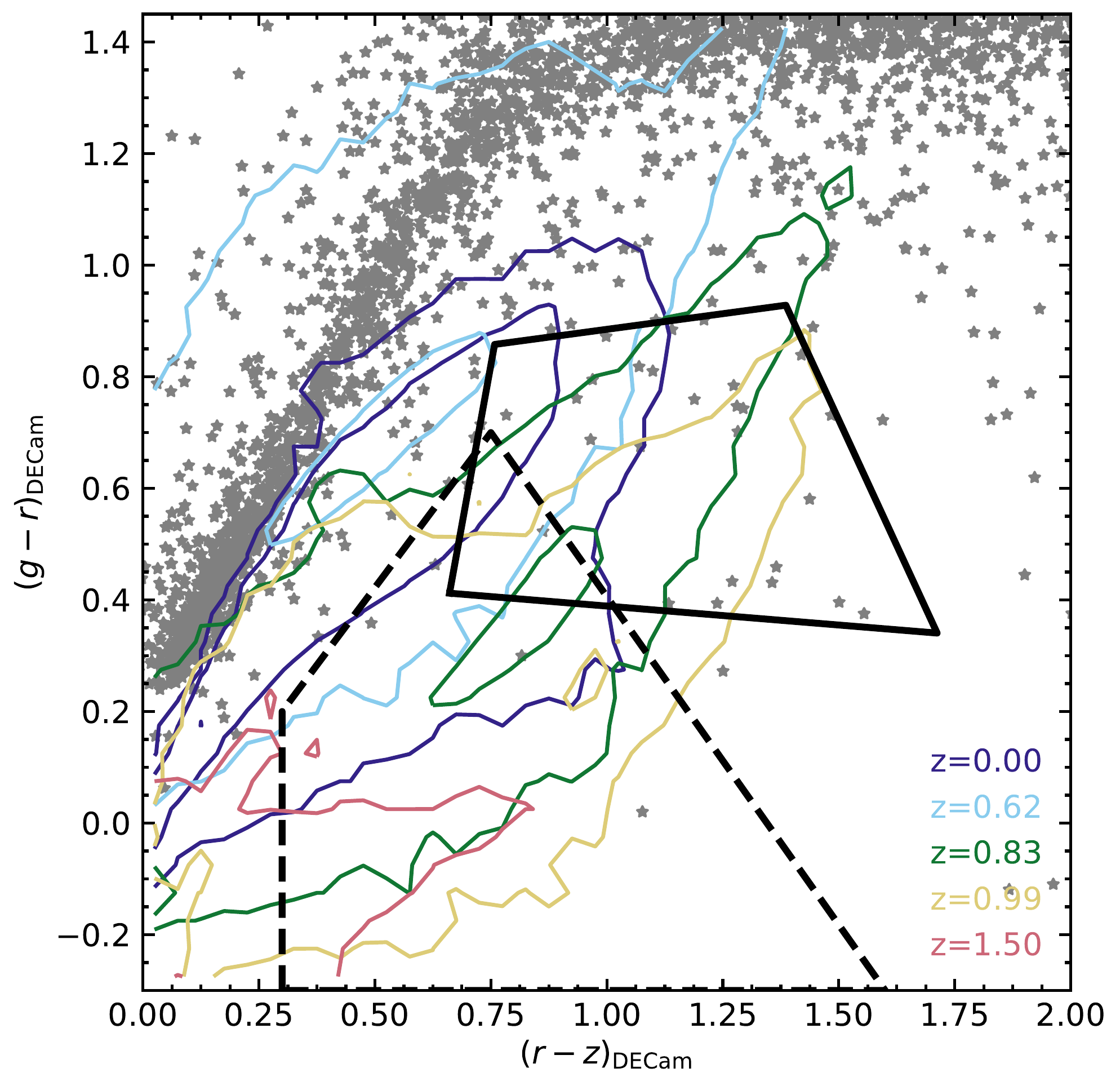}
\caption{\label{fig:colours} DECam (g-r) vs (r-z) parameter space with the isodensity lines at log$_{10}(\Phi/Mpc^{-3}h^{3}\rm{dlog}_{10}L)=-4.5, -1.5, 1$ for model galaxies, with ${\rm Flux}_{\rm [OII]}>8\times10^{-17}{\rm erg\, s}^{-1}{\rm cm}^{-2}$ at the redshifts indicated in the legend. The polygon with solid lines shows the eBOSS-SGC colour selection and the one with dashed lines that for DESI, as summarized in Table \ref{tbl:obs}. Note that the flux limit difference between the eBOSS-SGC and DESI selections is about a 20 per cent and thus, the distribution of model galaxies in this plot is very similar for both cuts, ${\rm Flux}_{\rm [OII]}>8\cdot10^{-17}{\rm erg\, s}^{-1}{\rm cm}^{-2}$   and ${\rm Flux}_{\rm [OII]}>10^{-16}{\rm erg\, s}^{-1}{\rm cm}^{-2}$. The location of stars are shown by the grey symbols.
} 
\end{figure}

Fig. \ref{fig:colours} presents the location of model galaxies with ${\rm Flux}_{\rm [OII]}>8\cdot10^{-17}{\rm erg\, s}^{-1}{\rm cm}^{-2}$ at redshifts $z=0.62$, $0.83$, $1.$, $1.5$ in the $(g-r)_{\rm DECam}$ vs. $(r-z)_{\rm DECam}$, colour-colour space. These distributions are compared to the location of stars~\citep{stars}, grey filled symbols in Fig. \ref{fig:colours} and to the regions delimited by the eBOSS-SGC~\citet{raichoor17} and DESI colour cuts~\citet{desi1}. These colour cuts are summarized in Table~\ref{tbl:obs}. The colours of model galaxies are roughly consistent with the regions defined for eBOSS-SGC \citet{comparat15eboss} and DESI \citep{desi1} to select ELGs in the range $0.6<z<1.7$. Further details on the colour cuts can be found in \citet{gp18}.

\section{Resolution and threshold checking}\label{App:check}

\begin{table}
\caption{The fraction in the simulation volume of each large-scale structures with different mesh numbers. Similar fractions were found in \citet{Cui2019}.}
    \label{tab:web_fraction}
    \centering
    \begin{tabular}{c|c|c|c|c|c}
       mesh number  & method & Knot & Filament & Sheet & Void \\
        $256^3$ & \vweb & 0.026 & 0.209 & 0.469 & 0.296 \\
        $256^3$ & \pweb & 0.027 & 0.252 & 0.490 & 0.227 \\
        \hline
        $512^3$ & \vweb & 0.029 & 0.227 & 0.474 & 0.270 \\
        $512^3$ & \pweb & 0.021 & 0.242 & 0.510 & 0.226 \\
    \end{tabular}
\end{table}
\begin{figure*}
 \includegraphics[width=0.8\textwidth]{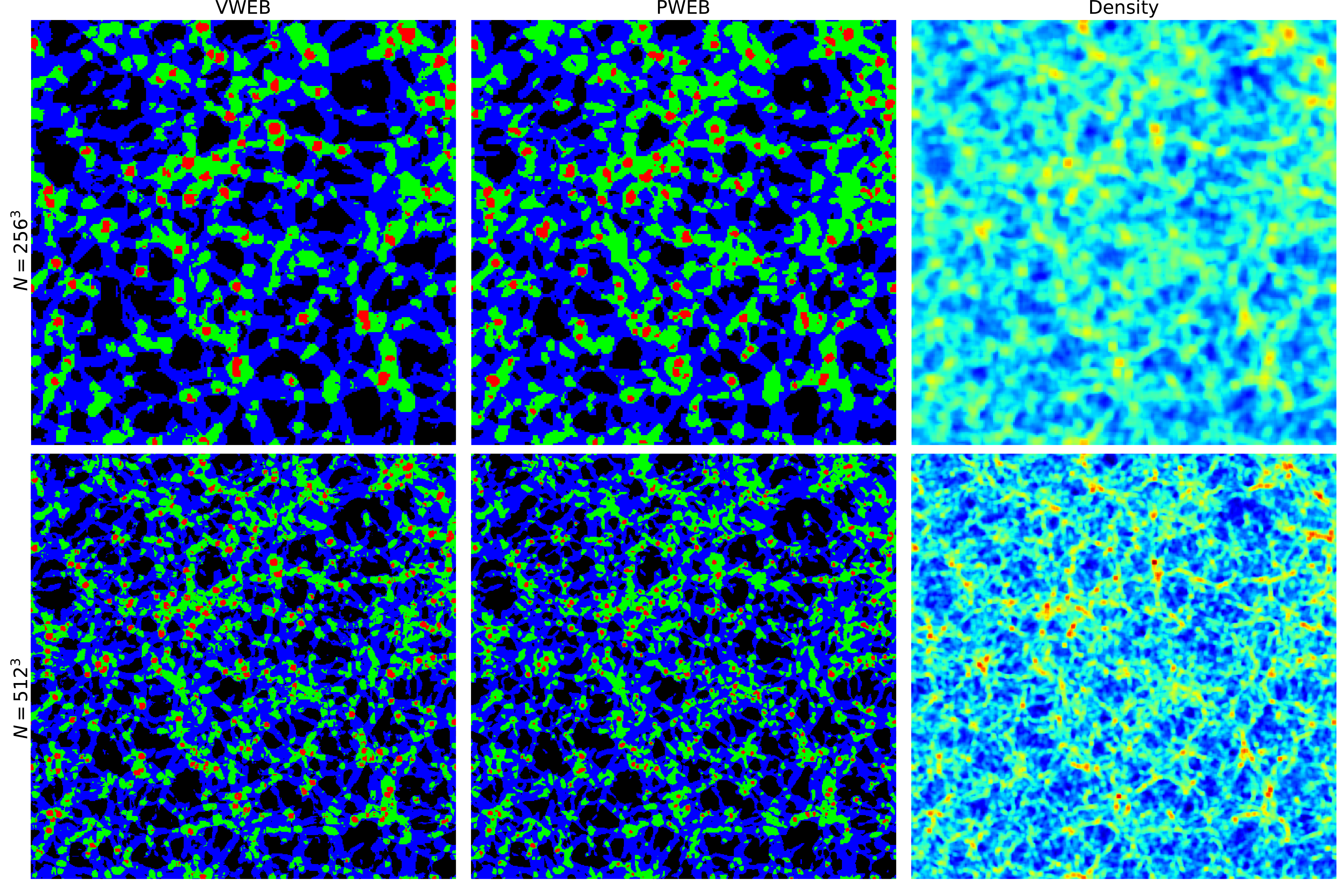}
 \caption{
 The projected structures for a slice of the simulation at $z=1$ with $256^3$ meshes (upper panels, $\sim 2 \Mpc$ thickness) and $512^3$ meshes (lower panels, $\sim 1 \Mpc$ thickness). The large-scale structures: knots, filaments, sheets and voids regions are shown as red, green, blue and black colours, respectively. From left-hand panel to right-hand panel, we show the results from \vweb and \pweb as well as the density fields. The thresholds applied are $\lambda_{th} = 0.1$ for \vweb and $\lambda_{th} = 0.005s^{-2}$ for \pweb meshes.}\label{fig:com_mesh}
\end{figure*}

In Fig.~\ref{fig:com_mesh}, we compare the results with two different mesh numbers $256^3$ (corresponding to a cell size of $\sim 2 \Mpc$) and $512^3$ ($\sim 1 \Mpc$ cell size). Both meshes adopt the referenced thresholds: $\lambda^V_{th} = 0.1$ for \vweb and $\lambda^P_{th} = 0.005s^{-2}$ for \pweb. The smoothing length in all cases is set to $5h^{-1}$Mpc (see \S~\ref{sec:cosmicweb}). It is clear that more details are revealed with the finer meshes. However, we do not go beyond the $512^3$ number of meshes, as finer mesh cells will have less particles which will provide noisier fields. We confirm here that with these two thresholds we have very similar volume fractions (see Tabel~\ref{tab:web_fraction} for details) between \vweb and \pweb classified large-scale structures. Similar fractions are also found assuming $\lambda^P_{th} = 0.00346s^{-2}$. Furthermore, we can see that these classified large-scale structures with both methods match the density fields shown in the right-hand side panels well.

We further investigated the effects of varying the two thresholds within two times of the reference values. They are either too large -- $\lambda^V_{th} = 0.05$ ($\lambda^P_{th} = 0.0025s^{-2}$) -- with more knots regions, or too small -- $\lambda^V_{th} = 0.2$ ($\lambda^P_{th} = 0.1s^{-2}$) -- with more space is occupied by Void. However, it is interesting to see that our main conclusions are basically unchanged.

\section{Large-scale environment with \pweb}\label{App:pweb}
\begin{figure*}
    \begin{subfigure}[b]{\textwidth}
    \begin{center}
    \includegraphics[width=.3\textwidth]{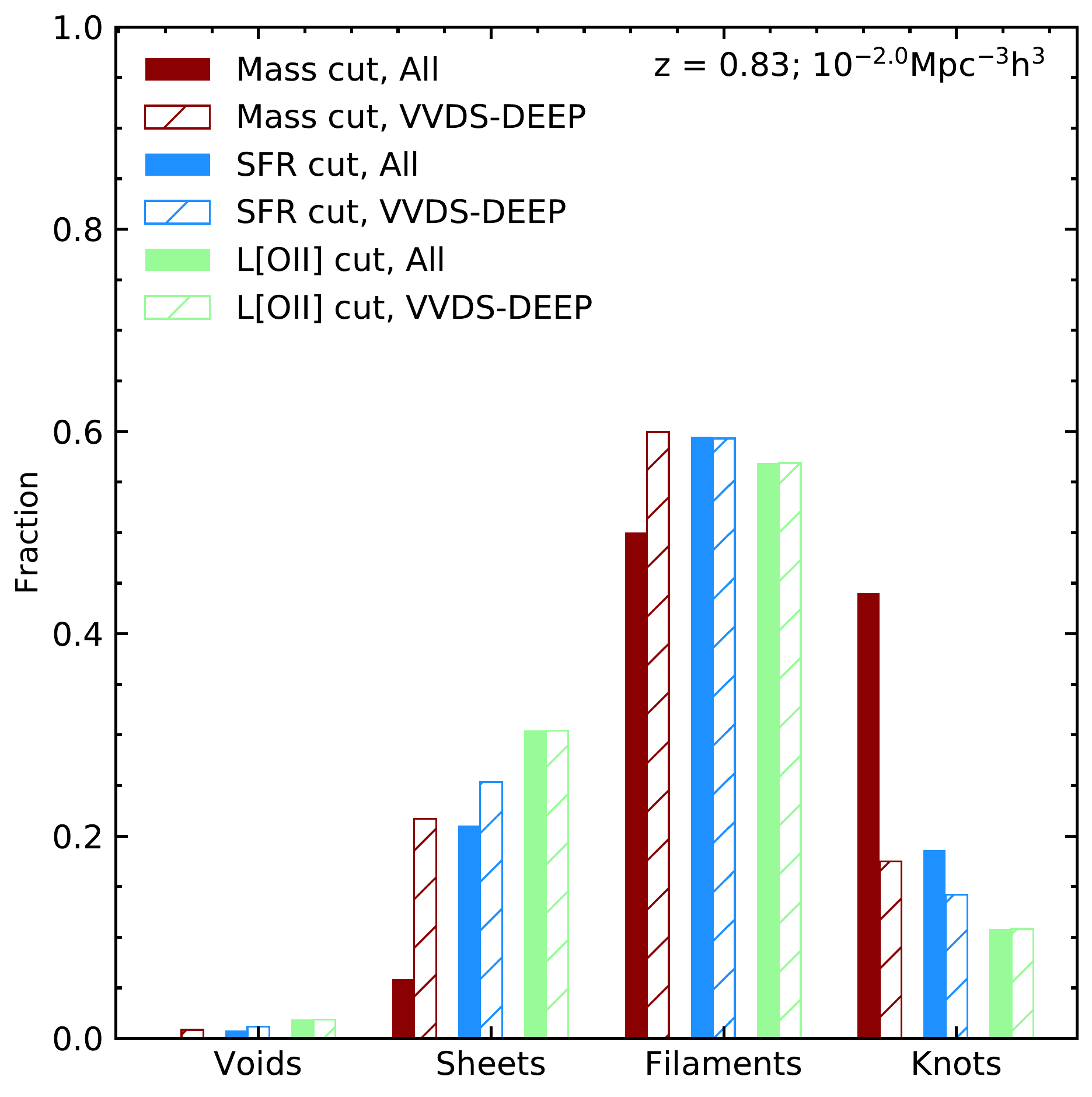}
    \includegraphics[width=.3\textwidth]{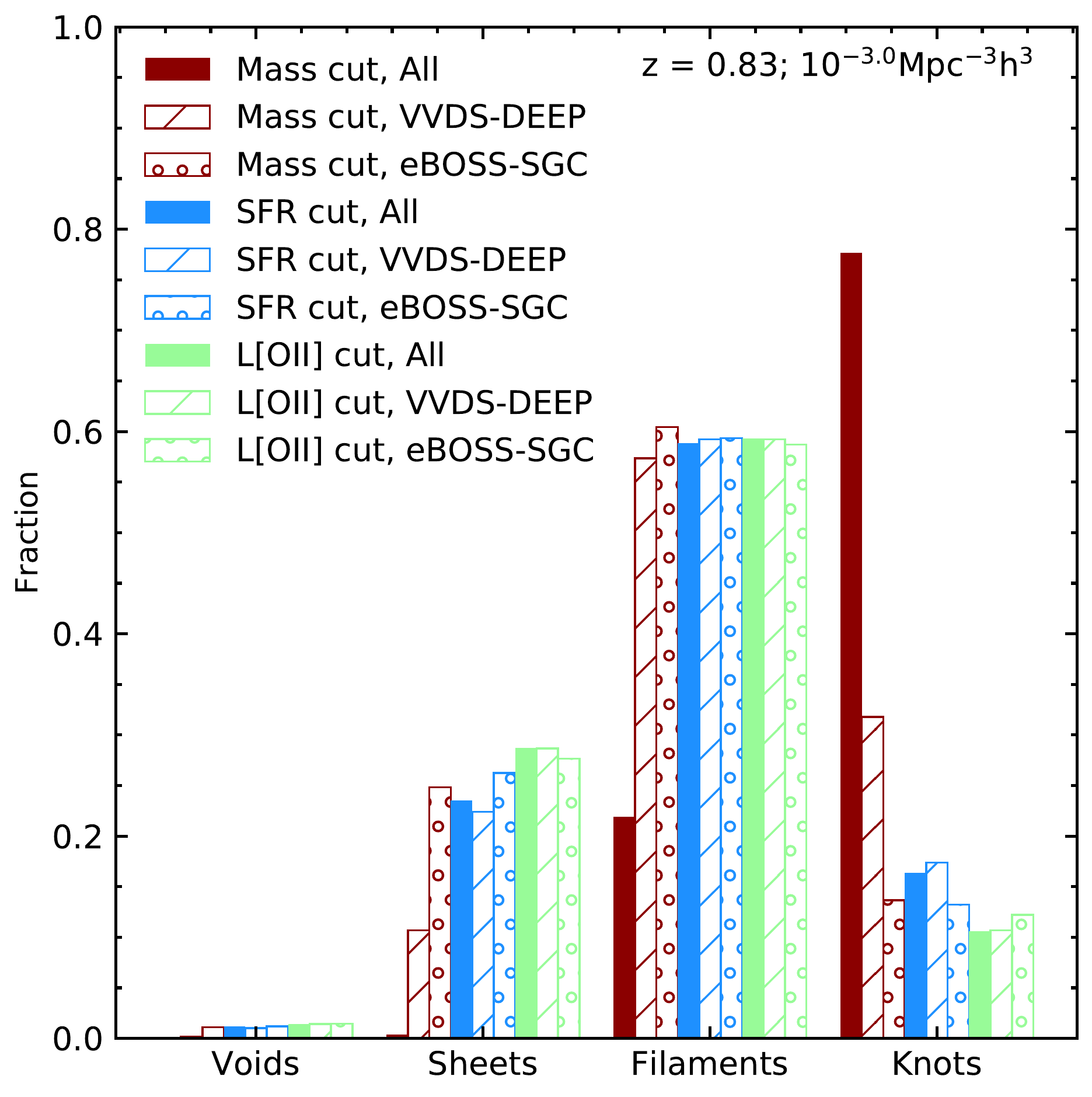}
    \includegraphics[width=.3\textwidth]{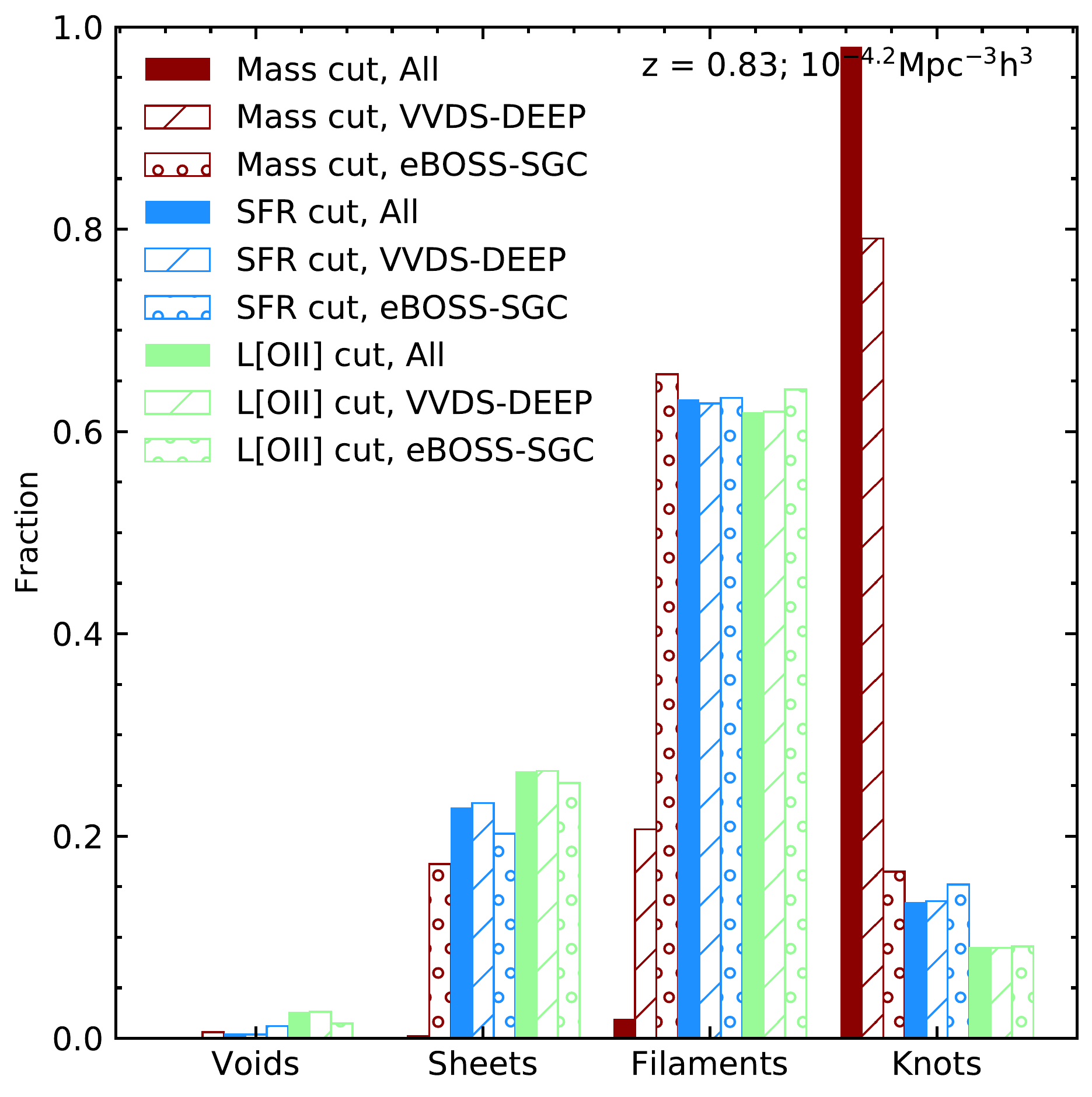}
    \end{center}
    \end{subfigure}

    \begin{subfigure}[b]{\textwidth}
    \begin{center}
    \includegraphics[width=.3\textwidth]{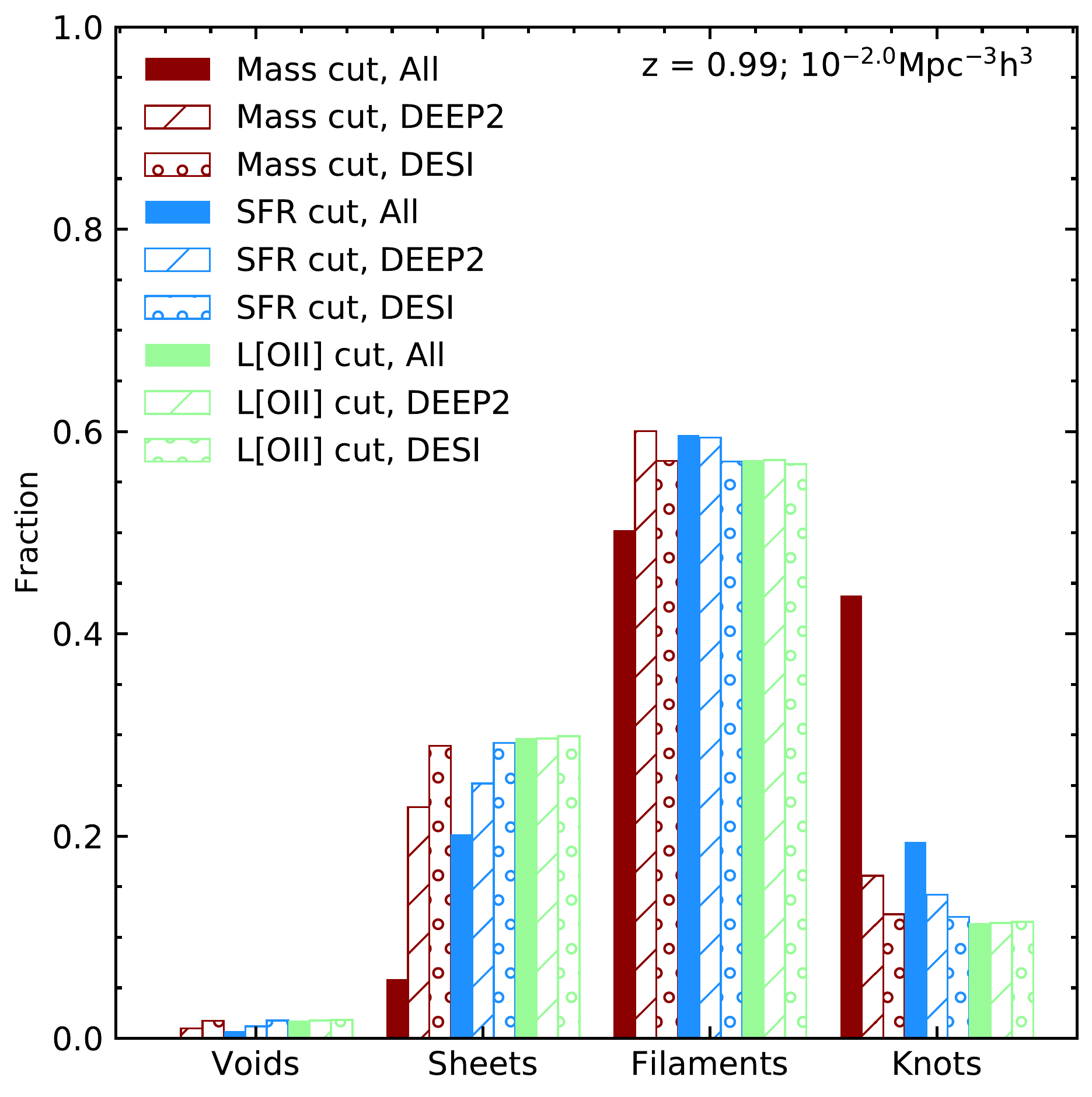}
    \includegraphics[width=.3\textwidth]{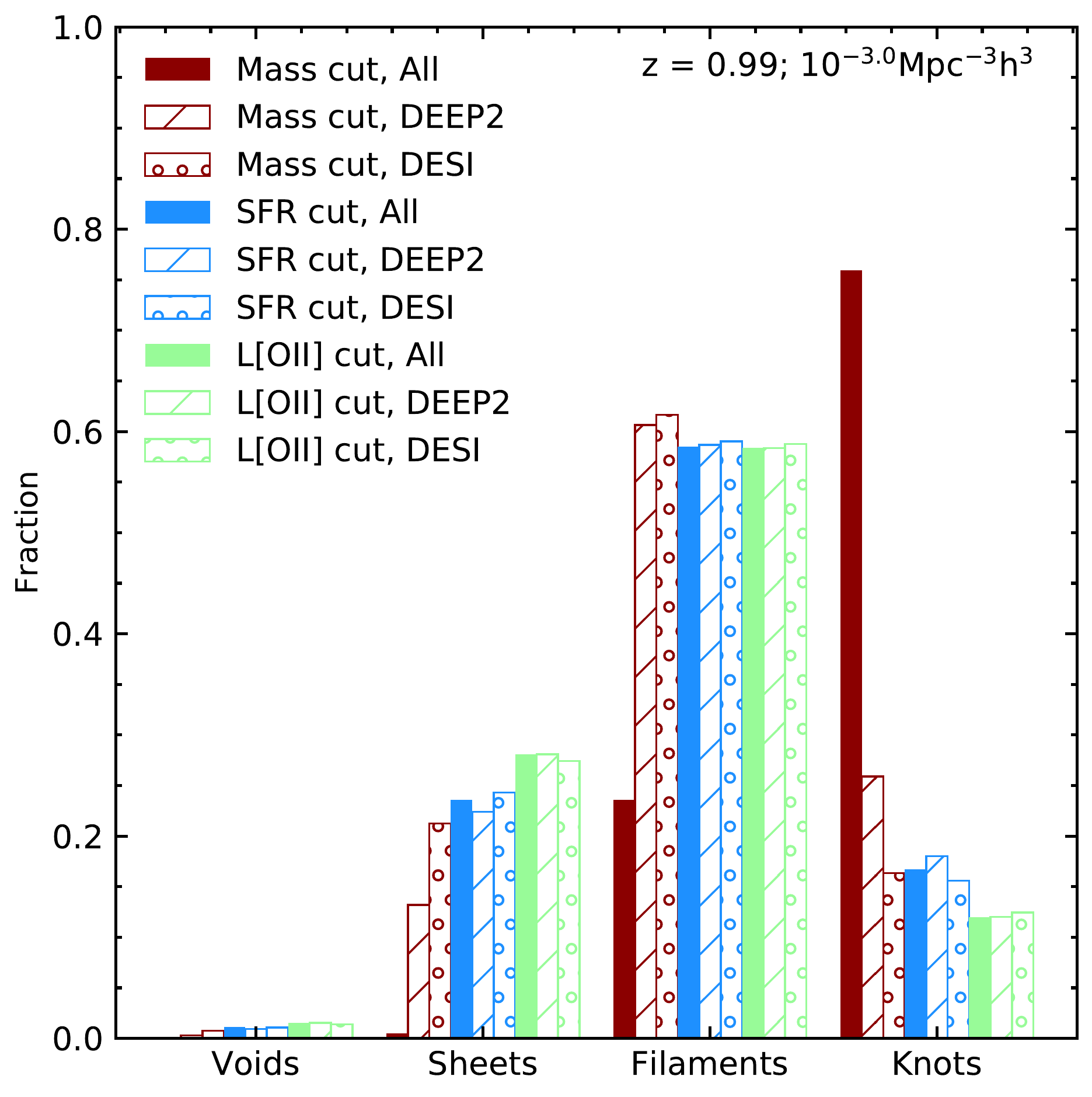}
    \includegraphics[width=.3\textwidth]{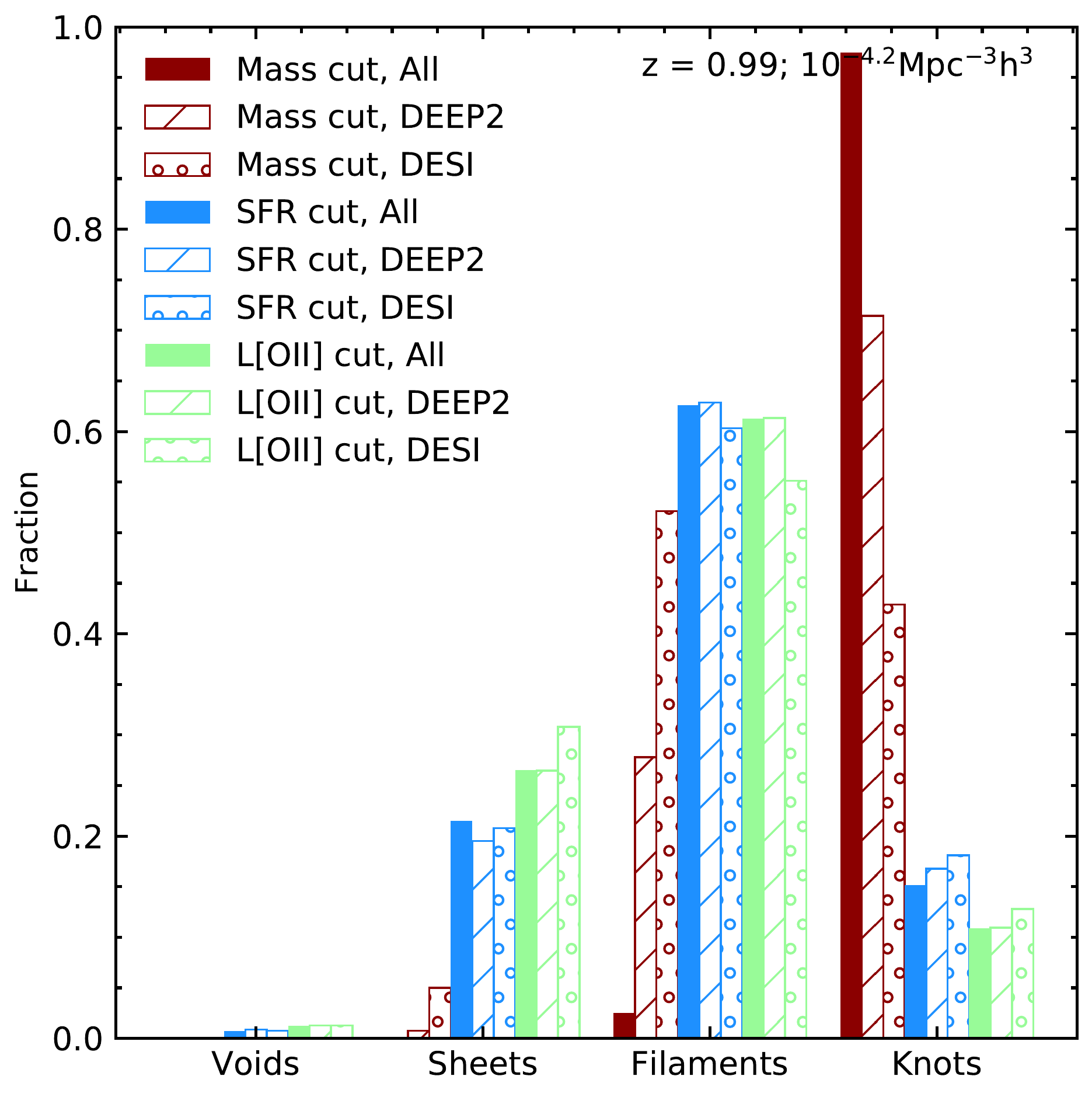}
    \end{center}
    \end{subfigure}
\caption{Histograms with the fraction of galaxies in voids, sheets, filaments and knots, similar to Fig.~\ref{fig:environ}. In this case the \LSE has been classified using \pweb (see \S~\ref{sec:LSE}). 
} 
\end{figure*}\label{fig:envPweb}

Fig.~\ref{fig:envPweb} shows the fraction of galaxies in voids, sheets, filaments and knots when the \LSE is classified using the \pweb algorithm described in \S~\ref{sec:LSE}, with a $0.005s^{-2}$ threshold.  Very similar fractions are found when imposing $0.00346s^{-2}$ and $0.01s^{-2}$ thresholds. This figure is qualitatively equivalent to Fig.~\ref{fig:environ}, in terms of global trends. However, quantitatively there are differences that become more pronounced for the samples with the lowest number density, in particular for the mass selected ones.

\bsp	
\label{lastpage}
\end{document}